\documentclass[reprint,superscriptaddress,amssymb,amsmath,aps,prx,longbibliography]{revtex4-2}

\usepackage{graphicx}
\usepackage{dcolumn}
\usepackage{bm}
\usepackage{siunitx}
\usepackage{upgreek}
\usepackage[breaklinks,colorlinks, citecolor=blue, urlcolor=blue]{hyperref}

\begin{document}

\title{Optomechanical~ground-state~cooling~in~a~continuous and~efficient~electro-optic~transducer}

\author{B.~M.~Brubaker}\thanks{These authors contributed equally}
\email[direct correspondence to ]{maxwell.urmey@colorado.edu }
\affiliation{JILA, National Institute of Standards and Technology and the University of Colorado, Boulder, Colorado 80309, USA}
\affiliation{Department of Physics, University of Colorado, Boulder, Colorado 80309, USA}
\author{J.~M.~Kindem}\thanks{These authors contributed equally}
\email[direct correspondence to ]{maxwell.urmey@colorado.edu }
\affiliation{JILA, National Institute of Standards and Technology and the University of Colorado, Boulder, Colorado 80309, USA}
\affiliation{Department of Physics, University of Colorado, Boulder, Colorado 80309, USA}
\author{M.~D.~Urmey}\thanks{These authors contributed equally}
\email[direct correspondence to ]{maxwell.urmey@colorado.edu }
\affiliation{JILA, National Institute of Standards and Technology and the University of Colorado, Boulder, Colorado 80309, USA}
\affiliation{Department of Physics, University of Colorado, Boulder, Colorado 80309, USA}
\author{S.~Mittal}
\affiliation{JILA, National Institute of Standards and Technology and the University of Colorado, Boulder, Colorado 80309, USA}
\affiliation{Department of Physics, University of Colorado, Boulder, Colorado 80309, USA}
\author{R.~D.~Delaney}
\affiliation{JILA, National Institute of Standards and Technology and the University of Colorado, Boulder, Colorado 80309, USA}
\affiliation{Department of Physics, University of Colorado, Boulder, Colorado 80309, USA}
\author{P.~S.~Burns}
\affiliation{JILA, National Institute of Standards and Technology and the University of Colorado, Boulder, Colorado 80309, USA}
\affiliation{Department of Physics, University of Colorado, Boulder, Colorado 80309, USA}
\author{M.~R.~Vissers}
\affiliation{National Institute of Standards and Technology, Boulder, Colorado 80305, USA}
\author{K.~W.~Lehnert}
\affiliation{JILA, National Institute of Standards and Technology and the University of Colorado, Boulder, Colorado 80309, USA}
\affiliation{Department of Physics, University of Colorado, Boulder, Colorado 80309, USA}
\affiliation{National Institute of Standards and Technology, Boulder, Colorado 80305, USA}
\author{C.~A.~Regal}
\affiliation{JILA, National Institute of Standards and Technology and the University of Colorado, Boulder, Colorado 80309, USA}
\affiliation{Department of Physics, University of Colorado, Boulder, Colorado 80309, USA}

\date{\today}

\begin{abstract}
\noindent The demonstration of a quantum link between microwave and optical frequencies would be an important step towards the realization of a quantum network of superconducting processors.
A major impediment to quantum electro-optic transduction in all platforms explored to date is noise added by thermal occupation of modes involved in the transduction process, and it has proved difficult to realize low thermal occupancy concurrently with other desirable features like high duty cycle and high efficiency. In this work, we present an efficient and continuously operating electro-optomechanical transducer whose mechanical mode has been optically sideband-cooled to its quantum ground state. The transducer achieves a maximum efficiency of 47\% and minimum input-referred added noise of 3.2 photons in upconversion. Moreover, the thermal occupancy of the transducer's microwave mode is minimally affected by continuous laser illumination with power more than two orders of magnitude greater than that required for optomechanical ground-state cooling.
\end{abstract}

\maketitle

\section{Introduction}

The past two decades have seen remarkable progress in the control of increasingly complex quantum systems, towards the ultimate goal of realizing a general-purpose quantum computer \cite{ladd_quantum_2010}. Superconducting electrical circuits operating at microwave frequencies and temperatures below 100~mK have emerged as a leading platform for quantum computing \cite{arute_quantum_2019,campagne_quantum_2020}, but microwave photons have energies much smaller than thermal energy at ambient temperature, and thus cannot be used to mediate the exchange of quantum information between distant quantum processors. Optical photons, conversely, are uniquely well-suited for the transmission of quantum information over long distances at room temperature \cite{liao_satellite-relayed_2018,chen_sending-or-not-sending_2020}. A quantum state-preserving interface linking microwave and optical frequencies, or quantum transducer, is thus a critical element in a future quantum internet \cite{wehner_quantum_2018}. 

Of the various figures of merit for quantum transduction~\cite{zeuthen_figures_2020}, perhaps the most important is the input-referred added noise $N_\text{add} = N_\text{out}/\eta_\text{t}$, where $N_\text{out}$ is the noise measured at the transducer output in photon units and $\eta_\text{t}$ is the transducer efficiency. The utility of $N_\text{add}$ as a performance metric stems from the fact that it can be compared directly to the signal to be transduced. Although high efficiency is not strictly a requirement for all transduction protocols~\cite{zeuthen_figures_2020}, achieving $N_\text{add} \lesssim 1$ photon requires high efficiency insofar as $N_\text{out}$ cannot be made extremely small. Thermal occupancy of any mode involved in the transduction process will contribute to $N_\text{add}$, and thus an ideal transducer should operate with every participating mode in its quantum ground state.

In all systems that have been explored as potential platforms for quantum electro-optic transduction \cite{lambert_coherent_2020,lauk_perspectives_2020}, ground-state operation is made challenging by the presence of laser light required to bridge the energy gap between microwave and optical photons. Laser illumination can lead to bulk heating of various materials used in transducers \cite{mirhosseini_superconducting_2020,hease_bidirectional_2020}, and the superconducting microwave circuits integral to most transducer architectures are especially susceptible. High-energy optical photons impinging on the superconductor will break Cooper pairs, increasing the thermal occupancy of the microwave mode and adding noise to the upconverted quantum state \cite{witmer_silicon-organic_2020,mirhosseini_superconducting_2020}.

One approach to mitigating the detrimental effects of laser illumination is to pulse transducer operation, at the cost of a reduced duty cycle. 
Using this method, several recent experiments have demonstrated ground-state operation  \cite{forsch_microwave--optics_2020,hease_bidirectional_2020,fu_ground-state_2020,stockill_ultra-low-noise_2021} and even added noise $N_\text{add,up} < 1$ in upconversion \cite{mirhosseini_superconducting_2020,sahu_quantum-enabled_2021}. However, achieving this performance has generally required low duty cycles $< 10^{-3}$ and low efficiency $\eta_\text{t} < 10^{-5}$, with two exceptions with better performance in only one of these metrics \cite{hease_bidirectional_2020,sahu_quantum-enabled_2021}. Duty cycle and efficiency are both important figures of merit for certain applications, such as the generation of entangled microwave/optical photon pairs~\cite{barzanjeh_entangling_2011,zhong_proposal_2020,rau_entanglement_2021}, but thus far continuous operation with high efficiency has been demonstrated only in an electro-optomechanical platform that used a low-frequency mechanical mode to mediate transduction \cite{higginbotham_harnessing_2018}. This transducer suffered from multiple sources of added noise including the residual thermal occupancy of the mechanical mode.

\begin{figure}[!t]
    \centering
    \includegraphics{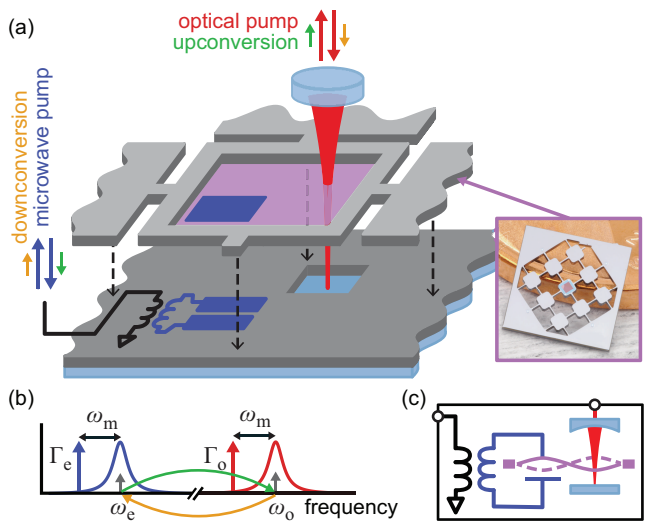}
    \caption{
    \textbf{Transducer design and operation.}
    (a) Illustration of transducer (not to scale). A vibrational mode of a silicon nitride membrane (purple) couples to a microwave-frequency inductor-capacitor circuit (dark blue) and a Fabry-P\'erot optical cavity mode (red) confined between two high-reflectivity mirrors (light blue).
   The membrane, microwave circuit, and one of the cavity mirrors are integrated into a stack of three silicon chips. Microwave and optical pumps (blue and red arrows) enhance the electromechanical and optomechanical interactions, respectively, and allow upconversion of a microwave signal (green arrow) or downconversion of an optical signal (orange arrow). The inset shows the membrane (false-colored purple) suspended on a phononically engineered silicon substrate (gray).
   (b) Transducer frequency scales. The microwave and optical pumps are red-detuned by the mechanical frequency $\omega_\text{m}$ from microwave and optical resonances $\omega_\text{e}$ and $\omega_\text{o}$, respectively. The electromechanical and optomechanical damping rates $\Gamma_\text{e}$ and $\Gamma_\text{o}$ scale with the corresponding pump powers. (c) Simplified diagram in which the transducer is represented as a two-port network.}
    \label{fig:schematic}
\end{figure}

In this work, we demonstrate continuous and efficient electro-optic transduction mediated by a mechanical mode that has been optomechanically sideband cooled to its quantum ground state, reaching a minimum occupancy of 0.34~phonons.
We operate this transducer in upconversion, and achieve input-referred added noise as low as $N_\text{add,up} = 3.2$ photons and a maximum efficiency $\eta_\text{t} = 47\%$.
The residual added noise is dominated by fluctuations introduced by the strong microwave pump that couples the mechanical mode to a superconducting microwave circuit, and could be reduced by increasing the vacuum electromechanical coupling rate. Notably, however, laser light near the superconducting circuit is not a significant contribution to the added noise, even with power more than two orders of magnitude greater than that required for optomechanical ground-state cooling. Our results lay the groundwork for quantum transduction and continuous entanglement generation using a membrane-based electro-optic interface with improved microwave circuit performance.

\section{Transducer Design and Operation}

The transducer, illustrated in Fig.~\ref{fig:schematic}(a), consists of microwave and optical resonators dispersively coupled to a single vibrational mode of a silicon nitride membrane suspended on a silicon chip \cite{andrews_bidirectional_2014,higginbotham_harnessing_2018}. The silicon chip substrate is patterned into a phononic shield (\cite{yu_phononic_2014,tsaturyan_demonstration_2014}, inset in Fig.~\ref{fig:schematic}(a)) with a bandgap centered on the resonant frequency of the membrane mode of interest, $\omega_\text{m}/2\pi = 1.45$~MHz. Our phononic shield has relatively few unit cells to enable integration into the transducer, but it nonetheless reduces acoustic radiation from the membrane into the substrate. The improved isolation results in a lower density of substrate modes near the membrane mode used for transduction and a low intrinsic energy dissipation rate $\gamma_\text{m}/2\pi = 113$~mHz.

The microwave resonator is a superconducting niobium titanium nitride (NbTiN) inductor-capacitor (LC)~circuit in a flip-chip configuration, whose capacitance is modulated by the motion of the membrane~\cite{andrews_bidirectional_2014}. The circuit has resonant frequency $\omega_\text{e}/2\pi = 7.938$~GHz and is coupled to a transmission line at rate $\kappa_\text{e,ext}/2\pi = 1.42$~MHz. Its total linewidth $\kappa_\text{e}$ increases from 1.64~MHz to 2.31~MHz as the microwave power incident on the circuit increases.

The optical resonator is a single-sided Fabry-P\'erot cavity (resonant frequency $\omega_\text{o}/2\pi = 277$~THz) formed by a curved mirror on a superpolished fused-silica substrate opposite a Bragg mirror stack deposited on a silicon chip bonded to the flip-chip assembly. This cavity architecture ensures robust alignment of the membrane to the cavity axis, suppressing scattering losses. As a result, the total cavity linewidth $\kappa_\text{o}/2\pi = 2.68$~MHz is dominated by its external coupling $\kappa_\text{o,ext}/2\pi = 2.12$~MHz through the curved mirror.

The transducer is operated in an optical-access dilution refrigerator with a 40~mK base plate temperature. Device design and fabrication details are provided in Appendix~\ref{sec:device}, and device characterization is discussed further in Appendix~\ref{sec:characterization}.

\begin{figure*}
    \centering
    \includegraphics{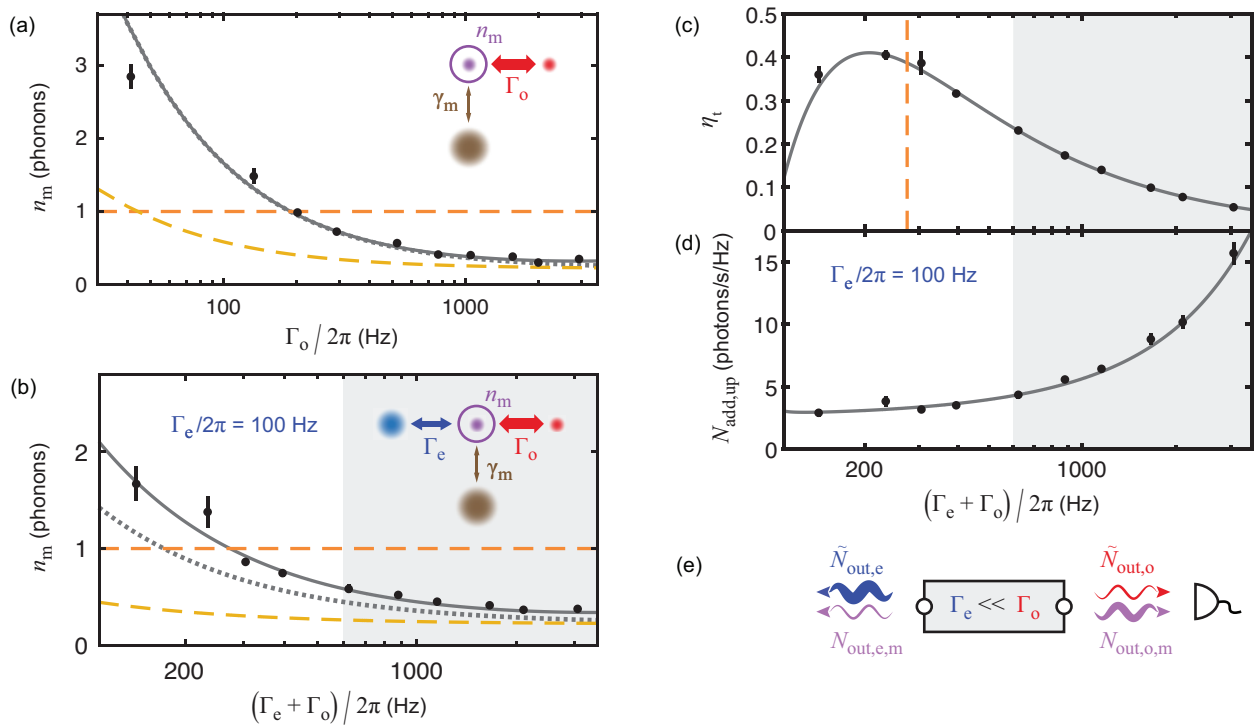}
    \caption{
    \textbf{Ground-state cooling of electro-optomechanical transducer.} 
    (a) Membrane mode occupancy $n_\text{m}$ vs.\ optomechanical damping rate $\Gamma_\text{o}$ with electromechanical damping $\Gamma_\text{e} = 0$. 
    Points are data, solid gray line is a fit, and dotted gray line shows the predicted behavior in the absence of technical noise. 
    Orange dashed line indicates $n_\text{m} = 1$, and yellow dashed line indicates the backaction limit (increase at low $\Gamma_\text{o}$ due to backaction from auxiliary optical cavity lock beam). 
    Inset illustrates optomechanical sideband cooling, where external blurred discs represent baths (with size indicating occupancy) to which the membrane mode (purple circle) is coupled by the corresponding damping rates (arrows, with width indicating magnitude). 
    The membrane mode occupancy (purple blurred disk) is determined by the weighted average of bath occupancies.
    (b) $n_\text{m}$ vs.\ $\Gamma_\text{e} + \Gamma_\text{o}$, sweeping $\Gamma_\text{o}$ at fixed electromechanical damping $\Gamma_\text{e}/2\pi = 100$~Hz, with curves color-coded as in panel~(a). 
    Inset illustrates relative bath occupancies and damping rates for electro-optomechanical sideband cooling in the gray shaded region where $\Gamma_\text{e} \ll \Gamma_\text{o}$. (c) Transducer efficiency $\eta_\text{t}$ vs.\ $\Gamma_\text{e} + \Gamma_\text{o}$, sweeping $\Gamma_\text{o}$ as in (b), with fit. 
    Orange dashed line marks the damping at which $n_\text{m} = 1$. (d) Upconversion added noise $N_\text{add,up}$ vs.\ $\Gamma_\text{e} + \Gamma_\text{o}$, with theory curve. 
    (e) Schematic representation of transducer output noise in the $\Gamma_\text{e} \ll \Gamma_\text{o}$ regime shaded gray in panels~(b), (c), and (d).
    The box represents the transducer, and wavy arrows represent contributions to the microwave and optical output noise around mechanical resonance, from membrane mode occupancy (purple) and technical noise (red/blue), with thickness indicating noise density. Photodetector indicates optical readout. All error bars represent one standard deviation.
    }
    \label{fig:gscool}
\end{figure*}

To operate the transducer, we apply microwave and optical pumps red-detuned by $\omega_\text{m}$ from the respective resonators  (Fig.~\ref{fig:schematic}(b)). The microwave pump generates an electromechanical beamsplitter interaction wherein phonon excitations of the membrane mode can decay into propagating microwave photons at a rate $\Gamma_\text{e}$ proportional to the incident microwave pump power $P_\text{e}$, and the optical pump generates an analogous optomechanical damping rate $\Gamma_\text{o}$ proportional to the incident power $P_\text{o}$ (see Ref.~\cite{aspelmeyer_cavity_2014} and Appendix~\ref{sec:theory}). These beamsplitter interactions enable bidirectional transduction between microwave and optical frequencies over a  bandwidth given by $\Gamma_\text{T} = \Gamma_\text{e} + \Gamma_\text{o} + \gamma_\text{m}$~\cite{andrews_bidirectional_2014}. The transmission efficiency on resonance between the external ports of the microwave and optical resonators (Fig.~\ref{fig:schematic}(c)) is given by
\begin{equation}
    \eta_\text{t} = \eta_\text{M}\frac{4\Gamma_\text{e}\Gamma_\text{o}}{\Gamma_\text{T}^2},
    \label{eq:eta}
\end{equation}
where the matched efficiency $\eta_\text{M} = \epsilon\frac{\kappa_\text{o,ext}}{\kappa_\text{o}}\frac{\kappa_\text{e,ext}}{\kappa_\text{e}}$ is the maximum efficiency achievable in the presence of imperfect optical cavity modematching $\epsilon$ and loss in the electromagnetic resonators. The transducer achieves this maximum efficiency when $\Gamma_\text{e}=\Gamma_\text{o} \gg \gamma_\text{m}$. 

In thermal equilibrium at temperature $T_\text{eq}$, the membrane mode is coupled to a bath with occupancy $n_\text{th} = k_\text{B}T_\text{eq}/\hbar\omega_\text{m}$ by its intrinsic decay rate $\gamma_\text{m}$.  Optomechanical damping couples the membrane mode to a bath with occupancy $n_\text{o} = n_\text{min,o}\,+\,n_\text{eff,o} \ll n_\text{th}$, where the first term represents the quantum backaction limit \cite{peterson_laser_2016} from imperfect sideband resolution and the second term is an effective optical mode thermal occupancy due to pump power-dependent technical noise (see Appendix~\ref{sec:theory}). Analogous effects contribute to the electromechanical bath occupancy $n_\text{e}$. The membrane mode occupancy $n_\text{m}$ is given by the weighted average of the baths to which it is coupled,
\begin{equation}
    n_\text{m} = \frac{\gamma_\text{m}n_\text{th} + \Gamma_\text{e}n_\text{e} + \Gamma_\text{o}n_\text{o}}{\Gamma_\text{T}},
    \label{eq:gscool}
\end{equation}
leading to sideband cooling of the membrane mode (Fig.~\ref{fig:gscool}(a) and (b), insets).

The optomechanical interaction enables readout of membrane motion in addition to sideband cooling and transduction. Random motion of the membrane is imprinted on the reflected optical pump in the form of amplitude and phase fluctuations. These fluctuations manifest as a pair of sidebands with Lorentzian profiles centered at detunings $\pm\omega_\text{m}$ in the noise spectral density at the transducer's optical output port (see Appendix~\ref{sec:theory}), which we measure using balanced heterodyne detection. We can then infer the membrane mode occupancy $n_\text{m}$ from the ratio of upper and lower sideband amplitudes \cite{safavi-naeini_observation_2012,peterson_laser_2016} and measure the transducer's added noise in upconversion. The analogous electromechanical readout can also be used to measure $n_\text{m}$ as well as added noise in downconversion, albeit with lower signal-to-noise ratio. 

Although the transducer is inherently bidirectional~\cite{andrews_bidirectional_2014}, upconversion is likely to be more useful in practice, because superconducting processors enable the synthesis of arbitrary quantum states in the microwave domain~\cite{vlastakis_deterministically_2013}, and heralded entanglement of remote quantum processors can be achieved using upconversion and optical detection only~\cite{krastanov_optically_2021}. In this work we measure the transducer's added noise in upconversion only, and also use optomechanical readout for all measurements of the membrane mode occupancy.

\section{Optomechanical ground-state Cooling}

We first demonstrate optomechanical ground-state cooling of the membrane mode with the microwave pump absent (Fig.~\ref{fig:gscool}(a)), to show that our transducer realizes the successes of past membrane optomechanical systems \cite{peterson_laser_2016,underwood_measurement_2015} even with the additional design complexity required for electromechanical coupling to superconducting circuitry. We measure the optomechanical sideband ratio while varying the damping $\Gamma_\text{o}$, and correct for squashing effects due to laser phase noise (see Refs.~\cite{jayich_cryogenic_2012,safavi-naeini_laser_2013} and Appendices~\ref{sec:theory} and  \ref{sec:analysis}) to obtain the membrane mode occupancy $n_\text{m}$ at each value of $\Gamma_\text{o}$. We then fit this data to Eq.~\eqref{eq:gscool} with $\Gamma_\text{e} = 0$ and the equilibrium occupancy $n_\text{th}$ as a fit parameter. We also introduce an additional fit parameter $a_\text{o} = n_\text{eff,o}/\Gamma_\text{o}$ to model the optical mode occupancy due to phase noise, and include a fixed term to account for the effects of a weak near-resonant beam used to lock the optical cavity (see Appendix~\ref{sec:chain}). 

The fit (solid gray line) yields $n_\text{th} = 1000 \pm 90$, implying that the membrane mode equilibrates to $T_\text{eq} = 70$~mK in the presence of laser light, consistent with an independent calibration based on sweeping the temperature of the cryostat base plate (see Appendix~\ref{sec:analysis}). Phase noise parameterized by $a_\text{o} = (2.8 \pm 1.6)\times10^{-6}\text{ Hz}^{-1}$ causes the fit to deviate slightly at high damping from the behavior expected from thermal noise with the best-fit value of $n_\text{th}$ and imperfect sideband resolution alone (dotted gray line). The good agreement with the fit, without additional power-dependent terms, indicates that the laser-induced heating of the membrane mode responsible for its elevated equilibrium temperature $T_\text{eq}$ saturates quickly at very low power and does not preclude cooling below $n_\text{m} = 1$ (orange dashed line) for $\Gamma_\text{o}/2\pi > 190$~Hz. With large $\Gamma_\text{o}$, we reach a minimum mode occupancy of $n_\text{m} = 0.32$, within a factor of 1.5 of the backaction limit (yellow dashed line). 

To enable transduction, we now introduce a microwave pump with fixed electromechanical damping $\Gamma_\text{e}/2\pi = 100$~Hz, and again sweep $\Gamma_\text{o}$ in order to ground-state cool the membrane mode (Fig.~\ref{fig:gscool}(b)). We observe that $n_\text{m}$ deviates further from the expected behavior from thermal noise and imperfect sideband resolution alone (dotted gray line) than in the purely optomechanical case, indicating an additional source of technical noise. Fixing $n_\text{th}$ and $a_\text{o}$ at the best-fit values from the pure optomechanical ground-state cooling data, we find good agreement with Eq.~\eqref{eq:gscool} with effective thermal occupancy $n_\text{eff,e} = 0.8 \pm 0.1$, consistent with $n_\text{eff,e} = 0.77 \pm 0.01$  obtained from an independent heterodyne measurement of the noise in the transducer's microwave output spectrum (see Appendix~\ref{sec:paramnoise}). This good agreement, with $n_\text{eff,e}$ independent of laser power, indicates that the additional source of technical noise is coupled in via the microwave pump. With sufficiently high $\Gamma_\text{o}$, the membrane mode can again be cooled close to the backaction limit, reaching a minimum occupancy of $n_\text{m} = 0.34$.

\section{Transducer characterization}

We next consider how the  electro-optomechanical device performs as a transducer when the optomechanical damping $\Gamma_\text{o}$ is varied at constant electromechanical damping $\Gamma_\text{e}$, as in Fig.~\ref{fig:gscool}(b).
We measure the transducer efficiency $\eta_\text{t}$ using the method developed in Ref.~\cite{andrews_bidirectional_2014}, and fit the data to Eq.~\eqref{eq:eta} with the matched efficiency $\eta_\text{M}$ as the only fit parameter (Fig.~\ref{fig:gscool}(c)). The fit yields $\eta_\text{M,fit} = 43 \pm 1\%$, in tension with $\eta_\text{M,exp} = \epsilon\frac{\kappa_\text{o,ext}}{\kappa_\text{o}}\frac{\kappa_\text{e,ext}}{\kappa_\text{e}} = 55\%$ expected from independent measurements of the optical cavity modematching $\epsilon$ and the overcoupling fractions $\kappa_\text{ext}/\kappa$ of the two electromagnetic resonators. We suspect the measured values of $\eta_\text{t}$ were artificially suppressed (see Appendix~\ref{sec:analysis}) and that the true matched efficiency during these measurements was $\eta_\text{M,exp}$ rather than $\eta_\text{M,fit}$. Nonetheless, adopting the fit value as a conservative measure of transducer performance, we find $\eta_\text{t} = 38\%$ at the damping for which the membrane mode occupancy is reduced below $n_\text{m} = 1$.

\begin{figure}
    \centering
    \includegraphics{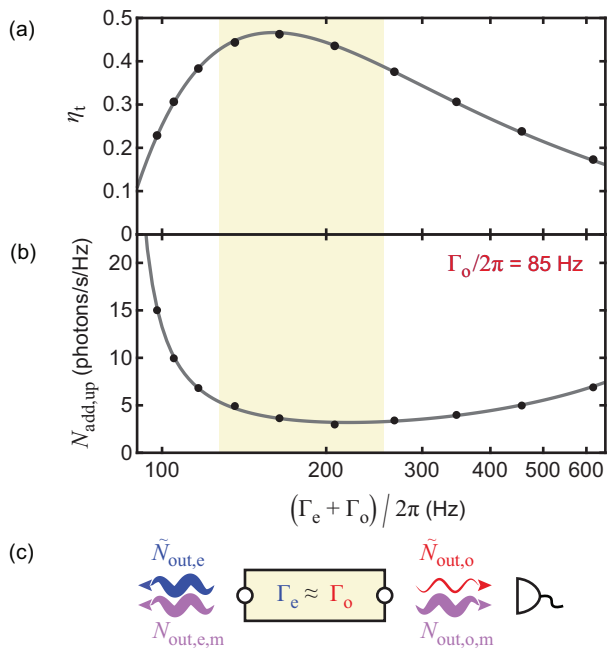}
    \caption{\textbf{Minimizing input-referred added noise.} 
    (a) Transducer efficiency $\eta_\text{t}$ vs.\  $\Gamma_\text{e} + \Gamma_\text{o}$, sweeping electromechanical damping $\Gamma_\text{e}$ at fixed optomechanical damping $\Gamma_\text{o}/2\pi = 85$~Hz, with fit. (b) Upconversion added noise $N_\text{add,up}$ vs.\ $\Gamma_\text{e} + \Gamma_\text{o}$ at $\Gamma_\text{o}/2\pi = 85$~Hz, with fit. (c) Schematic representation of transducer output noise in the $\Gamma_\text{e} \approx \Gamma_\text{o}$ regime shaded yellow in panels~(a) and (b), where the box represents the transducer, wavy arrows represent contributions to the microwave and optical output noise near mechanical resonance, and the photodetector indicates optical readout, as in Fig.~\ref{fig:gscool}(e). Error bars representing one standard deviation are smaller than the size of the points.
    }
    \label{fig:nadd}
\end{figure}

We can infer the transducer's input-referred upconversion added noise $N_\text{add,up}$ from the above measurements of the membrane mode occupancy $n_\text{m}$ and the transducer efficiency $\eta_\text{t}$. We also take into account an additional contribution to $N_\text{add,up}$ from white noise $\tilde{N}_\text{out,o}$ measured at the transducer's optical output port, which arises from the technical noise responsible for the effective optical mode occupancy $n_\text{eff,o}$ (see Appendix~\ref{sec:theory}), and plot the results in Fig.~\ref{fig:gscool}(d). To understand the observed increase in added noise when $\Gamma_\text{o} \gg \Gamma_\text{e}$, it is helpful to consider a schematic representation (Fig.~\ref{fig:gscool}(e)) of the noise emerging from the transducer in this regime, shaded gray in Figs.~\ref{fig:gscool}(b), (c), and (d). When the optomechanical damping $\Gamma_\text{o}$ dominates, almost all the noise arising from thermal occupancy of the membrane mode is routed to the transducer's optical output port. This contribution $N_\text{out,o,m}$ to the optical output noise density asymptotes at high $\Gamma_\text{o}$. But to obtain the input-referred added noise $N_\text{add,up}$ we must divide by the transducer efficiency, which decreases with increasing $\Gamma_\text{o}$, resulting in increased added noise. The fact that $\tilde{N}_\text{out,o}$ increases with increasing $\Gamma_\text{o}$ further exacerbates the added noise.

More precisely, the transducer's input-referred upconversion added noise is given in units of photons/s/Hz (or more simply photons) by
\begin{align}
    N_\text{add,up} &= \frac{N_\text{out,o,m} + \tilde{N}_\text{out,o}}{\mathcal{A}_\text{e}\mathcal{A}_\text{o}\,\eta_\text{t}} \nonumber \\ &= \frac{\gamma_\text{m}n_\text{th} + \Gamma_\text{e}n_\text{e} + \Gamma_\text{o}n_\text{o}}{\mathcal{A}_\text{e}\frac{\kappa_\text{e,ext}}{\kappa_\text{e}}\Gamma_\text{e}} + \frac{\tilde{N}_\text{out,o}}{\mathcal{A}_\text{e}\mathcal{A}_\text{o}\,\eta_\text{t}},
    \label{eq:nadd}
\end{align}
where $\mathcal{A}_\text{e} = 1 + n_\text{min,e}$ is the transducer gain from imperfect electromechanical sideband resolution, and $\mathcal{A}_\text{o}$ is defined analogously. The electromechanical and optomechanical bath occupancies $n_\text{e}$ and $n_\text{o}$ depend implicitly on the damping rates $\Gamma_\text{e}$ and $\Gamma_\text{o}$, respectively, because of technical noise. In the absence of technical noise in the optomechanical system, $n_\text{o} \rightarrow n_\text{min,o}$ and the last term vanishes. Then it is clear that increasing the optomechanical damping $\Gamma_\text{o}$ can only ever increase $N_\text{add,up}$, and that quantum-enabled upconversion would require electromechanical ground-state cooling.

As the added noise is lowest at small $\Gamma_\text{o}$ in Fig.~\ref{fig:gscool}(d), we now fix $\Gamma_\text{o}/2\pi = 85$~Hz and vary $\Gamma_\text{e}$ to minimize $N_\text{add,up}$. We begin by measuring transducer efficiency as a function of  $\Gamma_\text{e}$ (Fig.~\ref{fig:nadd}(a)). We observe that the microwave mode linewidth $\kappa_\text{e}$ increases with increasing microwave pump power $P_\text{e}$, an effect likely related to the elevated thermal occupancy $n_\text{eff,e}$ observed in the electro-optomechanical experiment in Fig.~\ref{fig:gscool}(b).
We account for this power-dependence in fitting the efficiency measurements (see Appendix~\ref{sec:analysis}), and obtain a peak efficiency $\eta_\text{t,max} = 47 \pm 1\%$ at $\Gamma_\text{e}/2\pi = 75$~Hz, close to the expected vale of $49\%$.

We then measure noise in the transducer's optical output spectrum while sweeping the electromechanical damping $\Gamma_\text{e}$ over this same range (Fig~\ref{fig:nadd}(b)), and combine the results with the efficiency measurements to obtain $N_\text{add,up}$ as a function of electromechanical damping. Fitting the data, we find that at $\Gamma_\text{e}/2\pi = 135$~Hz the added noise reaches a minimum value of $N_\text{add,up} = 3.2 \pm
0.1$~photons/s/Hz, an order of magnitude improvement relative to previously reported measurements on a similar device~\cite{higginbotham_harnessing_2018}. In this fit, we parameterize the effective microwave mode occupancy as $n_\text{eff,e} = a_\text{e}\Gamma_\text{e} + b_\text{e}$ and again including an additional fixed term to account for the optical cavity lock beam (see Appendix~\ref{sec:analysis}). The fit yields $n_\text{th} = 980 \pm 30$ and $a_\text{e}$ and $b_\text{e}$ consistent with independent microwave noise measurements (see Appendix~\ref{sec:paramnoise}). 

The dominant contributions to the added noise at the minimum are 1.0~photons/s/Hz from residual thermal motion of the membrane and 1.4~photons/s/Hz from the effective occupancy of the microwave mode, with several smaller sources responsible the remaining 0.8~photons/s/Hz (see Appendix~\ref{sec:analysis}). Fig.~\ref{fig:nadd}(c) shows a schematic representation of transducer noise around the added noise minimum, where $\Gamma_\text{e} \approx \Gamma_\text{o}$ (yellow shaded region in Figs.~\ref{fig:nadd}(a) and (b)). The noise arising from membrane mode thermal occupancy is divided roughly equally between the two transducer ports, but the mode is not in its quantum ground state due to the smaller total damping: $n_\text{m} = 1.5$~phonons when $N_\text{add,up}$ is minimized. Eq.~\eqref{eq:nadd} indicates that in the absence of technical noise, the added noise would continue to decrease with increasing electromechancial damping even in the $\Gamma_\text{e} \gg \Gamma_\text{o}$ regime where the efficiency is small. In our experiment, the power-dependent effective microwave mode occupancy $n_\text{eff,e}$ causes $n_\text{m}$ to increase at large $\Gamma_\text{e}$, resulting in an optimum value of $\Gamma_\text{e}$ at which $N_\text{add,up}$ is minimized.

Finally, we consider how the transducer's downconversion added noise $N_\text{add,down}$ (not measured in this work) would scale with microwave and optical pump powers. Added noise is in general not the same in downconversion as in upconversion. In the absence of microwave technical noise, optomechanical ground-state cooling would be a sufficient condition for quantum-enabled downconversion, independent of the strength of the microwave pump. However, the technical noise responsible for the effective microwave mode thermal occupancy $n_\text{eff,e}$ also generates a white noise contribution $\tilde{N}_\text{out,e}$ to the microwave noise density at the external port of the microwave circuit. As indicated in Figs.~\ref{fig:gscool}(e) and \ref{fig:nadd}(c), this excess noise is significantly larger than the analogous optical technical noise $\tilde{N}_\text{out,o}$, and it precludes quantum-enabled downconversion with this transducer.

\section{Effects of Laser Illumination}

\begin{figure}
    \centering
    \includegraphics{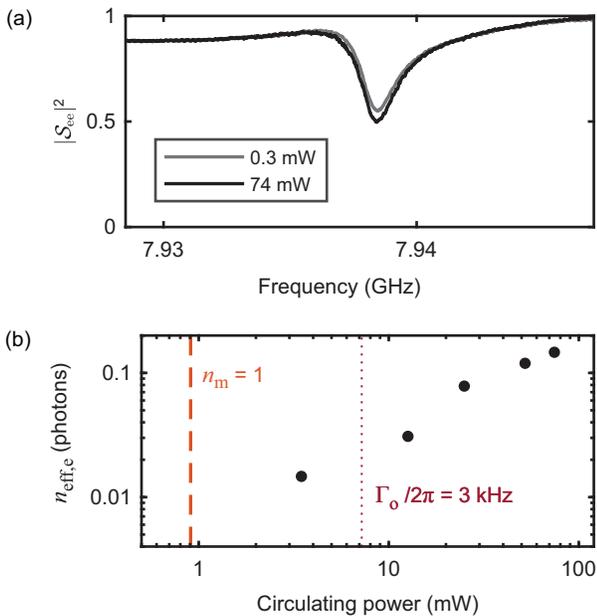}
    \caption{\textbf{Effects of laser illumination on microwave circuit.} (a) Microwave circuit power reflection measurements with 0.3~mW of circulating lock light (gray) and with 74~mW of circulating pump light (black).
    This pump power is a factor of 10 greater than that required for the highest damping rate $\Gamma_\text{o}$ shown in Fig.~\ref{fig:gscool}, and two orders of magnitude greater than that required for optomechanical ground-state cooling.
    (b)~Laser-induced microwave circuit noise vs.\ circulating power in the optical cavity. The noise remains negligible up to very large circulating power.
    }
    \label{fig:lasernoise}
\end{figure}

As indicated by the data presented in Figs.~\ref{fig:gscool} and \ref{fig:nadd}, the transducer performance is chiefly limited by the pump power-dependent effective thermal occupancy $n_\text{eff,e}$ of the superconducting microwave circuit. 
The continuously applied optical pump, however, had no discernible effect on the microwave circuit during the experiments described above, in striking contrast to other platforms in which pulsed operation is required to avoid microwave circuit heating \cite{mirhosseini_superconducting_2020,fu_ground-state_2020,stockill_ultra-low-noise_2021,sahu_quantum-enabled_2021}. Indeed, laser illumination has very little effect on the microwave circuit even with much larger optical pump power. The gray curve in Fig.~\ref{fig:lasernoise}(a) shows the circuit's power reflection coefficient $\left|\mathcal{S}_\text{ee}(\omega)\right|^2$ when the only light in the optical cavity is $0.3$~mW of circulating power from the lock beam, while the black curve shows $\left|\mathcal{S}_\text{ee}(\omega)\right|^2$ in the presence of $74 \pm 8$ mW of circulating pump power. The laser-induced frequency shift and increase in internal loss are barely perceptible.

By measuring the noise emerging from the external port of the superconducting circuit over a bandwidth much broader than $\Gamma_\text{T}$, we can directly probe laser-induced heating of the microwave mode as a function of the circulating power in the optical cavity (Fig.~\ref{fig:lasernoise}(b)), with no microwave pump present. In the transducer ground-state cooling measurements shown in Fig.~\ref{fig:gscool}(b), the maximum optomechanical damping of 3~kHz was obtained with 7~mW of circulating power, indicating that damping (and thus transducer bandwidth) could be further increased by more than an order of magnitude while maintaining $n_\text{eff,e} < 0.15$. This insensitivity of the superconducting circuit to optical illumination is a consequence of our modular transducer design, and has proven advantageous in work integrating superconducting qubits with electro-optic transducers~\cite{delaney_non-destructive_2021}.

\section{Conclusion}

The device improvements necessary for achieving quantum-enabled operation of a membrane-based electro-optomechanical transducer are readily attainable. In particular, the effects of the noise introduced by the microwave pump were exacerbated by an atypically small vacuum electromechanical coupling rate $g_\text{e} = 2\pi\times 1.6$~Hz for the transducer used in this work, arising from modifications to the flip-chip fabrication procedure to accommodate the phononic shield (see Appendix~\ref{sec:characterization}). Simply reproducing couplings realized in other devices \cite{burns_reducing_2019} would reduce the microwave pump power required to obtain a given electromechanical damping $\Gamma_\text{e}$ by more than an order of magnitude. 

With this improvement alone, we expect $N_\text{add,up}<1$ for transduction with $\Gamma_\text{e} = \Gamma_\text{o} \approx 2\pi \times 1$~kHz. The transducer bandwidth $\Gamma_\text{T}/2\pi \approx 2$~kHz would be small compared to the bandwidth potentially available in integrated transducers, but nonetheless compatible with superconducting qubit lifetimes~\cite{reagor_quantum_2016}. Moreover, the transducer would operate continuously, requiring no dead time to recover from laser illumination, and reduction of microwave pump power-dependent noise would enable bandwidth to be further increased by at least an order of magnitude. Integrating such a transducer with an optical measurement setup that enables the detection of single MHz-frequency phonons \cite{galinskiy_phonon_2020} would put the heralded upconversion of non-Gaussian quantum states within reach.

Our electro-optomechanical device also has other applications in which its high efficiency and continuous operation would compensate for its low bandwidth. For instance, with reduced technical noise such a device could be operated as a source of entangled microwave/optical photon pairs \cite{barzanjeh_entangling_2011,zhong_proposal_2020,rau_entanglement_2021}, in which case an important figure of merit is the rate at which entanglement is generated, proportional to the product of bandwidth, efficiency, and duty cycle. By this metric a device with 2~kHz bandwidth and $50\%$ efficiency operated continuously would perform as well as a device with 20~MHz bandwidth, $10\%$ efficiency and a duty cycle of $6\times10^{-4}$ \cite{sahu_quantum-enabled_2021}. 

\section*{Acknowledgements} 
We thank Hans Green, Kim Hagen, Jim Beall and Kat Cicak for technical and fabrication assistance, Luca Talamo and Kazemi Adachi for helpful discussions, and Josh Combes, Tasshi Dennis, and Akira Kyle for comments on the manuscript. This work was supported by funding from ARO CQTS Grant No.\ 67C1098620, NSF Grant No.\ PHYS 1734006, and AFOSR MURI Grant No.\ FA9550-15-1-0015.

\appendix

\section{Experimental setup}
\label{sec:chain}

\begin{figure*}[!t]
    \centering
    \includegraphics{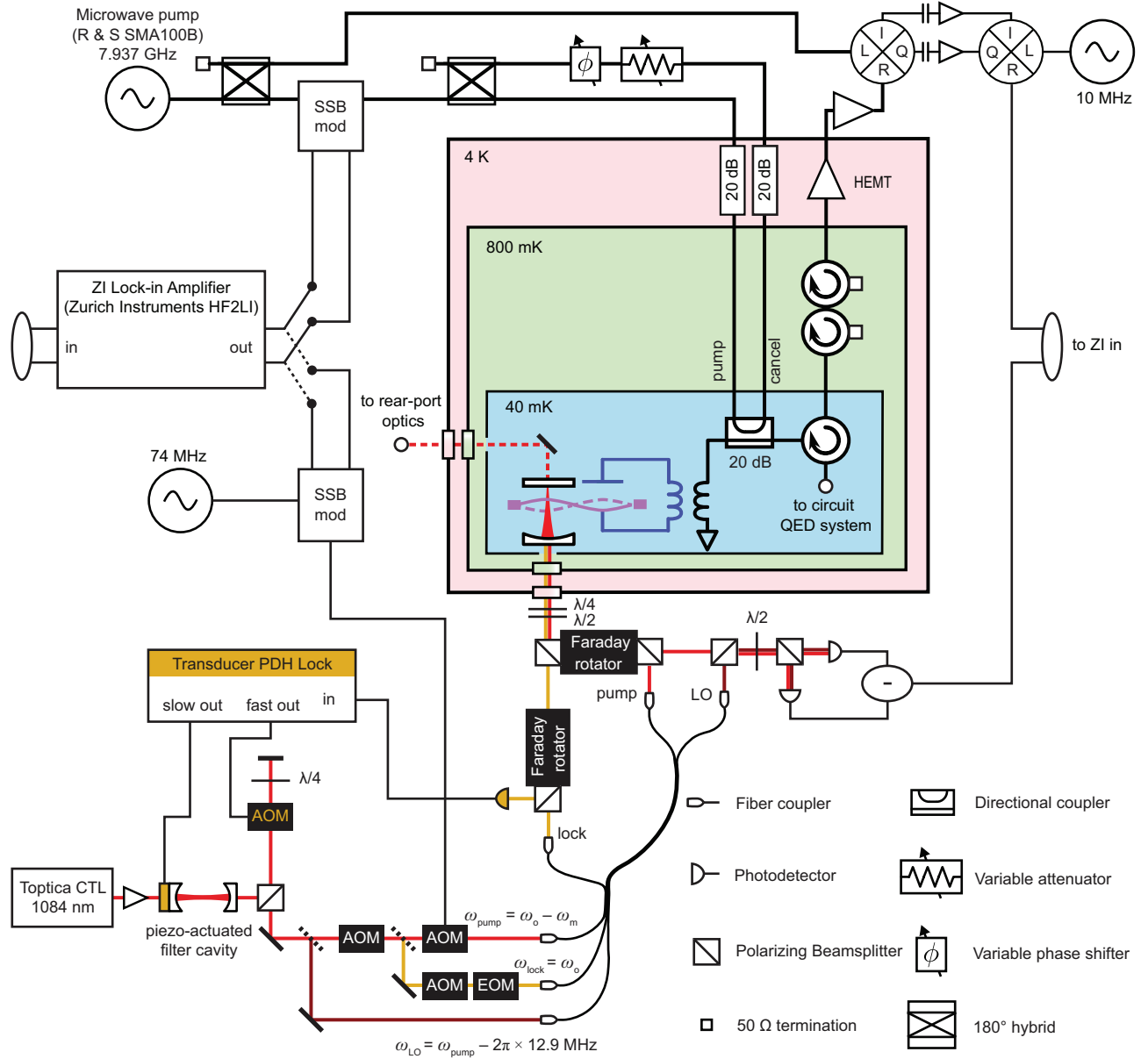}
    \caption{
    \textbf{Experimental setup.} Diagram illustrating the system used to control and probe the transducer, comprising microwave (top) and optical (bottom) subsystems along with the lock-in amplifier used for data acquisition (left). Elements used to lock the optomechanical cavity are shown in yellow.
    }
    \label{fig:AChain}
\end{figure*}

The transducer is operated in a CryoConcept Horizontal 200 dilution refrigerator with free-space optical access and $^4$He precooling to minimize vibrations (previously used in Ref.~\cite{higginbotham_harnessing_2018}). Optical access to the device is provided by fused silica windows that filter room-temperature thermal radiation to avoid heating of the cryostat base plate~\cite{kuhn_free-space_2014}, and the cryostat reaches a base temperature of $T_\text{bp} = 40$~mK. A back port enables cavity transmission measurements for beam alignment and membrane imaging. The transducer device shares space on the cryostat base plate with a circuit~QED module used in Ref.~\cite{delaney_non-destructive_2021} but not in this work.

The overall experimental setup is shown in Fig.~\ref{fig:AChain}. The microwave pump, sourced by an a signal generator with ultralow phase noise (Rohde \& Schwartz SMA100B with option SMAB-B711), is injected into the cryostat along with a phase-coherent cancellation tone. The amplitude and phase of the cancellation tone are adjusted to minimize the power routed towards the microwave measurement chain, to prevent saturation of the cryogenic preamplifier (a LNF~LNC4\_8C HEMT) by the high-power microwave pump. Demodulation of microwave signals is performed by an effective heterodyne detector comprising a homodyne measurement followed by AC-coupled amplification and modulation of the baseband signal up to 10 MHz.

The most significant change relative to the optical control and measurement system of Ref.~\cite{higginbotham_harnessing_2018} is the use of a widely tunable external cavity diode laser (Toptica Photonics CTL) in place of an Innolight Mephisto Nd:YAG laser to source the optical pump. Wavelength tuning is necessary to compensate for the absence of piezoelectric actuators to adjust the length of the optomechanical cavity. We operate the laser at a wavelength of 1084.4~nm, where there is a local maximum in the transducer's vacuum optomechanical coupling $g_\text{o}$ (see Appendix~\ref{sub:cavity}). The CTL is placed in an acoustically isolated enclosure to reduce its sensitivity to environmental disturbances, and locked to a filter cavity with 40~kHz linewidth~\cite{purdy_cavity_2012} via feedback to the diode current and piezo actuators on the laser cavity (not shown in Fig.~\ref{fig:AChain}) using the Pound-Drever-Hall (PDH) technique. Transmission through the filter cavity reduces the laser phase noise at detuning $\omega_\text{m}$ from the carrier by 37~dB. We use a single-frequency fiber amplifier (Nufern Nuamp NUA-1064-PB-0005-C1) to amplify the CTL output beam before the filter cavity.

The beam transmitted through the filter cavity is then frequency-shifted by a double-pass acousto-optic modulator (AOM) and split three ways: one beam (red in Fig.~\ref{fig:AChain}) provides the optical pump to operate the transducer, a second (yellow) is used to lock the frequency of the optomechanical cavity to that of the laser, and a third (maroon) provides a local oscillator (LO) for balanced heterodyne detection. The relative detunings of the three beams are controlled by three additional AOMs. The incident pump beam and lock beam are orthogonally polarized to route the beams emerging from the cryostat to separate detectors. The detuning of the pump beam from the cavity mode is thus given by $\Delta_\text{o} = \left(\omega_\text{pump} - \omega_\text{lock} + \Delta_\text{lock}\right) \pm \Delta_\text{B}$, where $\Delta_\text{lock}$ is the detuning of the locking beam from the cavity mode,  $\Delta_\text{B}$ is the cavity birefringence, and the sign depends on whether the lock beam addresses the higher- or lower-frequency cavity mode. The LO beam is detuned from the pump beam by $\left(\omega_\text{pump} - \omega_\text{LO}\right)/2\pi = 12.9$~MHz. 

The frequency of the incident lock beam is locked to a TEM$_{00}$ mode of the optical cavity using PDH feedback. An electro-optic modulator (EOM) imprints phase modulation sidebands on the lock beam, and the PDH error signal obtained by demodulating the detected lock beam at the sideband frequency is used to apply feedback in parallel to the double-pass AOM and a piezoelectric actuator on one of the filter cavity mirrors. The AOM channel has high bandwidth but limited range, while the slow feedback to the filter cavity length eliminates drift of the laser frequency relative to the frequency of the cavity mode. The incident lock beam power is $P_\text{lock} = 20$~nW, and the PDH error signal bias is adjusted to keep the lock beam slightly red-detuned from the cavity, with $\Delta_\text{lock}/2\pi$ ranging from $-30$~kHz to $-80$~kHz, to avoid optomechanical instability.

A Zurich Instruments HF2LI lock-in amplifier is used for all data acquisition. To characterize the transducer and measure its efficiency, we use the HF2LI to synthesize swept MHz-frequency tones for single-sideband (SSB) modulation of the transducer pumps and demodulate the heterodyne detector output signals to recover these tones  (see Appendix~\ref{sub:mechanical}). For noise measurements, we use the HF2LI to digitize time traces of the noise at the heterodyne detector outputs, and then compute the noise variance or power spectral density. See Ref.~\cite{higginbotham_harnessing_2018} for further details of signal processing with the HF2LI.

\section{Transducer device}
\label{sec:device}

\subsection{Design and Layout}
\label{sub:design}
As shown in Fig.~\ref{fig:schematic}(a), the membrane, the microwave circuit, and one of the optical cavity mirrors are integrated into a stack of three 380~\si{\micro\meter}-thick silicon chips. We refer to the chips hosting the membrane, the greater part of the circuit, and the mirror as the top, middle, and bottom chips, respectively. The fabrication procedure is described in Appendix~\ref{sub:fab}. Here we discuss the device design at a higher level.

\begin{figure}[!t]
    \centering
    \includegraphics{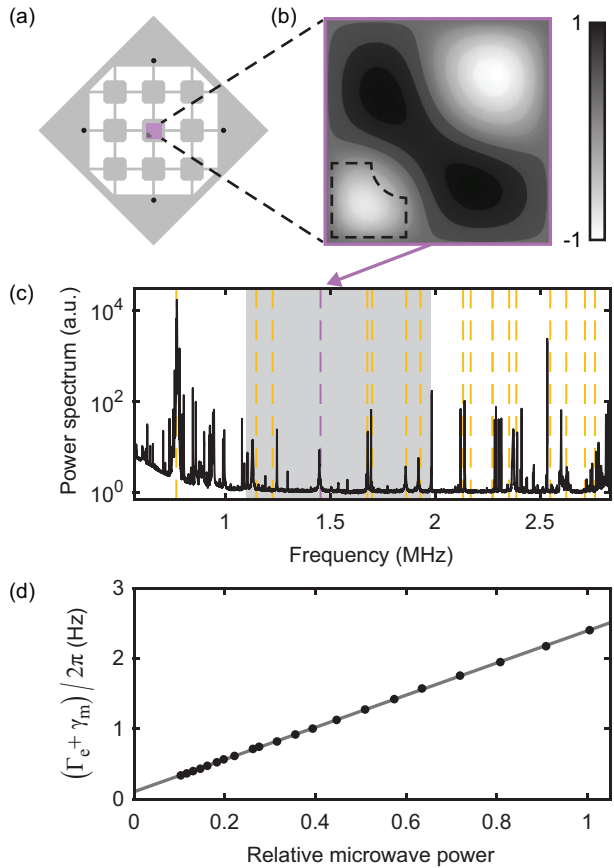}
    \caption{
    \textbf{Mechanical design and characterization.}
    (a)~Phononic shield design. Black circles (not to scale) indicate location of 20~\si{\micro\meter} diameter posts that determine the nominal capacitor pad separation. 
    (b)~Normalized displacement of the 1.45 MHz vibrational mode from finite-element simulation of the membrane and capacitor pad (dashed line).
    (c)~Optically measured mechanical spectrum. The shaded region indicates the bandgap obtained from a simulation with periodic boundary conditions. Expected frequencies of membrane modes (dashed lines) are obtained from finite-element simulation. 
    (d)~Total damping obtained from electromechanical ring-down measurements vs.\ microwave pump power. Extrapolation to zero pump power yields an intrinsic mechanical damping rate of $\gamma_\text{m}/2\pi = 108$~mHz.
    }
    \label{fig:AMechanics}
\end{figure}

The silicon substrate of the top chip is patterned into a few-cell phononic shield, made up of 900~\si{\micro\meter} $\times$ 900~\si{\micro\meter} masses joined by 100~\si{\micro\meter} $\times$ 550~\si{\micro\meter} tethers, with the 100~nm-thick silicon nitride membrane supported on the central mass (Fig.~\ref{fig:AMechanics}(a)). The transverse dimensions of the membrane (500~\si{\micro\meter} $\times$ 500~\si{\micro\meter}) determine the resonant frequency $\omega_\text{m}/2\pi = 1.45$~MHz of the vibrational mode with two antinodes in each direction (Fig.~\ref{fig:AMechanics}(b)), which we label the (2,2) mode \cite{andrews_2015}. The mass and tether dimensions are chosen to center the phononic bandgap on $\omega_\text{m}$, and finite-element simulations of the shield with periodic boundary conditions predict a bandgap between $1.10$ and $1.98$~MHz. 

The overall size of the top chip is constrained by the flip-chip electromechanical circuit architecture: larger chips make it difficult to control the capacitor pad separation $d$ that determines $G_\text{e}$, the coefficient relating the microwave mode frequency shift to membrane displacement. This constraint limits the number of unit cells in the phononic shield, but the design nevertheless yields a vibrational spectrum with relatively few substrate modes in the vicinity of the (2,2) mode (Fig.~\ref{fig:AMechanics}(c)). Membranes embedded in similar phononic crystal structures with more unit cells have been shown to thermalize to $T\lesssim100$~mK in the presence of laser light~\cite{kampel_improving_2017}.

A NbTiN pad on one corner of the membrane together with a split pad on the middle chip form a parallel plate capacitor, and the two halves of the split pad are connected through a loop inductor to complete the electromechanical circuit. The middle chip portion of the circuit contributes some parasitic capacitance that dilutes the sensitivity of the circuit's resonant frequency to membrane motion. The electromechanical frequency shift per unit membrane displacement is given by
\begin{equation}
    G_\text{e} = p\frac{\omega_\text{e}}{2d},
    \label{eq:G_e}
\end{equation}
where $p$ is the ratio of the circuit's motionally modulated capacitance to its total capacitance, which decreases with increasing pad separation $d$. Four posts arrayed in a 3.6~mm square on the middle chip set the target capacitor pad separation $d_0 = 300$~nm and support the top chip, which is affixed to the middle chip using cryogenic epoxy. The actual capacitor gap $d$ can differ from $d_0$ because the epoxy contracts upon thermal cycling. For this device, we infer $d=830$~nm from electromechanical measurements (see Appendix~\ref{sub:mechanical} for further discussion). Regions of the top chip and middle chip substrates are recessed to limit the area with small separation between the top chip and middle chip. 

The bottom chip serves as a substrate for a high-reflectivity Bragg mirror (7~ppm power transmission coefficient). The top and bottom layers of this Bragg mirror are both SiO$_2$, which results in a strong bond between the mirror coating and the substrate and enables bonding of the mirror chip to the underside of the middle chip (see Appendix~\ref{sub:fab}). A hole in the middle chip opposite the non-metalized portion of the membrane allows laser light to pass through to the mirror. A higher-throughput curved mirror (190~ppm power transmission coefficient, 1~cm radius of curvature) is positioned opposite the mirror chip at a distance of $L=2.3$~mm to form a stable single-sided Fabry-P\'erot optical cavity (Fig.~\ref{fig:AOptSim}(a)), whose mode intersects the corner of the membrane opposite the capacitor pad. 

The bonding of the bottom chip to the middle chip fixes the distance between the flat mirror and the membrane at the middle chip thickness $t=380$~\si{\micro\meter} and ensures that the mirror and the membrane do not significantly misalign as the transducer is cooled to cryogenic temperatures. The coefficient $G_\text{o}$ relating the optical cavity frequency shift to membrane displacement is determined by the longitudinal position of the membrane in the standing wave of the cavity mode (see Appendix~\ref{sub:cavity}). 

The vacuum electromechanical and optomechanical couplings that enter into the system Hamiltonian are given respectively by $g_\text{e} = G_\text{e}x_\text{zp,e}$ and $g_\text{o} = G_\text{o}x_\text{zp,o}$, where $x_\text{zp}$ denotes the zero-point motion of the membrane mode. The zero-point motion $x_\text{zp,o}$ for the optomechanical interaction is modified relative to the na\"{i}ve value $x_\text{zp} = \sqrt{\hbar/2\omega_\text{m}m}$ for a harmonic oscillator with mass $m$ by a factor that accounts for the transverse displacement profile of the membrane mode (Fig.~\ref{fig:AMechanics}(b)) integrated over the spatial extent of the beam spot. The zero-point motion $x_\text{zp,e}$ for the electromechanical interaction is likewise modified by the integral of membrane motion over the area of the capacitor pad. For the perturbed (2,2) mode we use for transduction, $x_\text{zp,e} < x_\text{zp,o}$ because motion of the membrane quadrant occupied by the capacitor pad is reduced relative to the motion at the optical spot due to the extra mass of the NbTiN film. Using a finite-element simulation of the membrane with the normalization convention of Ref.~\cite{andrews_2015}, we obtain
$x_\text{zp,e} = 0.5$~fm and $x_\text{zp,o} = 0.9$~fm.

\subsection{Fixed membrane-mirror cavity architecture}
\label{sub:cavity}

\begin{figure}[!t]
    \centering
    \includegraphics{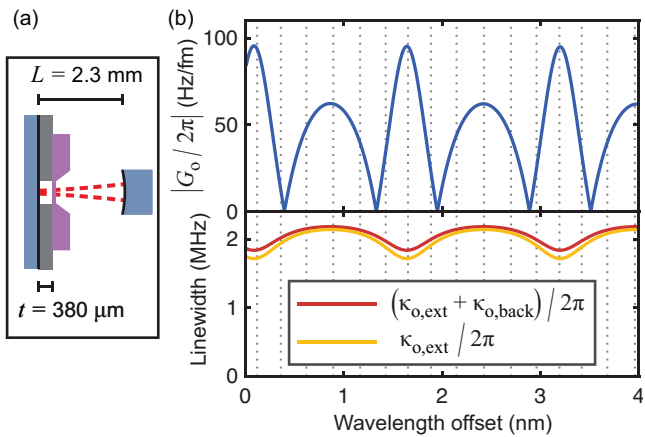}
    \caption{
    \textbf{Optical cavity simulation.}
    (a) Optical cavity length scales. 
    (b) Simulated optomechanical cavity parameters. The free spectral range associated with the $t=380$~\si{\micro\meter} spacing between the membrane and mirror chip sets the periodicity of the optical frequency shift per membrane displacement $G_\text{o}$ (top) and cavity mirror loss rates $\kappa_\text{o,ext}$ and $\kappa_\text{o,back}$ (bottom) that would be obtained if the optical cavity had a piezoelectrically actuated curved mirror. Gray dotted lines indicate operating parameters available to a cavity without piezoelectric actuation like the one used in this work. 
    }
    \label{fig:AOptSim}
\end{figure}

Previous optomechanical cavity designs~\cite{andrews_bidirectional_2014,higginbotham_harnessing_2018} used two piezoelectric actuators to provide the requisite two degrees of freedom for \textit{in situ} control of the frequency shift per membrane displacement $G_\text{o}$ and continuous tuning of the perturbed cavity frequency to match that of a fixed-frequency laser. The presence of two planar surfaces susceptible to misalignment in such cavities resulted in large scattering losses when the membrane was positioned at regions of higher intracavity field intensity, limiting operation to membrane positions with low $G_\text{o}$~\cite{higginbotham_harnessing_2018}. The piezoelectric actuators on these cavities also provided a channel through which electrical noise could couple in to the optomechanical system. In this work, we instead used a cavity with two fixed mirrors together with a tunable laser.

To understand the operation of an optomechanical cavity with two fixed mirrors, it is illustrative to first consider the case of an otherwise identical cavity with piezoelectric actuation of the curved mirror opposite the silicon chip stack.
Because the distance between the flat mirror and the membrane is fixed, the only way to translate the membrane relative to the cavity mode standing wave (and thus tune $G_\text{o}$ \textit{in situ}) is to vary the laser wavelength. The curved mirror can then be actuated to maintain cavity resonance and probe $G_\text{o}$ continuously as a function of laser wavelength.
Tuning the laser also changes the optical cavity decay rates $\kappa_\text{o,ext}$, $\kappa_\text{o,back}$, and $\kappa_\text{o,loss}$ associated with the input mirror, the weakly coupled back mirror, and absorption/scattering losses, respectively.
$G_\text{o}$, $\kappa_\text{o,ext}$, and $\kappa_\text{o,back}$ can be calculated using a one-dimensional transfer matrix model.
All three quantities are periodic in the free spectral range $\Delta\lambda_\text{FSR,mem}=\lambda^2/2t=1.55$~nm associated with the distance $t=380$~\si{\micro\meter} between the membrane and flat mirror.
The loss rate $\kappa_\text{o,loss}$ is difficult to model because it depends on the alignment between the membrane and the cavity mirrors, the membrane-perturbed frequencies of higher-order modes, and the location of lossy scatterers. 

Fig.~\ref{fig:AOptSim}(b) shows the results of the transfer matrix model simulation using the distances $t$ and $L$ and the mirror coating transmission coefficients of our cavity. In the course of tuning the resonant wavelength of the cavity by $\Delta\lambda_\text{FSR,mem}$, the optomechanical frequency shift $G_\text{o}$ exhibits two nulls, corresponding to the membrane sitting at the node and antinode of the optical field intensity, and two local maxima, where the intensity gradient is maximized. The larger local maximum corresponds to light building up preferentially in the shorter section of the cavity, between the membrane and the mirror deposited on the bottom chip, while at the lower local maximum, the intracavity light is concentrated in the longer cavity section between the membrane and curved mirror. The features have different widths, with the wider one corresponding to light building up on the side of the mirror that is actuated (the curved mirror in the simulation shown in Fig.~\ref{fig:AOptSim}). The decay rate $\kappa_\text{o,ext}$ through the higher-throughput curved mirror is maximized when intracavity light is concentrated in the longer cavity section, while the decay rate $\kappa_\text{o,back}$ through the lower-throughput flat mirror is maximized when light is concentrated in the shorter section. 

We chose to deposit the higher-transmission coating on the curved mirror farther from the membrane, designating it the input mirror, in order keep the sum $\kappa_\text{o,ext} + \kappa_\text{o,back}$ (equal to the cavity linewidth $\kappa_\text{o}$ in the absence of loss) relatively constant as a function of wavelength. Depending on the laser wavelength, the optical field will preferentially build up on either one side of the membrane or the other, and the change between the two cavity section lengths counteracts the change between the mirror throughputs. This insensitivity of $\kappa_\text{o,ext} + \kappa_\text{o,back}$ to wavelength allows flexibility in operating the transducer, because in the absence of additional loss the optomechanical sideband resolution is similar at the two local maxima in $G_\text{o}$. The option to use the lower local maximum is important because there the cavity mode is less susceptible to loss from any misalignment between the membrane and flat mirror on the bottom chip \cite{purdy_cavity_2012}. 

Eliminating piezoelectric actuation of the curved mirror restricts operation to a discrete set of wavelengths at which the laser is resonant with a longitudinal mode of the optical cavity, separated by roughly the unperturbed optical cavity free spectral range $\Delta\lambda_\text{FSR} = \lambda ^2/2L=260$~pm (gray dashed lines in Fig.~\ref{fig:AOptSim}(b)). Because the membrane sits at a different position in the intracavity standing wave at each of these wavelengths, we retain the ability to choose an operating point with high coupling and low loss \textit{in situ}. In this work we operate our transducer at a wavelength near the lower of the two local maxima in  $G_\text{o}$.  

\subsection{Device Fabrication}
\label{sub:fab}
Three separate silicon wafers are processed in order to construct the device. The bottom chip wafer has a high-reflectivity SiO$_{2}$/TaO$_{5}$ Bragg mirror coating deposited using ion beam sputtering by FiveNine Optics. In order to ensure the wafer is flat at room temperature, a requirement for bonding the bottom chip and middle chip wafers, a 3~\si{\micro\meter} SiO$_{2}$ stress compensation layer is deposited on the back of the wafer as well. The portion of the electromechanical circuit on the middle chip wafer is fabricated from a 200~nm layer of sputtered NbTiN and patterned using a fluorine etch. The posts that define the target capacitor pad separation are then patterned from a 300~nm aluminum film, and the optical access hole in the middle chip is etched using a deep reactive ion etch (DRIE).

The bottom chip and middle chip wafers are bonded using a hydrophilic Si-SiO$_{2}$ bonding process \cite{waferBonding}. Surface preparation of the bonding interfaces is done with two atmospheric plasma treatments: one oxygen-rich plasma to descum the surface and one hydrogen-rich plasma to create the desired surface chemistry. The wafers are then pressed together and annealed at 150~$^\circ$C. After bonding, the SiO$_{2}$ stress-compensation layer is removed with a HF vapor etch to make a symmetric silicon/mirror coating/silicon stack. The bonded wafers are then diced to create 8~mm $\times$ 8~mm chips. 

A 100~nm high-stress layer of stoichiometric Si$_3$N$_4$ is grown on the top chip wafer using low-pressure chemical vapor deposition (LPCVD). The capacitor pad is then patterned from a 30~nm film of sputtered NbTiN using a 5:2:1 H$_{2}$O:H$_{2}$O$_{2}$:NH$_{4}$OH solution. A DRIE step followed by immersion in potassium hydroxide (KOH) is used to define and release the membrane. Lastly, a final DRIE step patterns the phononic shield (see Appendix~\ref{sub:design}) and dices the wafer into 6~mm $\times$ 6~mm chips. The membranes are cleaned in nanostrip at 60~$^\circ$C and then in HF. A top chip is then flipped over onto a middle chip/bottom chip stack using a Westbond manual die bonder, and small amounts of Stycast 2850 cryogenic epoxy are applied to affix the top chip and middle chip together.

A superpolished fused silica substrate for the curved mirror is fabricated by Perkins Precision Developments. A high-reflectivity SiO$_{2}$/TaO$_{5}$ Bragg mirror is then deposited on this substrate by FiveNine Optics. The fused silica substrate and three-chip stack are supported by an Invar enclosure in order to minimize thermal contraction upon cryogenic cooling. External coupling to the microwave circuit is achieved through its mutual inductance with a loop coupler at the end of a coplanar waveguide on the middle chip. This coplanar waveguide is wiredbonded to a flexible printed circuit board with an SMA connector at the opposite end supported by the invar structure.

\section{Theory}
\label{sec:theory}

\subsection{Hamiltonian formalism}
\label{sub:hamiltonian}
The electro-optomechanical system is described by the Hamiltonian
\begin{align}
    \hat{H}/\hbar = \omega_\text{o}& \hat{a}^\dagger \hat{a} + \omega_\text{e}\hat{b}^\dagger \hat{b} + \omega_\text{m}\hat{c}^\dagger \hat{c} \nonumber \\ &   + \left(g_\text{o}\hat{a}^\dagger \hat{a} + g_\text{e}\hat{b}^\dagger \hat{b}\right)\left(\hat{c}+\hat{c}^\dagger\right),
    \label{eq:H0}
\end{align}
where $\hat{a}$, $\hat{b}$ and $\hat{c}$ are respectively the optical, microwave and mechanical mode annihilation operators, and  $g_\textrm{o}$ and $g_\textrm{e}$ are the vacuum optomechanical and electromechanical coupling rates~\cite{aspelmeyer_cavity_2014}.  

Optical and microwave pumps detuned from the corresponding electromagnetic resonances by $\Delta_\text{o}$ and $\Delta_\text{e}$ produce phase space displacements $\bar{a}$ and $\bar{b}$ for the optical and microwave modes, respectively. Linearizing Eq.~\eqref{eq:H0} about $\bar{a}$ and $\bar{b}$ yields 
\begin{align}
    \hat{H}_\textrm{lin}/\hbar = -\Delta_\text{o}&\hat{a}^\dagger\hat{a} - \Delta_\text{e}\hat{b}^\dagger \hat{b} + \omega_\text{m}\hat{c}^\dagger \hat{c} + g_\text{o}\bar{a}(\hat{a}^\dagger + \hat{a})(\hat{c} + \hat{c}^\dagger) \nonumber \\ &+ g_\text{e}\bar{b}(\hat{b}^\dagger + \hat{b})(\hat{c} + \hat{c}^\dagger),
    \label{eq:Hlin}
\end{align}
in the rotating frame of the pump fields. $\hat{H}_\textrm{lin}$ contains both beamsplitter terms of the form $\hat{a}^\dagger\hat{c}$ and two-mode squeezing terms of the form $\hat{a}^\dagger\hat{c}^\dagger$. The beamsplitter terms can be  resonantly enhanced by red-detuning the pumps by approximately the mechanical mode frequency ($\Delta_\text{o} \approx \Delta_\text{e} \approx -\omega_\text{m}$). Then Eq.~\eqref{eq:Hlin} becomes the Hamiltonian for three harmonic oscillators at the same frequency that can exchange quanta via the beamsplitter interactions at rates $g_\text{o}|\bar{a}|$ and $g_\text{e}|\bar{b}|$. 

We then extend the Hamiltonian model to include decay rates represented by ports on the three resonators. Specifically, we add one port to the mechanical mode (representing energy dissipation at rate $\gamma_\text{m}$), two ports to the microwave mode (representing coupling to a transmission line at a rate $\kappa_\text{e,ext}$ and internal loss at rate $\kappa_\text{e,int}$), and two ports to the optical mode (representing coupling through the higher-throughput curved mirror at rate $\kappa_\text{o,ext}$ and the sum $\kappa_\text{o,int}+\kappa_\text{o,back}$ of scattering/absorption losses and transmission through the lower-throughput back mirror). In the weak coupling regime ($g_\text{o} \ll \kappa_\text{o} = \kappa_\text{o,ext} + \kappa_\text{o,int}+\kappa_\text{o,back}$), the optomechanical interaction then contributes net damping to the mechanical mode given by
\begin{equation}
    \Gamma_\text{o} = g_\text{o}^2\left|\bar{a}\right|^2\left(\frac{\kappa_\text{o}}{\kappa_\text{o}^2/4 + \left(\omega_\text{m}+\Delta_\text{o}\right)^2} - \frac{\kappa_\text{o}}{\kappa_\text{o}^2/4 + \left(\omega_\text{m}-\Delta_\text{o}\right)^2}\right),
    \label{eq:Gamma_o}
\end{equation}
where $\left|\bar{a}\right| = \sqrt{(P_\text{o}/\hbar\omega_\text{p,o})\,\epsilon_\text{PC}\,\kappa_\text{o,ext}/\left(\kappa_\text{o}^2/4 + \Delta_\text{o}^2\right)}$, $\omega_\text{p,o} = \omega_\text{o} - \Delta_\text{o}$ is the optical pump frequency, and $\epsilon_\text{PC}$ is the modematching of the incident pump beam to the optical cavity mode. The second term in Eq.~\eqref{eq:Gamma_o} represents anti-damping from imperfect suppression of the two-mode squeezing terms $\hat{a}^\dagger\hat{c}^\dagger$ and $\hat{a}\hat{c}$ in Eq.~\eqref{eq:Hlin}; it is suppressed relative to the first term at optimal detuning $\Delta_\text{o} = -\omega_\text{m}$ in the resolved sideband limit $\kappa_\text{o} \ll 4\omega_\text{m}$. Likewise, the electromechanical damping rate for $g_\text{e} \ll \kappa_\text{e} = \kappa_\text{e,ext} + \kappa_\text{e,int}$ is
\begin{equation}
    \Gamma_\text{e} = g_\text{e}^2\left|\bar{b}\right|^2\left(\frac{\kappa_\text{e}}{\kappa_\text{e}^2/4 + \left(\omega_\text{m}+\Delta_\text{e}\right)^2} - \frac{\kappa_\text{e}}{\kappa_\text{e}^2/4 + \left(\omega_\text{m}-\Delta_\text{e}\right)^2}\right),
    \label{eq:Gamma_e}
\end{equation}
where $\left|\bar{b}\right| = \sqrt{(P_\text{e}/\hbar\omega_\text{p,e})\,\kappa_\text{e,ext}/\left(\kappa_\text{e}^2/4 + \Delta_\text{e}^2\right)}$ and $\omega_\text{p,e} = \omega_\text{e} - \Delta_\text{e}$ is the microwave pump frequency. 

\subsection{Transducer Efficiency}
\label{sub:eta_theory}
The transmission of a narrowband signal from the external port of the microwave resonator through the transducer to the external port of the optical cavity is governed by the scattering parameter 
\begin{equation}
    \mathcal{S}_\text{oe}(\omega) = \sqrt{\mathcal{A}_\text{e}\mathcal{A}_\text{o}\frac{\kappa_\text{e,ext}}{\kappa_\text{e}}\frac{\kappa_\text{o,ext}}{\kappa_\text{o}}}\frac{\sqrt{\Gamma_\text{e}\Gamma_\text{o}}}{\Gamma_\text{T}/2 - i(\omega-\omega_\text{m})},
    \label{eq:Soe}
\end{equation}
where $\omega$ is the detuning of the input signal from the microwave pump, $\mathcal{A}_\text{o} = -\left((\kappa_\text{o}/2)^2 + (\Delta_\text{o} - \omega_\text{m})^2\right)/4\Delta_\text{o}\omega_\text{m}$ is a gain factor arising from the optomechanical two-mode squeezing terms in Eq.~\eqref{eq:Hlin}, and the definition of $\mathcal{A}_\text{e}$ is analogous~\cite{andrews_bidirectional_2014}. These factors are undesirable because any transducer gain is necessarily accompanied by added noise~\cite{caves_quantum_1982}.
With optimally red-detuned pumps, the gain factors reduce to $\mathcal{A}_\text{o} = 1 + (\kappa_\text{o}/4\omega_\text{m})^2$ and $\mathcal{A}_\text{e} = 1 + (\kappa_\text{e}/4\omega_\text{m})^2$, and the transducer gain approaches unity in the resolved sideband limit. The transmission of a narrowband signal modematched to the optical cavity through the transducer to the external port of the microwave resonator, $\mathcal{S}_\text{eo}(\omega)$, is formally identical to $\mathcal{S}_\text{oe}(\omega)$.

Following Ref.~\cite{andrews_bidirectional_2014}, we use a conservative definition of the transducer efficiency that accounts for energy lost in each of the three modes involved in the transduction process as well as imperfect optical cavity modematching (not included in Eq.~\eqref{eq:Soe}). To properly account for modematching, we must adapt the formalism of Ref.~\cite{andrews_bidirectional_2014} to our heterodyne detection scheme. In downconversion the relevant modematching factor is the modematching $\epsilon_\text{PC}$ of the incident pump beam to the cavity, while in upconversion the relevant factor is the modematching $\epsilon_\text{CL}$ of the light emerging from the cavity to the LO beam. We thus redefine the optical input and output ports to absorb modematching factors, so that $\mathcal{S}_\text{oe}(\omega_\text{m})\rightarrow\sqrt{\epsilon_\text{CL}}\mathcal{S}_\text{oe}(\omega_\text{m})$ and $\mathcal{S}_\text{eo}(\omega_\text{m})\rightarrow\sqrt{\epsilon_\text{PC}}\mathcal{S}_\text{eo}(\omega_\text{m})$. The upconversion efficiency for an incident signal on resonance is then defined as $\eta_\text{t,up} = \left|\mathcal{S}_\text{oe}(\omega_\text{m})\right|^2/\mathcal{A}_\text{o}\mathcal{A}_\text{e}$ and the downconversion efficiency is defined as $\eta_\text{t,down} = \left|\mathcal{S}_\text{eo}(\omega_\text{m})\right|^2/\mathcal{A}_\text{o}\mathcal{A}_\text{e}$. We take our figure of merit to be the bidirectional efficiency $\eta_\text{t} = \sqrt{\eta_\text{t,up}\eta_\text{t,down}}$, given by Eq.~\eqref{eq:eta} in the main text. For $\Gamma_\text{e} = \Gamma_\text{o} \gg \gamma_\text{m}$, $\eta_\text{t}$ achieves its maximum value $\eta_\text{M} = \epsilon\frac{\kappa_\text{o,ext}}{\kappa_\text{o}}\frac{\kappa_\text{e,ext}}{\kappa_\text{e}}$, where we have defined the bidirectional modematching $\epsilon=\sqrt{\epsilon_\text{PC}\epsilon_\text{CL}}$.

\subsection{Optomechanical sideband cooling}
\label{sub:sideband}

As noted in the main text, the same beamsplitter processes that mediate transduction also cool the membrane mode. For definiteness, we consider the optomechanical interaction in isolation. Optomechanical cooling arises from the imbalance between the rates of Stokes and anti-Stokes scattering processes when the pump is red-detuned from the cavity mode. The Stokes process, wherein a pump photon is scattered to lower energy by emitting a phonon into the membrane mode, is suppressed relative to the complementary anti-Stokes process by the reduced density of states away from cavity resonance. The rates of the anti-Stokes and Stokes processes are given respectively by $n_\text{m}\Gamma_{\text{o},+}$ and $(n_\text{m}+1)\Gamma_{\text{o},-}$, where $n_\text{m}$ is the phonon occupancy of the membrane mode, and $\Gamma_{\text{o},+}$ and $\Gamma_{\text{o},-}$ are the anti-Stokes and Stokes contributions to the net optomechanical damping rate $\Gamma_\text{o}$ (i.e., the positive and negative terms in Eq.~\eqref{eq:Gamma_o})~\cite{aspelmeyer_cavity_2014}. 

If the mechanical mode were completely decoupled from its thermal environment, it would reach equilibrium when the Stokes and anti-Stokes scattering rates become equal, resulting in a final occupancy
    \begin{equation}
    n_\text{min,o} = \frac{\bar{r}_\text{o}}{1-\bar{r}_\text{o}} = \frac{\left(\kappa_\text{o}/2\right)^2 + \left(\Delta_\text{o} + \omega_\text{m}\right)^2}{-4\Delta_\text{o}\omega_\text{m}},
    \label{eq:nmin}
\end{equation}
where $\bar{r}_\text{o} = \Gamma_{\text{o},-}/\Gamma_{\text{o},+}$. To generalize to the case where the mechanical mode is coupled to more than one bath, note that $\Gamma_{\text{o},-} = \Gamma_\text{o}n_\text{min,o}$: thus heating by Stokes scattering can be described as a coupling to a bath 
$n_\text{min,o}$ at rate $\Gamma_\text{o}$, to which the membrane mode equilibrates when $\Gamma_\text{o}$ overwhelms all other contributions to the total mechanical mode damping $\Gamma_\text{T}$. This lower bound on the mode occupancy arising from residual Stokes scattering is called the quantum backaction limit. 

The derivation of the electromechanical backaction limit is entirely analogous to the above optomechanical derivation. In the absence of technical noise, the competition between the membrane mode's coupling to its thermal environment through its intrinsic dissipation rate $\gamma_\text{m}$ and its coupling to optomechanical and electromechanical backaction baths results in Eq.~\eqref{eq:gscool} for the final occupancy of the membrane mode. The transducer gain factors introduced in Appendix~\ref{sub:eta_theory} are related to the backaction bath occupancies by $\mathcal{A}_\text{o} = 1 + n_\text{min,o}$ and $\mathcal{A}_\text{e} = 1 + n_\text{min,e}$.

\subsection{Optical output spectrum}
\label{sub:output_spectrum}
To infer the membrane mode occupancy and study the transducer's noise performance, we make heterodyne measurements of the noise density at the optical cavity output port. Here we describe the relevant features of the optical output spectrum in the absence of technical noise, and in Appendix~\ref{sub:technoise} we generalize the expressions to include the effects of technical noise. The scattering processes described above generate Lorentzian peaks in the optical output spectral density at detunings $\pm\omega_\text{m}$ from the pump frequency. In units of photons/s/Hz referred to the transducer's optical output port, the spectrum around the upper (anti-Stokes) sideband has the form
\begin{equation}
    S_{\text{o},+}(\omega) = 1 + \epsilon_\text{CL}\frac{\kappa_\text{o,ext}}{\kappa_\text{o}}\frac{\Gamma_\text{o}\Gamma_\text{T}\left(1+n_\text{min,o}\right)n_\text{m}}{\Gamma_\text{T}^2/4 + \left(\omega - \omega_\text{m}\right)^2},
    \label{eq:usb}
\end{equation}
and the spectrum around the lower (Stokes) sideband is
\begin{equation}
    S_{\text{o},-}(\omega) = 1 + \epsilon_\text{CL}\frac{\kappa_\text{o,ext}}{\kappa_\text{o}}\frac{\Gamma_\text{o}\Gamma_\text{T}\,n_\text{min,o}\left(n_\text{m}+1\right)}{\Gamma_\text{T}^2/4 + \left(\omega + \omega_\text{m}\right)^2},
    \label{eq:lsb}
\end{equation}
where the background terms correspond to the sum of vacuum noise and the added noise of an ideal heterodyne detector. The upper and lower sideband amplitudes $N_{\text{out,o},\pm} = S_{\text{o},\pm}(\pm\omega_\text{m})-1$ are proportional to the optomechanical anti-Stokes and Stokes scattering rates, respectively, and thus the ratio of measured sideband amplitudes $r_\text{o} = N_{\text{out,o},-}/N_{\text{out,o},+}$ can be used to infer the mode occupancy by inverting
\begin{equation}
    r_\text{o} = \frac{n_\text{min,o}}{n_\text{min,o} + 1}\frac{n_\text{m} + 1}{n_\text{m}}.
    \label{eq:sidebandtherm}
\end{equation}
At high temperatures, $n_\text{m} \approx n_\text{m} + 1$, and comparing Eq.~\eqref{eq:sidebandtherm} to Eq.~\eqref{eq:nmin} reveals that $r_\text{o}\rightarrow \bar{r}_\text{o}$. As $n_\text{m}$ decreases, $r_\text{o}$ increases, asymptoting towards 1 as $n_\text{m}$ approaches its minimum value $n_\text{min,o}$.

When both pumps are present, the beamsplitter interactions transduce incident microwave signals detuned by $+\omega_\text{m}$ from the microwave pump to outgoing signals at detuning $+\omega_\text{m}$ from the optical pump, and vice versa. Thus the transducer noise that competes with an upconverted signal at the optical output port is $N_{\text{out,o},+}$. The input-referred upconversion added noise is then given by
\begin{equation}
    N_\text{add,up} = \frac{N_{\text{out,o},+}}{\mathcal{A}_\text{e}\mathcal{A}_\text{o}\,\eta_\text{t,up}} = \frac{n_\text{m}}{\mathcal{A}_\text{e}\frac{\kappa_\text{e,ext}}{\kappa_\text{e}}}\frac{\Gamma_\text{T}}{\Gamma_\text{e}},
    \label{eq:nadd_supp}
\end{equation}
This equation indicates that, up to small effects from imperfect sideband resolution and microwave mode internal loss, the ability to ground-state cool the mechanical mode with purely electromechanical damping is a necessary and sufficient condition for $N_\text{add,up} < 1$. Assuming comparable electromechanical and optomechanical sideband resolution, $n_\text{m}$ will remain approximately constant at constant total damping $\Gamma_\text{T}$, but trading off electromechanical damping for optomechanical damping while holding $\Gamma_\text{T}$ constant can only increase the added noise. Moreover, $N_\text{add,up}$ is independent of all other parameters of the optomechanical system. Qualitatively, this is because after the signal to be upconverted enters the mechanical mode, it copropagates with any noise routed to the optical output port, so the effects of the optical cavity parameters on the signal and noise cancel.

The form of the upper and lower electromechanical sideband amplitudes $N_{\text{out,e},+}$ and $N_{\text{out,e},-}$ in the microwave output spectrum is analogous to the optomechanical case described above. In the absence of technical noise, the input-referred downconversion added noise is given by
\begin{equation}
    N_\text{add,down} = \frac{N_{\text{out,e},+}}{\mathcal{A}_\text{e}\mathcal{A}_\text{o}\,\eta_\text{t,down}} = \frac{n_\text{m}}{\epsilon_\text{PC}\mathcal{A}_\text{o}\frac{\kappa_\text{o,ext}}{\kappa_\text{o}}}\frac{\Gamma_\text{T}}{\Gamma_\text{o}},
    \label{eq:nadd_down}
\end{equation}
independent of electromechanical system parameters. Without technical noise, the ability to ground-state with the purely optomechanical damping is a necessary and sufficient condition for $N_\text{add,down} < 1$.

\subsection{Effects of technical noise}
\label{sub:technoise}
We define technical noise broadly to encompass all pump power-dependent effects on the transducer noise performance other than heating of the thermal bath $n_\text{th}$ to which the mechanical mode is coupled. Noise sources that fall into this category include amplitude and phase fluctuations of the incident microwave or optical pumps at detuning $\omega_\text{m}$, fluctuations of resonator parameters at frequency $\omega_\text{m}$ (e.g., fluctuations of the superconducting circuit resonant frequency due to two-level systems~\cite{gao_noise_2007}), and heating of the superconducting circuit by either of the pumps. All of these noise sources present either as an effective microwave mode thermal occupancy $n_\text{eff,e}$ coupled to the membrane mode at rate $\Gamma_\text{e}$ or as an effective optical mode occupancy $n_\text{eff,o}$ coupled to the membrane mode at rate $\Gamma_\text{o}$~\cite{aspelmeyer_cavity_2014,jayich_cryogenic_2012}. Thus the formalism presented in the main text can be used to model the effects of technical noise on sideband cooling without regard to the physical origin of the noise.

When readout of mechanical motion is performed using a pump coupled to a bath with nonvanishing effective occupancy $n_\text{eff}$, two other effects of technical noise on the output spectrum must be considered. First, technical noise will generate an additional contribution to $N_\text{add}$ distinct from its direct effect on $n_\text{m}$. Second, technical noise from pump or parameter fluctuations can modify the amplitudes of the Stokes and anti-Stokes peaks in the output spectrum, and to correctly infer $n_\text{m}$ we must correct for this effect.

In the remainder of this section, we restrict our focus to technical noise arising from amplitude and phase fluctuations of the incident optical pump, following the formalism of Ref.~\cite{jayich_cryogenic_2012}, as optomechanical readout is used for all the data presented in Figs.~\ref{fig:gscool} and \ref{fig:nadd}. The optical mode effective thermal occupancy due to amplitude and phase fluctuations is given by
\begin{widetext}
\begin{equation}
    n_\text{eff,o} = \frac{1}{4}\epsilon_\text{PC}\frac{\kappa_\text{o,ext}}{\kappa_\text{o}}\mathcal{A}_\text{o}\frac{\kappa_\text{o}^2}{4}\left(\left|B_\text{x}(\omega_\text{m})\right|^2C_\text{xx} + \left|B_\text{y}(\omega_\text{m})\right|^2C_\text{yy} +2\text{Im}\left[B_\text{x}(\omega_\text{m})B_\text{y}^*(\omega_\text{m})\right]C_\text{xy}\right),
    \label{eq:n_effo}
\end{equation}
\end{widetext}
where $B_\text{x}(\omega) = e^{-i\phi}\chi_\text{o}(\omega) + e^{i\phi}\chi_\text{o}^*(-\omega)$ and $B_\text{y}(\omega) = e^{-i\phi}\chi_\text{o}(\omega) - e^{i\phi}\chi_\text{o}^*(-\omega)$ are amplitude and phase noise susceptibilities, $\phi=\text{tan}^{-1}\left(2\Delta_\text{o}/\kappa_\text{o}\right)$ is the optomechanical phase shift, $\chi_\text{o}(\omega) = 1/\left(\kappa_\text{o}/2 -i\left(\omega+\Delta_\text{o}\right)\right)$ is the optical cavity susceptibility, and $C_\text{xx}$, $C_\text{yy}$, and $C_\text{xy}$ are technical noise spectral densities. Specifically, $C_\text{xx}$ and $C_\text{yy}$ are two-sided amplitude and phase noise spectral densities, respectively, while $C_\text{xy}$ is the two-sided amplitude/phase cross-correlation spectral density, subject to the constraint $C_\text{xy}^2 \leq C_\text{xx}C_\text{yy}$. All three spectral densities are evaluated at detuning $\omega_\text{m}$ from the pump, and expressed in units of photons/s/Hz normalized to the shot noise of the beam incident on the optical cavity. They are thus all proportional to pump power, motivating the parameterization $n_\text{eff,o} = a_\text{o}\Gamma_\text{o}$ used in the main text (see Appendix~\ref{sub:gscool} for further discussion).

$C_\text{xx}$, $C_\text{yy}$, and $C_\text{xy}$ can all be modeled as white noise spectral densities in the narrow spectral window of interest around $\pm\omega_\text{m}$. They thus generate frequency-independent excess noise $\tilde{S}_{\text{o},+}$ $\big(\tilde{S}_{\text{o},-}\big)$ over the background in the optical output spectrum around the upper (lower) sideband. These excess noise spectral densities are given by
\begin{widetext}
\begin{equation}
\tilde{S}_{\text{o},\pm} = \frac{1}{4}\epsilon_\text{CL}\epsilon_\text{PC}\left(\left[\left|\rho\right|^2 + \left|\kappa_\text{o,ext}\chi_\text{o}(\pm\omega_\text{m})-1\right|^2\right]\left(C_\text{xx} + C_\text{yy}\right) - 2\text{Re}\left[\rho^*\left(\kappa_\text{o,ext}\chi_\text{o}(\pm\omega_\text{m}) - 1\right)\left(C_\text{xx} + 2iC_\text{xy} - C_\text{yy} \right)\right]\right)
\label{eq:S_tilde}
\end{equation}
\end{widetext}
where $\rho = 1 - \kappa_\text{o,ext}/\left(\kappa_\text{o}/2 - i\Delta_\text{o}\right)$ is the attenuation of the reflected pump due to absorption in the cavity. Amplitude noise $C_\text{xx}$ would contribute to $\tilde{S}_{\text{o},\pm}$ even in the absence of a cavity, while $C_\text{yy}$ and $C_\text{xy}$ contribute because the cavity rotates phase fluctuations into the amplitude quadrature. For an overcoupled cavity at optimal detuning in the resolved sideband limit, phase noise contributes primarily to $\tilde{S}_{\text{o},+}$ and amplitude noise contributes primarily to $\tilde{S}_{\text{o},-}$. In the discussion of added noise in the main text, for which only noise in the upper sideband is relevant, we introduce the symbol $\tilde{N}_\text{out,o} = \tilde{S}_{\text{o},+}(+\omega_\text{m}) = \tilde{S}_{\text{o},+}$, to parallel the notation used for the Lorentzian contribution to the noise.

The final effect we must account for is a modification of the Stokes and anti-Stokes peaks in the optical output spectrum due to correlations between membrane motion and amplitude and phase fluctuations of the incident light. Eqs.~\eqref{eq:usb} and \eqref{eq:lsb} become
\begin{widetext}
\begin{align}
    S_{\text{o},+}(\omega) = 1 + \tilde{S}_{\text{o},+} + \epsilon_\text{CL}&\frac{\kappa_\text{o,ext}}{\kappa_\text{o}}\mathcal{A}_\text{o}\frac{\Gamma_\text{o}}{\Gamma_\text{T}^2/4 + \left(\omega - \omega_\text{m}\right)^2} \nonumber \\
    &\times\left[\Gamma_\text{T}\left(n_\text{m} - \epsilon_\text{PC}\frac{\kappa_\text{o}^2}{4}\text{Re}\big[\tilde{B}_+\big]\right) - 2\epsilon_\text{PC}\left(\omega-\omega_\text{m}\right)\frac{\kappa_\text{o}^2}{4}\text{Im}\big[\tilde{B}_+\big]\right]
    \label{eq:usb_noise}
\end{align}
and
\begin{align}
    S_{\text{o},-}(\omega) = 1 + \tilde{S}_{\text{o},-} + \epsilon_\text{CL}&\frac{\kappa_\text{o,ext}}{\kappa_\text{o}}\mathcal{A}_\text{o}\frac{\Gamma_\text{o}}{\Gamma_\text{T}^2/4 + \left(\omega + \omega_\text{m}\right)^2} \nonumber \\
    &\times\left[\Gamma_\text{T}\left(\frac{n_\text{min,o}}{\mathcal{A}_\text{o}}\left(n_\text{m}+1\right) + \epsilon_\text{PC}\frac{\kappa_\text{o}^2}{4}\text{Re}\big[\tilde{B}_-\big]\right) - 2\epsilon_\text{PC}\left(\omega+\omega_\text{m}\right)\frac{\kappa_\text{o}^2}{4}\text{Im}\big[\tilde{B}_-\big]\right],
    \label{eq:lsb_noise}
\end{align}
where
\begin{align}
    \tilde{B}_\pm = \frac{e^{-i\phi}}{4}&\left(\kappa_\text{o,ext}\left|\chi_\text{o}(\pm\omega_\text{m})\right|^2\left[ B_\text{x}(\mp\omega_\text{m})\left(C_\text{xx} + iC_\text{xy}\right) + B_\text{y}(\mp\omega_\text{m})\left(iC_\text{xy} - C_\text{yy}\right)\right]\right. \nonumber \\ 
    &\left.-\  \vphantom{\left|\chi_\text{o}(\pm\omega_\text{m})\right|^2}\chi_\text{o}^*(\pm\omega_\text{m})\left[\left(B_\text{x}(\mp\omega_\text{m})C_\text{xx} + iB_\text{y}(\mp\omega_\text{m})C_\text{xy}\right)\left(1+\rho\right) + \left(iB_\text{x}(\mp\omega_\text{m})C_\text{xy} - B_\text{y}(\mp\omega_\text{m})C_\text{yy}\right)\left(1-\rho\right)\right]\right).
\end{align}
\end{widetext}

Eqs.~\eqref{eq:usb_noise} and \eqref{eq:lsb_noise} reveal that these correlations both change the peak amplitudes $N_{\text{out,o},\pm}$ and add an anti-symmetric contribution to the lineshape of each peak, where the sign of these extra terms depends on the relative values of $C_\text{xx}$, $C_\text{yy}$, and $C_\text{xy}$. Eq.~\eqref{eq:nadd_supp} for the transducer's upconversion added noise is then modified to account for amplitude and phase noise by the substitution $N_{\text{out,o},+} \rightarrow N_{\text{out,o},+} + \tilde{N}_\text{out,o}$.

\section{Transducer characterization}
\label{sec:characterization}

Transducer device parameters are summarized in Table~\ref{tab:transducer}. Measurements of these parameters are detailed below.

\subsection{Microwave circuit characterization}
\label{sub:microwave}

\begin{table*}
\caption{Electro-optic transducer parameters. The range of values for the microwave circuit linewidth $\kappa_\text{e}$ and the electromechanical transducer gain $\mathcal{A}_\text{e}$ correspond to the range of pump powers used in Fig.~\ref{fig:nadd}. For the optical cavity modematchings and transducer efficiency, the first (second) numbers correspond to the data shown in Fig.~\ref{fig:gscool} (Fig.~\ref{fig:nadd}) } 
  \begin{center}
    \begin{tabular}{ |p{7.2cm}|p{2.0cm} |p{3.8cm}| }
    \hline
    \textbf{Parameter} & \textbf{Symbol} & \textbf{Value} \\ \hline
    Optical cavity frequency &
    $\omega_\textrm{o}$ & $\omega_\textrm{o}/2\pi = 277 $~THz \\ \hline 

    Optical cavity external coupling & $\kappa_\textrm{o,ext}$ & $\kappa_\textrm{o,ext}/2\pi = 2.12$~MHz \\ \hline
    
    Optical cavity linewidth &
    $\kappa_\textrm{o}$ & $\kappa_\textrm{o}/2\pi = 2.68$~MHz \\ \hline
    
    Microwave circuit frequency &
    $\omega_\textrm{e}$ & $\omega_\textrm{e}/2\pi = 7.938 $~GHz \\ \hline
    
    Microwave circuit external coupling &
    $\kappa_\textrm{e,ext}$ & 
    $\kappa_\textrm{e,ext}/2\pi = 1.42$~MHz \\ \hline
    
    Microwave circuit linewidth &
    $\kappa_\textrm{e}$ & 
    $\kappa_\textrm{e}/2\pi = 1.64 - 2.31$~MHz \\ \hline
    
    Mechanical mode frequency &
    $\omega_\textrm{m}$ & $\omega_\textrm{m}/2\pi = 1.451 $~MHz \\ \hline
    
    Intrinsic mechanical dissipation rate &
    $\gamma_\textrm{m}$ & $\gamma_\textrm{m}/2\pi = 113 $~mHz \\ \hline 
    
    Vacuum optomechanical coupling &
    $g_\textrm{o}$ & $g_\textrm{o}/2\pi = 60$~Hz \\ \hline
    
    Vacuum electromechanical coupling &
    $g_\textrm{e}$ & $g_\textrm{e}/2\pi = 1.6$~Hz \\ \hline
    
    Modematching of pump beam to cavity mode & $\epsilon_\text{PC}$ & $\epsilon_\text{PC} = 0.86~(0.80)$ \\ \hline

    Modematching of cavity mode to LO beam & $\epsilon_\text{CL}$ & $\epsilon_\text{CL} = 0.91~(0.79)$ \\ \hline

    Bidirectional modematching & $\epsilon = \sqrt{\epsilon_\text{PC}\epsilon_\text{CL}}$ & $\epsilon = 0.88~(0.79)$ \\ \hline

    Matched efficiency (peak efficiency) & $\eta_\text{M}~(\eta_\text{t,max})$ & $\eta_\text{M} = 55\%~(\eta_\text{t,max} = 49\%)$ \\ \hline

    Optomechanical transducer gain & $\mathcal{A}_\text{o}$ & $\mathcal{A}_\text{o} = 1.22$ \\ \hline

    Electromechanical transducer gain & $\mathcal{A}_\text{e}$ & $\mathcal{A}_\text{e} = 1.08 - 1.16$ \\ \hline

    \end{tabular}
  \end{center}
  \label{tab:transducer}
\end{table*}

We characterize the microwave circuit by measuring the reflection scattering parameter $\mathcal{S}_\text{ee}(\omega)$ encoding its amplitude and phase response as a function of detuning from resonance. When normalized to the off-resonance level, this scattering parameter is 
\begin{equation}
    \mathcal{S}_\text{ee}(\omega) = \frac{\kappa_\text{e,ext}}{\kappa_\text{e}/2 - i(\omega-\omega_\text{e})} - 1.
\end{equation}
We use a vector network analyzer to generate a weak swept probe tone and inject it into the fridge together with the strong pump tone at fixed detuning $\Delta_\text{e} = -\omega_\text{m}$ from $\omega_\text{e}$. Sweeping the microwave pump power $P_\text{e}$ and fitting $\mathcal{S}_\text{ee}(\omega)$ then allows us to measure the power-dependence of the microwave circuit's internal loss $\kappa_\text{e,int}$ (plotted in Fig.~\ref{fig:AMWNoise}(a) in Appendix~\ref{sec:paramnoise}) as well as the power-independent microwave circuit external coupling rate $\kappa_\text{e,ext}/2\pi = 1.42$~MHz. 

For the data shown in Fig.~\ref{fig:nadd}, the incident microwave pump power was swept over the range $0.2~\text{ nW} < P_\text{e} < 10 \text{ nW}$, and the circuit's total linewidth $\kappa_\text{e}/2\pi=\left(\kappa_\text{e,ext}+\kappa_\text{e,int}\right)/2\pi$ varied between 1.64~MHz and 2.31~MHz over this range. 

\subsection{Optical cavity characterization}
\label{sub:optical}

We also characterize the optical cavity by measuring its reflection response, and the relevant scattering parameter is formally analogous to $\mathcal{S}_\text{ee}(\omega)$. We tune the laser wavelength to put the lock beam close to resonance with the optomechanical cavity, and then sweep the frequency of the laser relative to that of the optomechanical cavity by ramping the voltage applied to a piezoelectric actuator on the filter cavity to which the laser is locked (see Appendix~\ref{sec:chain}). This scheme measures $\left|\mathcal{S}_\text{oo}(\omega)\right|^2$, with the horizontal axis calibrated using the phase modulation sidebands on the lock beam. Fitting the data then yields $\kappa_\text{o}/2\pi = 2.68\pm0.05$~MHz for the optical cavity linewidth.

To optimize and measure the modematching of the cavity mode to the pump or LO beams ($\epsilon_\text{PC}$ and $\epsilon_\text{CL}$, respectively; see Appendix~\ref{sub:eta_theory}), we inject an additional probe beam into the cavity through the low-transmission back mirror and steer the transmitted mode to a detector where it interferes with the beam of interest. We then equalize the power of the beams, measure the visibility $\mathcal{V}$ of the interference fringes that result from a relative detuning between the beams, and define the associated modematching as $\mathcal{V}^2$. 

The modematching factors could change slightly over the course of a cooldown as beam alignments drifted. The data shown in Fig.~\ref{fig:gscool} in the main text was acquired immediately after the modematchings were reoptimized and measured to be $\epsilon_\text{PC} = 0.86$ and $\epsilon_\text{CL} = 0.91$. Both before and after the acquisition of the data shown in Fig.~\ref{fig:nadd}, the modematchings were measured to be $\epsilon_\text{PC} = 0.80$ and $\epsilon_\text{CL} = 0.79$. Once $\epsilon_\text{PC}$ and the cavity linewidth $\kappa_\text{o}$ are known, the external coupling rate $\kappa_\text{o,ext}/2\pi = 2.12$~MHz is inferred from the depth of dip on resonance in a measurement of $\left|\mathcal{S}_\text{oo}(\omega)\right|^2$ in which the pump beam is used to probe the cavity. The modematching of the lock beam to the cavity mode was not directly measured. We infer $\epsilon_\text{lock} = 0.85$ using the depth of dip in the lock beam $\left|\mathcal{S}_\text{oo}(\omega)\right|^2$ measurement together with the measured value of $\kappa_\text{o,ext}$.

One final modematching factor that does not involve the cavity mode is also relevant to the calibration of the efficiency (see Appendix~\ref{sub:eta}). We define $\epsilon_\text{PL}$ as the modematching of the LO beam to the promptly reflected pump beam with the cavity unlocked. We measured this modematching to be $\epsilon_\text{PL} = 0.75$ for the data shown in Fig.~\ref{fig:gscool} and $\epsilon_\text{PL} = 0.79$ for the data shown in Fig.~\ref{fig:nadd}.

For the data shown in Fig.~\ref{fig:gscool}, the optical pump power incident on the cavity was swept over the range $20~\text{ nW} < P_\text{o} < 1.7 \text{ }\si{\micro{W}}$. For the data shown in Fig.~\ref{fig:lasernoise}, the relevant quantity is not the incident power $P_\text{o}$ but rather the intracavity circulating power
\begin{equation}
    P_\text{circ} = \Delta\nu_\text{FSR}\,\kappa_\text{o,ext}\left(\frac{\epsilon_\text{PC}P_\text{o}}{\kappa_\text{o}^2/4 + \Delta_\text{o}^2} + \frac{\epsilon_\text{lock}P_\text{lock}}{\kappa_\text{o}^2/4 + \Delta_\text{lock}^2}\right),
\end{equation}
where $\Delta\nu_\text{FSR} = c/2L$ is the free spectral range of the optical cavity in frequency units.
Uncertainty in $P_\text{circ}$ comes from uncertainty in the cavity length $L$ and uncertainty in the measured value of the optical insertion loss within the cryostat. 

\subsection{Optomechanical and electromechanical characterization}
\label{sub:mechanical}

To study the vibrational modes of the membrane, we exploit the phenomenon of optomechanically induced transparency (OMIT), wherein $\mathcal{S}_\text{oo}(\omega)$ is modified by optomechanical interference effects for a swept probe tone coherent with the pump field~\cite{aspelmeyer_cavity_2014}, as well as the analogous phenomenon of electromechanically induced transparency (EMIT). We generate these coherent probe tones using SSB modulation of either the microwave pump (for EMIT measurements) or the RF drive to the AOM that controls the optical pump beam detuning (for OMIT measurements; see Appendix~\ref{sec:chain}), with the pump frequency fixed in both cases. These measurements reveal the vibrational spectrum of the membrane over a wide frequency range (see Fig.~\ref{fig:AMechanics}(c)); we then identify the (2,2) mode and sweep over a narrower range for further characterization. The mechanical features in OMIT and EMIT generally have Fano lineshapes, with linewidth $\Gamma_\text{T}$. We can also obtain $\Gamma_\text{T}$ by transducing a swept signal to measure the mechanical mode's Lorentzian transmission profile (see Appendix~\ref{sub:eta_theory}), using SSB modulation of the microwave pump and optical heterodyne detection to measure $\mathcal{S}_\text{oe}(\omega)$ or SSB modulation of the optical pump and microwave heterodyne detection to measure $\mathcal{S}_\text{eo}(\omega)$.

The membrane mode quality factor $Q_\text{m}=\omega_\text{m}/\gamma_\text{m}$ of the (2,2) mode is sufficiently high that the frequency-domain characterization described above becomes unwieldy, so we interrogate the mechanical mode at low damping using a purely electromechancial time-domain measurement ($\Gamma_\text{o} = 0$). We apply a low-power microwave pump tone at fixed red detuning $\Delta_\text{e} = -\omega_\text{m}$ from $\omega_\text{e}$, turn on a temporary SSB probe tone at detuning $+\omega_\text{m}$ from the pump to ring up the mechanical mode, then turn off the probe tone and demodulate the microwave heterodyne signal at frequency $\omega_\text{m}$ to monitor the exponential decay of the membrane oscillations at rate $\Gamma_\text{T} = \Gamma_\text{e} + \gamma_\text{m}$. By repeating this measurement at different microwave pump powers $P_\text{e}$ we map out $\Gamma_\text{T}$ vs.\ $P_\text{e}$ (Fig.~\ref{fig:AMechanics}(d)). Extrapolating to $P_\text{e} = 0$ then yields the bare mechanical linewidth $\gamma_\text{m}/2\pi = 108$~mHz, corresponding to a quality factor $Q_\text{m}=1.3\times10^7$. Performing these ringdown measurements electromechanically rather than optomechanically allows us to avoid optomechanical damping from the lock beam (See Appendix~\ref{sub:misc}). Measurements of the temperature-dependence of $\gamma_\text{m}$, discussed further in Appendix~\ref{sub:calibration}, indicate that $\gamma_\text{m}/2\pi = 113$~mHz is actually the relevant value for the data presented in the main text because the membrane equilibrates to an elevated temperature in the presence of laser light.

We use this same low-power measurement of $\Gamma_\text{T}$ vs.\ $P_\text{e}$ and Eq.~\eqref{eq:Gamma_e} to infer the vacuum electromechanical coupling $g_\text{e}$ (the electromechanical damping $\Gamma_\text{e}$ does not scale linearly with $P_\text{e}$ at high power because of power-dependent microwave circuit loss).
The slope of the linear fit in Fig.~\ref{fig:AMechanics}(d) yields $g_\text{e} = 1.6 \pm 0.2$~Hz, with the uncertainty dominated by uncertainty in the cryostat insertion loss. Together with the zero-point motion obtained from simulation (see Appendix~\ref{sub:design}), this measurement implies that the electromechanical frequency shift per unit membrane displacement is $G_\text{e} = 3.2$~Hz/fm. Solving Eq.~\eqref{eq:G_e} self-consistently for the capacitor pad separation $d$ and the participation ratio $p(d)$ then yields $d=830$~nm and $p=0.67$. An analogous linear fit to $\Gamma_\text{T}$ vs.\ $P_\text{o}$ at $\Delta_\text{o} = -\omega_\text{m}$ in a purely optomechanical measurement yields $g_\text{o}/2\pi = 60 \pm 6$~Hz, with uncertainty again due to insertion loss.
The corresponding frequency shift $G_\text{o}/2\pi = 70 \pm 10$~Hz/fm is consistent with operation on the lower local maximum of optomechanical coupling in the top panel of Fig.~\ref{fig:AOptSim}(b), and the associated value of $\kappa_\text{o,ext}$ from the simulation (bottom panel) matches the measured value $\kappa_\text{o,ext}/2\pi = 2.12$~MHz.

The capacitor pad separation inferred from the electromechanical damping sweep is substantially larger than those previously observed in similar flip-chip devices. The device used in Ref.~\cite{higginbotham_harnessing_2018} had $d=300$~nm, and purely electromechanical devices tested since then have regularly achieved gaps in the 150 $-$ 200~nm range~\cite{burns_reducing_2019}. The large gap in this device was likely a consequence of modifications to the fabrication procedure required to integrate the phononic shield into the flip-chip architecture. In past devices without phononic shielding, the top chip was supported by two inner posts near the membrane as well as the four outer posts described in Appendix~\ref{sub:design} above, but these inner posts reduced the effectiveness of the phononic shield in simulations. 

We are presently exploring the effects of aligning inner posts with nodal lines of the membrane mode's transverse displacement profile to mitigate their impact on phononic shield performance. Another cause of unreliable pad separation is the thermal contraction of the epoxy used to affix the top chip to the middle chip, and we are investigating the Si-SiO$_{2}$ bonding technique used to affix the bottom and middle chips as an alternative to epoxy in future devices. Achieving $d=200$~nm would boost $g_\text{e}$ by a factor of 5, corresponding to an enhancement of the damping per incident microwave photon by a factor of 25 at low pump power (the improvement would be larger at higher power because the internal loss of the microwave circuit would be smaller at the power required to yield a given damping). Without any other device improvements, this increased electromechanical coupling would yield a membrane mode occupancy of $n_\text{m} = 0.4$~phonons and upconversion added noise $N_\text{add,up} = 0.9$~photons/s/Hz for $\Gamma_\text{e} = \Gamma_\text{o} = 2\pi\times1$~kHz.

\subsection{Characterization of other parameters}
\label{sub:misc}
For all the measurements described above, the microwave pump detuning $\Delta_\text{e}$ is set to the optimal value $-\omega_\text{m}$ by stepping up the pump frequency until it begins to ring up the membrane mode for $\Delta_\text{e}>0$, and then stepping the pump back down by $\omega_\text{m}$. We use a similar though slightly more involved procedure to set the pump and lock beam detunings $\Delta_\text{o}$ and $\Delta_\text{lock}$ to the desired values. First, with the pump beam blocked, we adjust the PDH error signal bias while monitoring ringing of the lock to set $\Delta_\text{lock}=0$. We then unblock the pump beam and adjust the AOM frequency that controls $\omega_\text{pump}$ (see Appendix~\ref{sec:chain}) while monitoring the ringing of the lock to set $\Delta_\text{o} = -\omega_\text{m}$. In doing so we measure the cavity birefringence to be $\Delta_\text{B}/2\pi = 2.4$~MHz. Finally, we adjust the PDH error signal bias to slightly red-detune the lock beam. The lock beam detuning can be set to $\sim10$~kHz precision by measuring how far $\omega_\text{pump}$ must be adjusted to maintain $\Delta_\text{o} = -\omega_\text{m}$. The lock beam optomechanical damping $\gamma_\text{lock}$ is determined by comparing the mechanical mode linewidth obtained from an $\mathcal{S}_\text{oe}$ measurement at low damping and $\Delta_\text{lock}=0$ to the value obtained from a measurement with the same $\Gamma_\text{e}$ and $\Gamma_\text{o}$ but $\Delta_\text{lock}\neq0$. The lock beam damping was $\gamma_\text{lock}/2\pi = 5$~Hz during the measurements shown in Fig.~\ref{fig:gscool} and $\gamma_\text{lock}/2\pi = 2$~Hz during the measurements shown in Fig.~\ref{fig:nadd}.

The values of $\gamma_\text{lock}$ and $\Delta_\text{lock}$ are important because any contribution to the optomechanical or electromechanical damping of the membrane mode also couples the mode to a backaction bath as described in Appendix~\ref{sub:sideband} above. This coupling cannot be neglected despite the low damping $\gamma_\text{lock}$ because the small detuning $\Delta_\text{lock}$ leads to a large backaction bath occupancy $n_\text{min,lock}$. Indeed, for $\Delta_\text{lock} \ll \omega_\text{m}$, the product $\gamma_\text{lock}n_\text{min,lock}$ is independent of detuning, and can only be reduced by reducing the incident lock power $P_\text{lock}$. To fit the data in Figs.~\ref{fig:gscool} and \ref{fig:nadd} we thus modify Eqs.~\eqref{eq:gscool} and \eqref{eq:nadd} to include a $\gamma_\text{lock}n_\text{min,lock}$ term in the numerator, fixed at $\gamma_\text{lock}n_\text{min,lock} = 2\pi\times40$~Hz by the independent measurements described above. This term results in the increase in the backaction limit at low damping in Fig.~\ref{fig:gscool}.

With optimally detuned pumps, the measured optical cavity linewidth $\kappa_\text{o}$ above yields $\mathcal{A}_\text{o} = 1.22$ for the optomechanical transducer gain (see Appendix~\ref{sub:eta_theory}), and a backaction-limited membrane mode occupancy $n_\text{min,o} = 0.22$ for optomechanical ground-state cooling. Likewise, power-dependence of the microwave mode linewidth implies an electromechanical transducer gain $\mathcal{A}_\text{e}$ that varies between 1.08 and 1.16 over the range of microwave pump power $P_\text{e}$ used in this work. The net transducer gain $\mathcal{A}_\text{e}\mathcal{A}_\text{o}$ thus varies between 1.32 and 1.42. 

For the data shown in Fig.~\ref{fig:gscool}, with fixed electromechanical damping $\Gamma_\text{e}/2\pi = 100$~Hz, the microwave mode linewidth was $\kappa_\text{e}/2\pi = 1.79$~MHz, yielding $\eta_\text{M,exp}=0.55$ for the expected value of the matched efficiency. For the data shown in Fig.~\ref{fig:nadd}(a), $\eta_\text{M}$ was not constant as a result of the power-dependent microwave loss. In these measurements the transducer efficiency $\eta_{t}$ was maximized at $\Gamma_\text{e}/2\pi=75$~Hz, where $\kappa_\text{e}/2\pi = 1.75$~MHz. The expected maximum efficiency $\eta_\text{t,max} = 49\%$ comes from evaluating Eq.~\eqref{eq:eta} at $\Gamma_\text{e}/2\pi=75$~Hz and $\Gamma_\text{o}/2\pi=85$~Hz using this value of the microwave circuit linewidth and other independently measured parameters.

\section{Analysis Details}
\label{sec:analysis}

\subsection{Transducer efficiency}
\label{sub:eta}

We measure the transducer efficiency $\eta_\text{t}$ using the method developed in Ref.~\cite{andrews_bidirectional_2014}, wherein measurements of upconversion, downconversion, and off-resonance reflection from both microwave and optical resonators can be used to calibrate out unknown path losses and gains. More precisely, a set of network analyzer measurements at fixed $\Gamma_\text{e}$ and $\Gamma_\text{o}$ yield
\begin{equation}
    \sqrt{\frac{\left(\alpha\left|\mathcal{S}_\text{oe}(\omega_\text{m})\right|^2\delta\right)\left(\gamma\left|\mathcal{S}_\text{eo}(\omega_\text{m}\right|^2\beta\right)}{\left(\alpha\left|\mathcal{S}_\text{ee,off}\right|^2\beta\right)\left(\epsilon_\text{PL}\gamma\left|\mathcal{S}_\text{oo,off}\right|^2\delta\right)}} = \frac{\mathcal{A}_\text{e}\mathcal{A}_\text{o}}{\sqrt{\epsilon_\text{PL}}}\eta_\text{t}, 
    \label{eq:eta_meas}
\end{equation}
where $\alpha$ is the insertion loss of the microwave path from the HF2LI lock-in amplifier output to the external port of the electromechanical circuit, $\beta$ is the gain of the microwave path from the electromechanical circuit output to the HF2LI input, $\gamma$ is the insertion loss of the RF and optical path from the HF2LI output to the external port of the optical cavity, $\delta$ is the efficiency of the optical path from the optical cavity output to the HF2LI input. With path losses and gains defined this way, the ``e'' and ``o'' ports coincide with the external ports of the microwave and optical resonators, so $\left|\mathcal{S}_\text{ee,off}\right|^2 = \left|\mathcal{S}_\text{oo,off}\right|^2 = 1$ by construction.
In our heterodyne detection scheme, a factor of the pump beam/LO modematching $\epsilon_\text{PL}$ appears in the measurement of the prompt reflection off the optical cavity, as also noted in Ref.~\cite{higginbotham_harnessing_2018}, and must be calibrated out with an independent measurement (see Appendix~\ref{sub:optical}).

During the acquisition of the data shown in Fig~\ref{fig:gscool}(c), a mixer in the RF~chain used for single sideband modulation of the optical pump (Fig.~\ref{fig:AChain}) was underdriven, which may have resulted in variation of $\gamma$ between measurements of $\left|\mathcal{S}_\text{eo}\right|^2$ and $\left|\mathcal{S}_\text{oo,off}\right|^2$, and thus a miscalibration of the efficiency at each data point. We suspect this was the origin of the discrepancy between the expected matched efficiency $\eta_\text{M,exp}$ and the value $\eta_\text{M,fit}$ obtained from fitting the data in Fig.~\ref{fig:gscool}(c). The alternative explanation is miscalibration of the parameters that determine $\eta_\text{M,exp}$, but this seems unlikely in light of the good agreement between the measured and expected values of $\eta_\text{t,max}$ obtained from the data in Fig.~\ref{fig:nadd}(a).

The fit to efficiency vs.\ damping must be modified when the electromechanical damping $\Gamma_\text{e}$ is swept, as in Fig.~\ref{fig:nadd}(a), because of the power-dependent microwave circuit loss. To process this data we define $\zeta_\text{t} = \eta_\text{t}/(\kappa_\text{e,ext}/\kappa_\text{e})$ and fit $\zeta_\text{t}$ vs.\ damping to Eq.~\eqref{eq:eta}, then compare $\zeta_\text{M}=0.59$ obtained from the fit to the expected value $\zeta_\text{M}=\epsilon\kappa_\text{o,ext}/\kappa_\text{o} = 0.63$. The theory curve shown in Fig.~\ref{fig:nadd}(a) is obtained by multiplying the fit to $\zeta_\text{t}$ vs.\ damping data by a polynomial fit to the power-dependent microwave mode overcoupling ratio $\kappa_\text{e,ext}/\kappa_\text{e}$. Because of the power-dependent loss, the efficiency is maximized at $\Gamma_\text{e} = 2\pi\times75$~Hz rather that at $\Gamma_\text{e} = \Gamma_\text{o} = 2\pi\times85$~Hz.

\subsection{Calibration of spectra and temperature sweeps }
\label{sub:calibration}

\begin{figure*}[!t]
    \centering
    \includegraphics{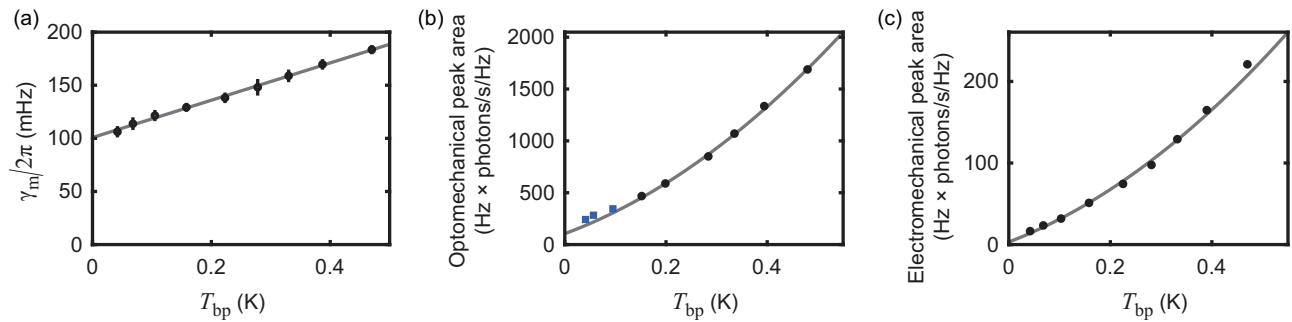}
    \caption{
    \textbf{Temperature sweeps.}
    (a) Intrinsic mechanical dissipation rate $\gamma_\text{m}$ vs.\ cryostat base plate temperature $T_\text{bp}$, with linear fit.
    (b) Area under mechanical peak in optical output spectrum vs.\ $T_\text{bp}$. The quadratic fit (gray) calibrates the optical measurement chain efficiency. Blue squares indicate data excluded from the fit, as the membrane mode thermalizes poorly to the base plate at low temperatures in the presence of the lock beam. (c) Area under mechanical peak in microwave output spectrum vs.\ $T_\text{bp}$ in the absence of optical illumination. All error bars represent one standard deviation.
    }
    \label{fig:ATempSweeps}
\end{figure*}

The measured optical heterodyne noise spectra are normalized to the sum of LO beam shot noise and heterodyne detector dark noise, obtained by blocking the pump beam incident on the heterodyne detector. These spectra are related to the transducer output-referred spectra (Eqs.~\eqref{eq:usb_noise} and \eqref{eq:lsb_noise}) by $S_{\text{det,o},\pm}(\omega) = \left(1-\xi_\text{o}\right) + \xi_\text{o}S_{\text{o},\pm}(\omega)$, where $\xi_\text{o}$ is the efficiency of the optical measurement chain, modeled as an effective beamsplitter between the transducer output and the input of an ideal heterodyne detector. Measured microwave heterodyne noise spectra are likewise normalized to the noise measured in the absence of the pump tone, given by the sum of vacuum noise, residual thermal noise at the base plate temperature $T_\text{bp} = 40$~mK, and the added noise of the microwave measurement chain. As in the optomechanical case, the spectra normalized this way are related to spectra normalized at the transducer's microwave output port by $S_{\text{det,e},\pm}(\omega) = \left(1-\xi_\text{e}\right) + \xi_\text{e}S_{\text{e},\pm}(\omega)$, where $\xi_\text{e}$ is the efficiency of the microwave measurement chain. The measurement chain efficiencies $\xi_\text{e}$ and $\xi_\text{o}$ appearing in these expressions can equivalently be expressed in terms of the added noise of the respective measurement chains referred to the transducer output, via $\xi_\text{e} = 1/\left(N_{\xi,\text{e}} + 1/2\right)$ and $\xi_\text{o} = 1/\left(N_{\xi,\text{o}} + 1/2\right)$.

As a cross-check on Stokes/anti-Stokes sideband ratio thermometry using Eq.~\eqref{eq:sidebandtherm}, we can use Eq.~\eqref{eq:usb} to infer the membrane mode occupancy $n_\text{m}$ from the upper sideband spectrum $S_{\text{det,o},+}(\omega)$ given independent measurements of $\xi_\text{o}$ and parameters in Table~\ref{tab:transducer}. To determine $\xi_\text{o}$, we sweep the cryostat base temperature $T_\text{bp}$ while measuring the optical noise spectrum, at low damping such that technical noise is negligible. At sufficiently high temperature, the local thermal environment of the membrane will equilibrate to $T_\text{bp}$, and thus we can replace the base-temperature equilibrium occupancy $n_\text{th}$ in Eq.~\eqref{eq:gscool} with $n_\text{bp} = k_\text{B}T_\text{bp}/\hbar\omega_\text{m}$. Then, if other parameters are independent of temperature, the measurement chain efficiency $\xi_\text{o}$ can be obtained from the slope of $N_{\text{det,o},+}(T_\text{bp})$, where $N_{\text{det,o},+}$ is the optomechanical upper sideband amplitude normalized at the detector input ($N_{\text{det,o},+} = \xi_\text{o}N_{\text{out,o},+}$ in the absence of technical noise). As $T_\text{bp}$ decreases, the membrane mode may decouple from the fridge base plate and equilibrate instead to a thermal bath at some elevated temperature $T_\text{eq} > T_\text{bp}$. The temperature sweep enables us to identify $T_\text{eq}$ as the temperature below which $N_{\text{det,o},+}(T_\text{bp})$ deviates from the expected linear behavior, and the corresponding thermal occupancy $n_\text{th} = k_\text{B}T_\text{eq}/\hbar\omega_\text{m}$ can be compared to the values obtained from fitting the data in Fig.~\ref{fig:gscool}(a) and Fig.~\ref{fig:nadd}(b).

While sweeping the base plate temperature $T_\text{bp}$, we repeated sets of electromechanical ringdown measurements of the sort shown in Fig.~\ref{fig:AMechanics}(d) and observed that the mechanical dissipation rate $\gamma_\text{m}$ increased linearly with temperature (Fig.~\ref{fig:ATempSweeps}(a)). Fitting to this data to $\gamma_\text{m} = a_\gamma T_\text{bp} + b_\gamma$ yields $a_\gamma/2\pi = 176$~mHz/K and $b_\gamma/2\pi = 101$~mHz. This behavior will generate quadratic terms in $N_{\text{det,o},+}(T_\text{bp})$ and $N_{\text{det,e},+}(T_\text{bp})$, as the membrane mode's coupling to the temperature-dependent bath is itself temperature-dependent.

The optomechanical temperature sweep data is shown in Fig.~\ref{fig:ATempSweeps}(b). We obtain peak amplitudes $N_{\text{det,o},+}$ and peak widths $\Gamma_\text{T}$ from Lorentzian fits to the spectra at different temperatures, and fit the temperature-dependence of the peak area rather than the peak amplitude to control for fluctuations in damping. We parameterize the temperature-dependence as $N_{\text{det,o},+}\Gamma_\text{T} = a_\xi(a_\gamma T_\text{bp}^2 + b_\gamma T_\text{bp}) + b_\xi$ with $a_\gamma$ and $b_\gamma$ fixed by the fit to the data in Fig.~\ref{fig:ATempSweeps}(a), where the theoretical values of the coefficients are 
\begin{equation}
a_\xi = 4\xi_\text{o}\epsilon_\text{CL}\frac{\kappa_\text{o,ext}}{\kappa_\text{o}}\mathcal{A}_\text{o}\frac{\Gamma_\text{o}}{\Gamma_\text{T}}\frac{k_\text{B}}{\hbar\omega_\text{m}}
\label{eq:tempsweep_a}
\end{equation}
and
\begin{equation}
b_\xi = 4\xi_\text{o}\epsilon_\text{CL}\frac{\kappa_\text{o,ext}}{\kappa_\text{o}}\mathcal{A}_\text{o}\frac{\Gamma_\text{o}}{\Gamma_\text{T}}\left(\gamma_\text{lock}n_\text{min,lock} + \Gamma_\text{o}n_\text{min,o}\right).
\label{eq:tempsweep_b}
\end{equation}
At low temperature, the data begins to deviate from quadratic temperature-dependence, indicating an elevated membrane mode equilibration temperature. We obtain the best fit when excluding the three lowest-temperature data points, marked in blue in Fig.~\ref{fig:ATempSweeps}(b). Then, since all other parameters are independently measured, the best-fit value of $a_\xi$ yields $\xi_\text{o} = 0.276 \pm 0.007$, while $b_\xi$, whose fractional uncertainty is much larger, yields $\xi_\text{o} = 0.27 \pm 0.07$. The mode equilibration temperature $T_\text{eq}$, defined as the temperature at which the area predicted by the fit is equal to the measured base-temperature peak area, is inferred to be $T_\text{eq} = 70\pm 10$~mK, with uncertainty again dominated by the fractional error in $b_\xi$. 

\begin{figure}[!t]
    \centering
    \includegraphics{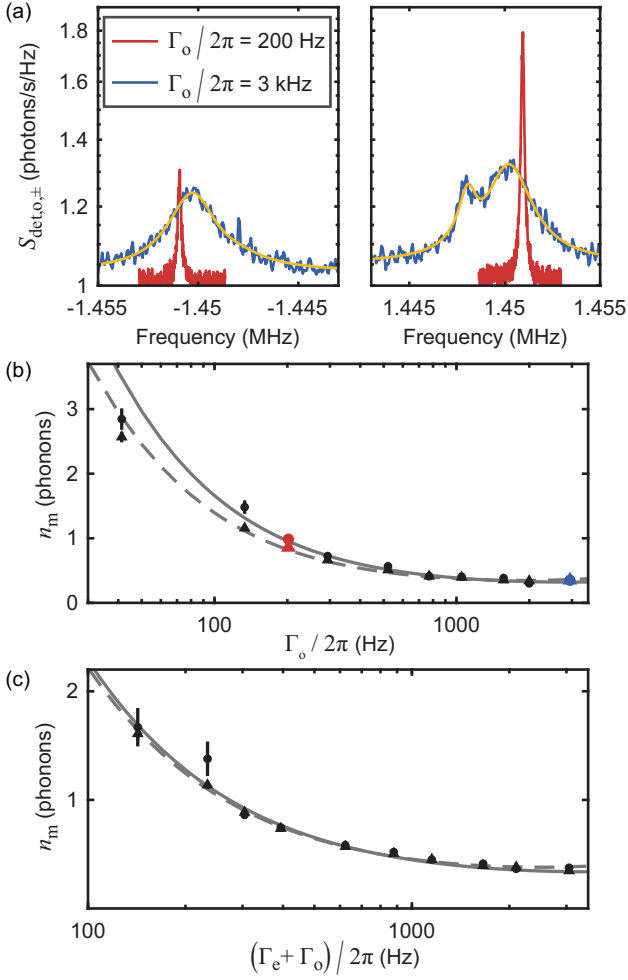}
    \caption{
    \textbf{Analysis of optomechanical spectra.}
    (a) Two examples of optomechanical spectra normalized to shot noise at the optical heterodyne detector input. The left (right) panel displays the lower (upper) sideband spectra. At high damping, we fit the spectra to the squared sum of two complex Lorentzian amplitudes (yellow line) to account for the $1.448$~MHz mode of the silicon substrate. 
    (b) Comparing two methods of inferring membrane mode occupancy $n_\text{m}$ from purely optomechanical damping sweep, with corresponding fits. Circles and solid line (also shown in Fig.~\ref{fig:gscool}(a)) are obtained using sideband asymmetry thermometry, while triangles and dashed line are obtained from the upper-sideband spectra $S_\text{det,o,+}$ and optical measurement chain efficiency $\xi_{o}$. Red and blue data points correspond to the spectra shown in panel (a). (c) Comparing the two methods of inferring $n_\text{m}$ for optomechanical damping sweep at fixed electromechanical damping (sideband asymmetry data and fit also shown in Fig.~\ref{fig:gscool}(b)). All error bars represent one standard deviation.
    }
    \label{fig:ASpectra}
\end{figure}

Finally, we explore how the values of the measurement chain efficiency $\xi_\text{o}$ and the equilibration temperature $T_\text{eq}$ depend on the decision to exclude low-temperature points from the fit. Excluding three points yields the best agreement between the values of $\xi_\text{o}$ inferred from Eqs.~\eqref{eq:tempsweep_a} and \eqref{eq:tempsweep_b}. When only the lowest point is excluded, Eq.~\eqref{eq:tempsweep_b} yields $\xi_\text{o} = 0.37 \pm 0.05$, two standard deviations away from the value obtained from $a_\xi$, which is very insensitive to the low-temperature data. Excluding two or four points does not change the fit very much, and the inferred equilibration temperature remains within the range quoted above. We thus take $\xi_\text{o} = 0.276 \pm 0.007$ and $T_\text{eq} = 70\pm 10$~mK as our final values for the measurement chain efficiency and equilibration temperature. The latter corresponds to $n_\text{th} = 1000 \pm 140$, which agrees well with the values obtained from independent fits to the data in Figs.~\ref{fig:gscool}(a) and \ref{fig:nadd}(b).

We follow an analogous procedure to measure the microwave measurement chain efficiency $\xi_\text{e}$ via the temperature-dependence of the electromechanical upper sideband amplitude $N_{\text{det,e},+}$. The analysis of the electromechanical temperature sweep data, shown in Fig.~\ref{fig:ATempSweeps}(c), is very similar to the optomechanical analysis and will not be discussed in detail. The fit yields $\xi_\text{e} = 0.029 \pm 0.001$, with no evidence for equilibration at an elevated temperature. We use this value of $\xi_\text{e}$ to calibrate broadband measurements of the noise emerging from the external port of the microwave circuit, with which we study the dependence of $n_\text{eff,e}$ on microwave and optical power (Figs.~\ref{fig:AMWNoise}(b) and \ref{fig:lasernoise}(b), respectively).

The electromechanical temperature sweep data indicates that laser light is responsible for the elevated equilibration temperature inferred from Figs.~\ref{fig:ATempSweeps}(b), \ref{fig:gscool}(a), and \ref{fig:nadd}(b). As noted in the main text, the fact that the optomechanical ground-state cooling data plotted in Fig.~\ref{fig:gscool}(a) agrees well with Eq.~\eqref{eq:gscool} without additional power-dependent terms indicates that membrane mode heating saturates at the low power used to lock the cavity, with no further scaling with the power of the pump beam. Similar low-power saturation was observed in membrane optomechanical systems in Refs.~\cite{higginbotham_harnessing_2018} and \cite{peterson_laser_2016}. The elevated $T_\text{eq}$ implies that the appropriate value of the intrinsic mechanical dissipation rate $\gamma_\text{m}$ for analyzing the data in the main text is the 70~mK value $\gamma_\text{m}/2\pi = 113$~mHz rather than the 40~mK value $\gamma_\text{m}/2\pi = 108$~mHz.

We can compare the temperature sweep calibration of the optical measurement chain efficiency $\xi_\text{o}$ to a more direct but less precise measurement. From power meter measurements we infer the efficiency of the optical path between the optical cavity output and the heterodyne detector input to be $\xi_\text{path} = 0.4$. The fractional contribution of LO beam shot noise (at power $P_\text{LO} = 1.3$~mW) to the sum of shot noise and heterodyne detector dark noise was measured to be $\xi_\text{dark} = 0.79$. The nominal photodetector quantum efficiency of $\sigma_\text{q}=0.87$ then yields $\xi_\text{o}=\xi_\text{path}\,\xi_\text{dark}\,\sigma_\text{q} = 0.27$, consistent with the temperature sweep measurements. 

On the microwave side, we measured the added noise of the microwave chain referred to the input of the HEMT amplifier at the 4~K stage of the cryostat by sweeping $T_\text{bp}$ and measuring the thermal noise of the effective 50\,$\Omega$ load seen by the HEMT. From this measurement we obtained $N_\text{HEMT}=8.5$~photons/s/Hz, equivalent to an efficiency of $1/(N_\text{HEMT}+1/2) = 0.11$ referred to the HEMT input. The discrepancy between this number and the measurement chain efficiency $\xi_\text{e}$ referred to the transducer's microwave output port implies a 5.8~dB loss between the transducer's microwave output port and the HEMT input. This is a rather large value, but the value inferred from measurements in Ref.~\cite{higginbotham_harnessing_2018} using the same cryostat was only slightly smaller, and more microwave connectors were present in the signal path here than in that earlier work.

\subsection{Ground-state cooling measurements}
\label{sub:gscool}

To process the data shown in Figs.~\ref{fig:gscool}(a) and (b), we first normalize the optomechanical upper- and lower-sideband spectra as described in Appendix~\ref{sub:calibration} above. Two such normalized spectra from the pure optomechanical data set are shown in Fig.~\ref{fig:ASpectra}(a). We observe that the white noise background around both sidebands increases linearly with optical pump power, as expected in the presence of amplitude and phase noise (see Appendix~\ref{sub:technoise}). We also observe a feature at frequency $\omega_\text{s}/2\pi = 1.448$~MHz emerging from the noise floor around the upper sideband at high power. We attribute this feature to a relatively massive vibrational mode of the silicon chip substrate, and account for it by fitting spectra to the squared sum of two complex Lorentzians with the amplitude of the second mode and the relative phase as additional fit parameters. This fit was motivated by observations of destructive interference between a membrane mode and a substrate mode in unpublished data from a previous device. For the present device, this coherent double-Lorentzian fit differs by less than 10\% from a fit to the incoherent sum of two squared Lorentzian amplitudes. The fits are not appreciably altered by including antisymmetric terms as in Eqs.~\eqref{eq:usb_noise} and \eqref{eq:lsb_noise}.

The upper and lower sideband amplitudes $N_{\text{det,o},\pm}$ obtained from these fits must then be corrected to undo the effects of amplitude and phase noise (Appendix~\ref{sub:technoise}) before we can infer the membrane mode occupancy. Correcting the peak amplitudes in this way in turn requires knowledge of the relative magnitudes of the shot noise-normalized technical noise spectral densities $C_\text{xx}$, $C_\text{yy}$, and $C_\text{xy}$, which are related to the spectral densities of fractional amplitude and phase fluctuations by $C_\text{xx} = 4\dot{N}_\text{o}\left(S_{\delta\text{A},\delta\text{A}}(\omega_\text{m})/2\right)$, $C_\text{yy} = 4\dot{N}_\text{o}\left(S_{\delta\phi,\delta\phi}(\omega_\text{m})/2\right)$, and $C_\text{xy} = 4\dot{N}_\text{o}\left(S_{\delta\text{A},\delta\phi}(\omega_\text{m})/2\right)$, where $\dot{N}_\text{o} = P_\text{o}/\hbar\omega_\text{p,o}$ is the photon flux of the pump incident on the optical cavity. Here $S_{\delta\text{A},\delta\text{A}}(\omega)$, $S_{\delta\phi,\delta\phi}(\omega)$, and $S_{\delta\text{A},\delta\phi}(\omega)$ are defined as one-sided double-sideband spectral densities in units of rad$^2$/Hz. The factor of 1/2 in each expression accounts for the transition from the two-sided spectral densities assumed by the theoretical framework in Appendix~\ref{sub:technoise} to the one-sided convention more commonly used by experimentalists. 

We made independent measurements of  $S_{\delta\text{A},\delta\text{A}}(\omega)$ and $S_{\delta\phi,\delta\phi}(\omega)$ shortly before the cooldown in which we acquired the data presented in this work, using a very similar optomechanical cavity in the same setup as an AM/$\Upphi$M transducer. We measured amplitude noise by comparing the sum and difference photocurrents in balanced direct detection of the pump beam promptly reflected off the cavity far from resonance~\cite{yu_quantum_2015}. We then repeated this measurement with the laser locked to the cavity and the pump beam near resonance to transduce amplitude noise to phase noise, and calibrated the cavity's AM/$\Upphi$M transduction coefficient using a phase modulation sideband of known modulation depth~\cite{safavi-naeini_laser_2013}. From these measurements we obtained $10\,\text{log}_{10}\left(S_{\delta\text{A},\delta\text{A}}(\omega)\right)\approx -155$~dBc/Hz and $10\,\text{log}_{10}\left(S_{\delta\phi,\delta\phi}(\omega)\right)\approx -136$~dBc/Hz
at frequencies near $\omega_\text{m}$. Using these numbers in Eq.~\eqref{eq:S_tilde} at the highest-power data point in Fig.~\ref{fig:gscool}, we reproduce the measured values of the off-resonance excess noise $\tilde{S}_{\text{o},\pm}$ to within a factor of 1.5 if we assume maximal positive correlations $C_\text{xy} = \sqrt{C_\text{xx}C_\text{yy}}$, lending credence to the conclusion that $C_\text{yy} \gg C_\text{xx},\ C_\text{xy}$. The results obtained from the subsequent analysis assuming phase noise only differ negligibly from the results obtained assuming the measured ratio between $S_{\delta\text{A},\delta\text{A}}(\omega_\text{m})$ and $S_{\delta\phi,\delta\phi}(\omega_\text{m})$ and maximal positive correlations, so we discuss the former analysis here for simplicity.

For $C_\text{yy} \gg C_\text{xx},\ C_\text{xy}$, the anti-symmetric contribution to both the Stokes and the anti-Stokes peak lineshapes is suppressed and the Lorentzian contribution is negative for both peaks, resulting in squashing. We undo this phase noise squashing by taking
\begin{widetext}
\begin{equation}
    N_{\text{det,o},\pm} \rightarrow N_{\text{det,o},\pm} \pm \frac{\partial\left(\epsilon_\text{CL}\epsilon_\text{PC}\frac{\kappa_\text{o,ext}}{\kappa_\text{o}}\kappa_\text{o}^2\mathcal{A}_\text{o}\frac{\Gamma_\text{o}}{\Gamma_\text{T}}\text{Re}\big[\tilde{B}_\pm\big]\right)}{\partial C_\text{yy}}\left(\frac{\partial\tilde{S}_{\text{o},\pm}}{\partial C_\text{yy}}\right)^{-1}\tilde{S}_{\text{det,o},\pm},
\label{eq:squash}
\end{equation}
\end{widetext}
where the two partial derivatives can be evaluated using independently measured transducer parameters, and $\tilde{S}_{\text{det,o},\pm}$ is the measured excess over the white noise background level of 1~photon/Hz/s. This method exploits the fact that the phase noise is responsible for both the squashing of the peak amplitudes and the increase in the white noise background to correct for squashing without assuming a specific value of $C_\text{yy}$. We then use the corrected peak amplitudes to infer the membrane mode occupancy $n_\text{m}$ via Eq.~\eqref{eq:sidebandtherm}, and fit the data as shown in Figs.~\ref{fig:gscool}(a) and (b), with an additional $\gamma_\text{lock}n_\text{min,lock}$ term in Eq.~\eqref{eq:gscool} to account for the lock backaction effects discussed in Appendix~\ref{sub:misc}. 

As a final check on the consistency of the sideband asymmetry analysis, we can use the value of the coefficient $a_\text{o}$ obtained from the fit shown in Fig.~\ref{fig:gscool}(a) to get an independent estimate of the phase noise spectral density. Setting the effective optical mode occupancy $n_\text{eff,o} = a_\text{o}\Gamma_\text{o}$ in Eq.~\eqref{eq:n_effo} and dropping the $C_\text{xx}$ and $C_\text{xy}$ terms, we obtain $10\,\text{log}_{10}\left(S_{\delta\phi,\delta\phi}(\omega_\text{m})\right) = -135^{\ +2}_{\ -3}$~dBc/Hz, consistent with the independent measurement.

As discussed in Appendix~\ref{sub:calibration}, we can also obtain the membrane mode occupancy $n_\text{m}$ from the normalized upper-sideband spectrum $S_\text{det,o,+}$, given the optical measurement chain efficiency $\xi_\text{o}$ and independent measurements of other transducer parameters. The $n_\text{m}$ values obtained this way and the corresponding fit are overlaid on the values obtained from the sideband asymmetry analysis in Fig.~\ref{fig:ASpectra}(b) and (c) for the purely optomechanical and electro-optomechanical damping sweeps respectively. Fitting to the purely optomechanical data to Eq.~\eqref{eq:gscool} yields $n_\text{th} = 750 \pm 50$ and $a_\text{o} = 5 \pm 1$, and fixing these values in the fit to the electro-optomechanical data yields $n_\text{eff,e} = 1.02 \pm 0.07$ for the effective microwave mode thermal occupancy. The origin of the discrepancy  between these values and those obtained from the sideband asymmetry analysis is unknown, but the analysis using only the upper sideband data is more susceptible to miscalibration of parameters. The equilibrium occupancy $n_\text{th}$ inferred from this fits is very sensitive to the first data point, and the discrepancy between the values of $n_\text{eff,e}$ obtained from the two electro-optomechanical fits is strongly correlated with the discrepancy between the values of $n_\text{th}$.

\subsection{Added noise measurements}
\label{sub:nadd}
The values of $N_\text{add,up}$ plotted in Fig.~\ref{fig:gscool}(d) are obtained from the inferred phonon occupancy data in Fig.~\ref{fig:gscool}(b) via Eq.~\eqref{eq:nadd_supp}, with an additional term accounting for the off-resonance level $\tilde{S}_{\text{o},\pm}$ of each spectrum. This off-resonance level (normalized at the transducer output) can be obtained from the directly measured spectra without need for independent measurement of the optical measurement chain efficiency $\xi_\text{o}$, because a measurement of $n_\text{m}$ calibrates the optical output spectra.

The analysis of the data shown in Fig.~\ref{fig:nadd}(b) is simpler than the ground-state cooling analysis discussed above, as both optical pump phase noise and interference from the substrate mode shown in Fig.~\ref{fig:ASpectra}(a) were negligible with the optomechanical damping fixed at $\Gamma_\text{o}/2\pi = 85$~Hz. However, the lower sideband amplitude $N_{\text{det,o},-}$ was a very small fractional excess over the white noise background at high microwave pump power, where $\Gamma_\text{o} \ll \Gamma_\text{e}$. We thus calibrate the added noise using the optical measurement chain efficiency $\xi_\text{o}$ obtained from the temperature sweep rather than sideband asymmetry. We fit the upper sideband amplitude $N_{\text{out,o},+} = N_{\text{det,o},+}/\xi_\text{o}$ as a function of damping to the expected behavior obtained by substituting Eq.~\eqref{eq:gscool} into Eq.~\eqref{eq:usb} and evaluating at frequency $\omega_\text{m}$:
\begin{equation}
    N_{\text{out,o},+} = 4\epsilon_\text{CL}\mathcal{A}_\text{o}\frac{\kappa_\text{o,ext}}{\kappa_\text{o}}\Gamma_\text{o}\frac{\gamma_\text{m}n_\text{th} + \Gamma_\text{e}n_\text{e} + \Gamma_\text{o}n_\text{o}}{\Gamma_\text{T}^2}.
    \label{eq:nout}
\end{equation}
As in the analysis of the data shown in Fig.~\ref{fig:gscool}, we also include a contribution from lock beam backaction in the fit. To obtain the data and the theory curve shown in Fig.~\ref{fig:nadd}(b), we divide by the product of efficiency $\eta_\text{t}$ and transducer gain $\mathcal{A}_\text{e}\mathcal{A}_\text{o}$. Calibrating the data with the sideband asymmetry analysis instead yields slightly lower added noise, so our choice to present the data calibrated using $\xi_\text{o}$ is also conservative.

At the optimal electromechanical damping rate $\Gamma_\text{e}/2\pi = 135$~Hz where added noise is minimized, the total added noise arises from the weighted sum of couplings to the various baths that determine $n_\text{m}$: 1.4~photons/s/Hz from $n_\text{eff,e}$, 1.0~photons/s/Hz from $n_\text{th}$, 0.4~photons/s/Hz from $n_\text{min,lock}$, 0.2~photons/s/Hz from $n_\text{min,o}$, and 0.1~photons/s/Hz from $n_\text{min,e}$. The final 0.1~photons/s/Hz comes from excess white noise referred to the transducer's microwave input.

\begin{figure}[!ht]
    \centering
    \includegraphics{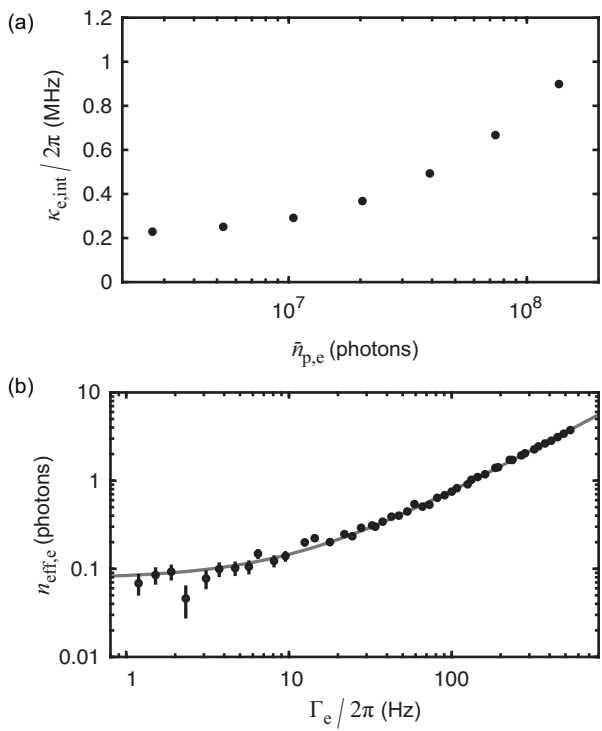}
    \caption{
    \textbf{Microwave pump effects on superconducting circuit.}
    (a) Internal loss of the microwave circuit $\kappa_\text{e,int}$ vs.\ $\bar{n}_\text{p,e}$, the resonator coherent state mean photon number.  
    (b) Effective microwave mode occupancy $n_\text{eff,e}$ vs.\ electromechanical damping $\Gamma_\text{e}$, with linear fit. All error bars represent one standard deviation.
    }
    \label{fig:AMWNoise}
\end{figure}

\section{Microwave pump effects on superconducting circuit}
\label{sec:paramnoise}

We observed two adverse effects of the microwave pump on the superconducting microwave circuit, whose origins are likely related. The power-dependence of the circuit's internal loss $\kappa_\text{e,int}$ is plotted in Fig.~\ref{fig:AMWNoise}(a) as a function of $\bar{n}_\text{p,e}$, the mean photon number of the coherent state induced in the microwave mode by the microwave pump, over the range of power used in Fig.~\ref{fig:nadd}. As noted in Appendix~\ref{sec:characterization}, this power-dependence results in power-dependent electromechanical transducer gain $\mathcal{A}_\text{e}$, modifies the expected behavior of transducer efficiency $\eta_\text{t}$ vs.\ electromechanical damping, and leads to sublinear scaling of damping vs.\ microwave pump power.

A more significant constraint on the transducer performance is the power-dependent microwave mode effective occupancy $n_\text{eff,e}$. To calibrate $n_\text{eff,e}$, we first refer the measured microwave noise spectral density $S_\text{det,e}$ to the transducer output by dividing by the measurement chain efficiency $\xi_\text{e}$, then divide by $4\kappa_\text{e,ext}/\kappa_\text{e}$ to translate the spectral density into a mode occupancy. Operationally, we are interested in the behavior of $n_\text{eff,e}$ as a function of electromechanical damping $\Gamma_\text{e}$, plotted in Fig.~\ref{fig:AMWNoise}(b). The noise is seen to scale as $n_\text{eff,e} = a_\text{e}\Gamma_\text{e} + b_\text{e}$, with $a_\text{e} = 1.1\times10^{-3}~\text{Hz}^{-1}$ and $b_\text{e} = 0.077$, with 1\% and 5\% fractional uncertainty in the slope and offset, respectively. Evaluated at $\Gamma_\text{e} = 2\pi\times 100$~Hz, as in the data shown in Fig.~\ref{fig:gscool}(b), the fit yields $n_\text{eff,e} = 0.77\pm0.01$, consistent with the value obtained from the Fig.~\ref{fig:gscool}(b) fit. Fitting the data in Fig.~\ref{fig:nadd}(b), in which $\Gamma_\text{e}$ is swept, yields $a_\text{e} = \left(1.17\pm0.05\right)\times10^{-3}~\text{Hz}^{-1}$ and $b_\text{e} = 0.1 \pm 0.1$ , consistent with the values inferred from direct measurements of the noise described above.

One possible contribution to $n_\text{eff,e}$ is the phase noise of the microwave pump. We used a Rohde \& Schwartz SMA100B signal generator to source the pump specifically for the ultra-low phase noise afforded by option SMAB-B711, and measured its DSB noise \textit{in situ} to be $10\,\text{log}_{10}\left(S_{\delta\phi,\delta\phi,\text{e}}(\omega)\right)=-149$~dBc/Hz at frequencies near $\omega_\text{m}$. Following the procedure used to infer the phase noise spectral density of the optical pump from the coefficient $a_\text{o}$, we find that microwave pump phase noise can account for about 25\% of the measured value of $a_\text{e}$. The remainder of the power-dependent noise must be attributed to the behavior the superconducting circuit itself. 

At present, the physical origin of this excess power-dependent noise is unknown, though it is likely related to the power-dependent internal loss shown in Fig.~\ref{fig:AMWNoise}(a). Possibilities include bulk heating of the circuit facilitated by low electron-phonon thermal conductivity, fluctuations of the circuit's resonant frequency or loss due to the participation of dielectric substrates in its electric field distribution, or trapped flux in the circuit, which is necessarily in close proximity to the optical cavity's invar support structure. The low-power asymptote $b_\text{e}$ is a surprising feature of the data that need not have the same physical origin as the high-power linear behavior. A spectral density of circuit parameter fluctuations that scales inversely with microwave pump power over some range is one possible origin for this nonzero intercept.

Finally, we note that microwave power-dependent noise was also observed in the device described in Ref.~\cite{higginbotham_harnessing_2018}, and motivated our decision to fabricate the superconducting circuit from NbTiN rather than Nb as in past work. In test circuits without released membranes that were otherwise identical to the flip-chip circuits used in our transducer, we observed reduction of the power-dependent noise by a factor of 4 in circuits fabricated from NbTiN. This improvement was largely preserved in our integrated electro-optomechanical device, and the microwave power-dependent noise remained a chief limitation primarily because of the large capacitor pad separation $d$ of the device used in this work. As noted in Appendix~\ref{sub:mechanical}, reproducing the values of $d$ obtained in previous devices would reduce the power required to obtain a given electromechanical damping by at least a factor of 25, which would greatly reduce the impact of power-dependent noise on transducer performance even without further reduction of the noise itself.


\begin{thebibliography}{43}%
\makeatletter
\providecommand \@ifxundefined [1]{%
 \@ifx{#1\undefined}
}%
\providecommand \@ifnum [1]{%
 \ifnum #1\expandafter \@firstoftwo
 \else \expandafter \@secondoftwo
 \fi
}%
\providecommand \@ifx [1]{%
 \ifx #1\expandafter \@firstoftwo
 \else \expandafter \@secondoftwo
 \fi
}%
\providecommand \natexlab [1]{#1}%
\providecommand \enquote  [1]{``#1''}%
\providecommand \bibnamefont  [1]{#1}%
\providecommand \bibfnamefont [1]{#1}%
\providecommand \citenamefont [1]{#1}%
\providecommand \href@noop [0]{\@secondoftwo}%
\providecommand \href [0]{\begingroup \@sanitize@url \@href}%
\providecommand \@href[1]{\@@startlink{#1}\@@href}%
\providecommand \@@href[1]{\endgroup#1\@@endlink}%
\providecommand \@sanitize@url [0]{\catcode `\\12\catcode `\$12\catcode
  `\&12\catcode `\#12\catcode `\^12\catcode `\_12\catcode `\%12\relax}%
\providecommand \@@startlink[1]{}%
\providecommand \@@endlink[0]{}%
\providecommand \url  [0]{\begingroup\@sanitize@url \@url }%
\providecommand \@url [1]{\endgroup\@href {#1}{\urlprefix }}%
\providecommand \urlprefix  [0]{URL }%
\providecommand \Eprint [0]{\href }%
\providecommand \doibase [0]{https://doi.org/}%
\providecommand \selectlanguage [0]{\@gobble}%
\providecommand \bibinfo  [0]{\@secondoftwo}%
\providecommand \bibfield  [0]{\@secondoftwo}%
\providecommand \translation [1]{[#1]}%
\providecommand \BibitemOpen [0]{}%
\providecommand \bibitemStop [0]{}%
\providecommand \bibitemNoStop [0]{.\EOS\space}%
\providecommand \EOS [0]{\spacefactor3000\relax}%
\providecommand \BibitemShut  [1]{\csname bibitem#1\endcsname}%
\let\auto@bib@innerbib\@empty
\bibitem [{\citenamefont {Ladd}\ \emph {et~al.}(2010)\citenamefont {Ladd},
  \citenamefont {Jelezko}, \citenamefont {Laflamme}, \citenamefont {Nakamura},
  \citenamefont {Monroe},\ and\ \citenamefont
  {O{\textquotesingle}Brien}}]{ladd_quantum_2010}%
  \BibitemOpen
  \bibfield  {author} {\bibinfo {author} {\bibfnamefont {T.~D.}\ \bibnamefont
  {Ladd}}, \bibinfo {author} {\bibfnamefont {F.}~\bibnamefont {Jelezko}},
  \bibinfo {author} {\bibfnamefont {R.}~\bibnamefont {Laflamme}}, \bibinfo
  {author} {\bibfnamefont {Y.}~\bibnamefont {Nakamura}}, \bibinfo {author}
  {\bibfnamefont {C.}~\bibnamefont {Monroe}},\ and\ \bibinfo {author}
  {\bibfnamefont {J.~L.}\ \bibnamefont {O{\textquotesingle}Brien}},\ }\bibfield
   {title} {\bibinfo {title} {Quantum computers},\ }\href
  {https://doi.org/10.1038/nature08812} {\bibfield  {journal} {\bibinfo
  {journal} {Nature}\ }\textbf {\bibinfo {volume} {464}},\ \bibinfo {pages}
  {45} (\bibinfo {year} {2010})}\BibitemShut {NoStop}%
\bibitem [{\citenamefont {Arute}\ \emph {et~al.}(2019)\citenamefont {Arute},
  \citenamefont {Arya}, \citenamefont {Babbush}, \citenamefont {Bacon},
  \citenamefont {Bardin}, \citenamefont {Barends}, \citenamefont {Biswas},
  \citenamefont {Boixo}, \citenamefont {Brandao}, \citenamefont {Buell},
  \citenamefont {Burkett}, \citenamefont {Chen}, \citenamefont {Chen} \emph
  {et~al.}}]{arute_quantum_2019}%
  \BibitemOpen
  \bibfield  {author} {\bibinfo {author} {\bibfnamefont {F.}~\bibnamefont
  {Arute}}, \bibinfo {author} {\bibfnamefont {K.}~\bibnamefont {Arya}},
  \bibinfo {author} {\bibfnamefont {R.}~\bibnamefont {Babbush}}, \bibinfo
  {author} {\bibfnamefont {D.}~\bibnamefont {Bacon}}, \bibinfo {author}
  {\bibfnamefont {J.~C.}\ \bibnamefont {Bardin}}, \bibinfo {author}
  {\bibfnamefont {R.}~\bibnamefont {Barends}}, \bibinfo {author} {\bibfnamefont
  {R.}~\bibnamefont {Biswas}}, \bibinfo {author} {\bibfnamefont
  {S.}~\bibnamefont {Boixo}}, \bibinfo {author} {\bibfnamefont {F.~G. S.~L.}\
  \bibnamefont {Brandao}}, \bibinfo {author} {\bibfnamefont {D.~A.}\
  \bibnamefont {Buell}}, \bibinfo {author} {\bibfnamefont {B.}~\bibnamefont
  {Burkett}}, \bibinfo {author} {\bibfnamefont {Y.}~\bibnamefont {Chen}},
  \bibinfo {author} {\bibfnamefont {Z.}~\bibnamefont {Chen}}, \emph {et~al.},\
  }\bibfield  {title} {\bibinfo {title} {Quantum supremacy using a programmable
  superconducting processor},\ }\href
  {https://doi.org/10.1038/s41586-019-1666-5} {\bibfield  {journal} {\bibinfo
  {journal} {Nature}\ }\textbf {\bibinfo {volume} {574}},\ \bibinfo {pages}
  {505} (\bibinfo {year} {2019})}\BibitemShut {NoStop}%
\bibitem [{\citenamefont {Campagne-Ibarcq}\ \emph {et~al.}(2020)\citenamefont
  {Campagne-Ibarcq}, \citenamefont {Eickbusch}, \citenamefont {Touzard},
  \citenamefont {Zalys-Geller}, \citenamefont {Frattini}, \citenamefont
  {Sivak}, \citenamefont {Reinhold}, \citenamefont {Puri}, \citenamefont
  {Shankar}, \citenamefont {Schoelkopf}, \citenamefont {Frunzio}, \citenamefont
  {Mirrahimi},\ and\ \citenamefont {Devoret}}]{campagne_quantum_2020}%
  \BibitemOpen
  \bibfield  {author} {\bibinfo {author} {\bibfnamefont {P.}~\bibnamefont
  {Campagne-Ibarcq}}, \bibinfo {author} {\bibfnamefont {A.}~\bibnamefont
  {Eickbusch}}, \bibinfo {author} {\bibfnamefont {S.}~\bibnamefont {Touzard}},
  \bibinfo {author} {\bibfnamefont {E.}~\bibnamefont {Zalys-Geller}}, \bibinfo
  {author} {\bibfnamefont {N.~E.}\ \bibnamefont {Frattini}}, \bibinfo {author}
  {\bibfnamefont {V.~V.}\ \bibnamefont {Sivak}}, \bibinfo {author}
  {\bibfnamefont {P.}~\bibnamefont {Reinhold}}, \bibinfo {author}
  {\bibfnamefont {S.}~\bibnamefont {Puri}}, \bibinfo {author} {\bibfnamefont
  {S.}~\bibnamefont {Shankar}}, \bibinfo {author} {\bibfnamefont {R.~J.}\
  \bibnamefont {Schoelkopf}}, \bibinfo {author} {\bibfnamefont
  {L.}~\bibnamefont {Frunzio}}, \bibinfo {author} {\bibfnamefont
  {M.}~\bibnamefont {Mirrahimi}},\ and\ \bibinfo {author} {\bibfnamefont
  {M.~H.}\ \bibnamefont {Devoret}},\ }\bibfield  {title} {\bibinfo {title}
  {Quantum error correction of a qubit encoded in grid states of an
  oscillator},\ }\href {https://doi.org/10.1038/s41586-020-2603-3} {\bibfield
  {journal} {\bibinfo  {journal} {Nature}\ }\textbf {\bibinfo {volume} {584}},\
  \bibinfo {pages} {368} (\bibinfo {year} {2020})}\BibitemShut {NoStop}%
\bibitem [{\citenamefont {Liao}\ \emph {et~al.}(2018)\citenamefont {Liao},
  \citenamefont {Cai}, \citenamefont {Handsteiner}, \citenamefont {Liu},
  \citenamefont {Yin}, \citenamefont {Zhang}, \citenamefont {Rauch},
  \citenamefont {Fink}, \citenamefont {Ren}, \citenamefont {Liu}, \citenamefont
  {Li}, \citenamefont {Shen}, \citenamefont {Cao}, \citenamefont {Li},
  \citenamefont {Wang} \emph {et~al.}}]{liao_satellite-relayed_2018}%
  \BibitemOpen
  \bibfield  {author} {\bibinfo {author} {\bibfnamefont {S.-K.}\ \bibnamefont
  {Liao}}, \bibinfo {author} {\bibfnamefont {W.-Q.}\ \bibnamefont {Cai}},
  \bibinfo {author} {\bibfnamefont {J.}~\bibnamefont {Handsteiner}}, \bibinfo
  {author} {\bibfnamefont {B.}~\bibnamefont {Liu}}, \bibinfo {author}
  {\bibfnamefont {J.}~\bibnamefont {Yin}}, \bibinfo {author} {\bibfnamefont
  {L.}~\bibnamefont {Zhang}}, \bibinfo {author} {\bibfnamefont
  {D.}~\bibnamefont {Rauch}}, \bibinfo {author} {\bibfnamefont
  {M.}~\bibnamefont {Fink}}, \bibinfo {author} {\bibfnamefont {J.-G.}\
  \bibnamefont {Ren}}, \bibinfo {author} {\bibfnamefont {W.-Y.}\ \bibnamefont
  {Liu}}, \bibinfo {author} {\bibfnamefont {Y.}~\bibnamefont {Li}}, \bibinfo
  {author} {\bibfnamefont {Q.}~\bibnamefont {Shen}}, \bibinfo {author}
  {\bibfnamefont {Y.}~\bibnamefont {Cao}}, \bibinfo {author} {\bibfnamefont
  {F.~Z.}\ \bibnamefont {Li}}, \bibinfo {author} {\bibfnamefont {J.~F.}\
  \bibnamefont {Wang}}, \emph {et~al.},\ }\bibfield  {title} {\bibinfo {title}
  {Satellite-{Relayed} {Intercontinental} {Quantum} {Network}},\ }\href
  {https://doi.org/10.1103/PhysRevLett.120.030501} {\bibfield  {journal}
  {\bibinfo  {journal} {Phys. Rev. Lett.}\ }\textbf {\bibinfo {volume} {120}},\
  \bibinfo {pages} {030501} (\bibinfo {year} {2018})}\BibitemShut {NoStop}%
\bibitem [{\citenamefont {Chen}\ \emph {et~al.}(2020)\citenamefont {Chen},
  \citenamefont {Zhang}, \citenamefont {Liu}, \citenamefont {Jiang},
  \citenamefont {Zhang}, \citenamefont {Hu}, \citenamefont {Guan},
  \citenamefont {Yu}, \citenamefont {Xu}, \citenamefont {Lin}, \citenamefont
  {Li}, \citenamefont {Chen}, \citenamefont {Li}, \citenamefont {You},
  \citenamefont {Wang}, \citenamefont {Wang}, \citenamefont {Zhang},\ and\
  \citenamefont {Pan}}]{chen_sending-or-not-sending_2020}%
  \BibitemOpen
  \bibfield  {author} {\bibinfo {author} {\bibfnamefont {J.-P.}\ \bibnamefont
  {Chen}}, \bibinfo {author} {\bibfnamefont {C.}~\bibnamefont {Zhang}},
  \bibinfo {author} {\bibfnamefont {Y.}~\bibnamefont {Liu}}, \bibinfo {author}
  {\bibfnamefont {C.}~\bibnamefont {Jiang}}, \bibinfo {author} {\bibfnamefont
  {W.}~\bibnamefont {Zhang}}, \bibinfo {author} {\bibfnamefont {X.-L.}\
  \bibnamefont {Hu}}, \bibinfo {author} {\bibfnamefont {J.-Y.}\ \bibnamefont
  {Guan}}, \bibinfo {author} {\bibfnamefont {Z.-W.}\ \bibnamefont {Yu}},
  \bibinfo {author} {\bibfnamefont {H.}~\bibnamefont {Xu}}, \bibinfo {author}
  {\bibfnamefont {J.}~\bibnamefont {Lin}}, \bibinfo {author} {\bibfnamefont
  {M.~J.}\ \bibnamefont {Li}}, \bibinfo {author} {\bibfnamefont
  {H.}~\bibnamefont {Chen}}, \bibinfo {author} {\bibfnamefont {H.}~\bibnamefont
  {Li}}, \bibinfo {author} {\bibfnamefont {L.}~\bibnamefont {You}}, \bibinfo
  {author} {\bibfnamefont {Z.}~\bibnamefont {Wang}}, \bibinfo {author}
  {\bibfnamefont {X.~B.}\ \bibnamefont {Wang}}, \bibinfo {author}
  {\bibfnamefont {Q.}~\bibnamefont {Zhang}},\ and\ \bibinfo {author}
  {\bibfnamefont {J.~W.}\ \bibnamefont {Pan}},\ }\bibfield  {title} {\bibinfo
  {title} {Sending-or-{Not}-{Sending} with {Independent} {Lasers}: {Secure}
  {Twin}-{Field} {Quantum} {Key} {Distribution} over 509 km},\ }\href
  {https://doi.org/10.1103/PhysRevLett.124.070501} {\bibfield  {journal}
  {\bibinfo  {journal} {Phys. Rev. Lett.}\ }\textbf {\bibinfo {volume} {124}},\
  \bibinfo {pages} {070501} (\bibinfo {year} {2020})}\BibitemShut {NoStop}%
\bibitem [{\citenamefont {Wehner}\ \emph {et~al.}(2018)\citenamefont {Wehner},
  \citenamefont {Elkouss},\ and\ \citenamefont {Hanson}}]{wehner_quantum_2018}%
  \BibitemOpen
  \bibfield  {author} {\bibinfo {author} {\bibfnamefont {S.}~\bibnamefont
  {Wehner}}, \bibinfo {author} {\bibfnamefont {D.}~\bibnamefont {Elkouss}},\
  and\ \bibinfo {author} {\bibfnamefont {R.}~\bibnamefont {Hanson}},\
  }\bibfield  {title} {\bibinfo {title} {Quantum internet: {A} vision for the
  road ahead},\ }\href {https://doi.org/10.1126/science.aam9288} {\bibfield
  {journal} {\bibinfo  {journal} {Science}\ }\textbf {\bibinfo {volume}
  {362}},\ \bibinfo {pages} {eaam9288} (\bibinfo {year} {2018})}\BibitemShut
  {NoStop}%
\bibitem [{\citenamefont {Zeuthen}\ \emph {et~al.}(2020)\citenamefont
  {Zeuthen}, \citenamefont {Schliesser}, \citenamefont {S\o{}rensen},\ and\
  \citenamefont {Taylor}}]{zeuthen_figures_2020}%
  \BibitemOpen
  \bibfield  {author} {\bibinfo {author} {\bibfnamefont {E.}~\bibnamefont
  {Zeuthen}}, \bibinfo {author} {\bibfnamefont {A.}~\bibnamefont {Schliesser}},
  \bibinfo {author} {\bibfnamefont {A.~S.}\ \bibnamefont {S\o{}rensen}},\ and\
  \bibinfo {author} {\bibfnamefont {J.~M.}\ \bibnamefont {Taylor}},\ }\bibfield
   {title} {\bibinfo {title} {Figures of merit for quantum transducers},\
  }\href {https://doi.org/10.1088/2058-9565/ab8962} {\bibfield  {journal}
  {\bibinfo  {journal} {Quantum Sci. Technol.}\ }\textbf {\bibinfo {volume}
  {5}},\ \bibinfo {pages} {034009} (\bibinfo {year} {2020})}\BibitemShut
  {NoStop}%
\bibitem [{\citenamefont {Lambert}\ \emph {et~al.}(2020)\citenamefont
  {Lambert}, \citenamefont {Rueda}, \citenamefont {Sedlmeir},\ and\
  \citenamefont {Schwefel}}]{lambert_coherent_2020}%
  \BibitemOpen
  \bibfield  {author} {\bibinfo {author} {\bibfnamefont {N.~J.}\ \bibnamefont
  {Lambert}}, \bibinfo {author} {\bibfnamefont {A.}~\bibnamefont {Rueda}},
  \bibinfo {author} {\bibfnamefont {F.}~\bibnamefont {Sedlmeir}},\ and\
  \bibinfo {author} {\bibfnamefont {H.~G.~L.}\ \bibnamefont {Schwefel}},\
  }\bibfield  {title} {\bibinfo {title} {Coherent {Conversion} {Between}
  {Microwave} and {Optical} {Photons}—{An} {Overview} of {Physical}
  {Implementations}},\ }\href {https://doi.org/10.1002/qute.201900077}
  {\bibfield  {journal} {\bibinfo  {journal} {Adv. Quantum Technol.}\ }\textbf
  {\bibinfo {volume} {3}},\ \bibinfo {pages} {1900077} (\bibinfo {year}
  {2020})}\BibitemShut {NoStop}%
\bibitem [{\citenamefont {Lauk}\ \emph {et~al.}(2020)\citenamefont {Lauk},
  \citenamefont {Sinclair}, \citenamefont {Barzanjeh}, \citenamefont {Covey},
  \citenamefont {Saffman}, \citenamefont {Spiropulu},\ and\ \citenamefont
  {Simon}}]{lauk_perspectives_2020}%
  \BibitemOpen
  \bibfield  {author} {\bibinfo {author} {\bibfnamefont {N.}~\bibnamefont
  {Lauk}}, \bibinfo {author} {\bibfnamefont {N.}~\bibnamefont {Sinclair}},
  \bibinfo {author} {\bibfnamefont {S.}~\bibnamefont {Barzanjeh}}, \bibinfo
  {author} {\bibfnamefont {J.~P.}\ \bibnamefont {Covey}}, \bibinfo {author}
  {\bibfnamefont {M.}~\bibnamefont {Saffman}}, \bibinfo {author} {\bibfnamefont
  {M.}~\bibnamefont {Spiropulu}},\ and\ \bibinfo {author} {\bibfnamefont
  {C.}~\bibnamefont {Simon}},\ }\bibfield  {title} {\bibinfo {title}
  {Perspectives on quantum transduction},\ }\href
  {https://doi.org/10.1088/2058-9565/ab788a} {\bibfield  {journal} {\bibinfo
  {journal} {Quantum Sci. Technol.}\ }\textbf {\bibinfo {volume} {5}},\
  \bibinfo {pages} {020501} (\bibinfo {year} {2020})}\BibitemShut {NoStop}%
\bibitem [{\citenamefont {Mirhosseini}\ \emph {et~al.}(2020)\citenamefont
  {Mirhosseini}, \citenamefont {Sipahigil}, \citenamefont {Kalaee},\ and\
  \citenamefont {Painter}}]{mirhosseini_superconducting_2020}%
  \BibitemOpen
  \bibfield  {author} {\bibinfo {author} {\bibfnamefont {M.}~\bibnamefont
  {Mirhosseini}}, \bibinfo {author} {\bibfnamefont {A.}~\bibnamefont
  {Sipahigil}}, \bibinfo {author} {\bibfnamefont {M.}~\bibnamefont {Kalaee}},\
  and\ \bibinfo {author} {\bibfnamefont {O.}~\bibnamefont {Painter}},\
  }\bibfield  {title} {\bibinfo {title} {Superconducting qubit to optical
  photon transduction},\ }\href {https://doi.org/10.1038/s41586-020-3038-6}
  {\bibfield  {journal} {\bibinfo  {journal} {Nature}\ }\textbf {\bibinfo
  {volume} {588}},\ \bibinfo {pages} {599} (\bibinfo {year}
  {2020})}\BibitemShut {NoStop}%
\bibitem [{\citenamefont {Hease}\ \emph {et~al.}(2020)\citenamefont {Hease},
  \citenamefont {Rueda}, \citenamefont {Sahu}, \citenamefont {Wulf},
  \citenamefont {Arnold}, \citenamefont {Schwefel},\ and\ \citenamefont
  {Fink}}]{hease_bidirectional_2020}%
  \BibitemOpen
  \bibfield  {author} {\bibinfo {author} {\bibfnamefont {W.}~\bibnamefont
  {Hease}}, \bibinfo {author} {\bibfnamefont {A.}~\bibnamefont {Rueda}},
  \bibinfo {author} {\bibfnamefont {R.}~\bibnamefont {Sahu}}, \bibinfo {author}
  {\bibfnamefont {M.}~\bibnamefont {Wulf}}, \bibinfo {author} {\bibfnamefont
  {G.}~\bibnamefont {Arnold}}, \bibinfo {author} {\bibfnamefont {H.~G.~L.}\
  \bibnamefont {Schwefel}},\ and\ \bibinfo {author} {\bibfnamefont {J.~M.}\
  \bibnamefont {Fink}},\ }\bibfield  {title} {\bibinfo {title} {Bidirectional
  {Electro}-{Optic} {Wavelength} {Conversion} in the {Quantum} {Ground}
  {State}},\ }\href {https://doi.org/10.1103/PRXQuantum.1.020315} {\bibfield
  {journal} {\bibinfo  {journal} {PRX Quantum}\ }\textbf {\bibinfo {volume}
  {1}},\ \bibinfo {pages} {020315} (\bibinfo {year} {2020})}\BibitemShut
  {NoStop}%
\bibitem [{\citenamefont {Witmer}\ \emph {et~al.}(2020)\citenamefont {Witmer},
  \citenamefont {McKenna}, \citenamefont {Arrangoiz-Arriola}, \citenamefont
  {Laer}, \citenamefont {Wollack}, \citenamefont {Lin}, \citenamefont {Jen},
  \citenamefont {Luo},\ and\ \citenamefont
  {Safavi-Naeini}}]{witmer_silicon-organic_2020}%
  \BibitemOpen
  \bibfield  {author} {\bibinfo {author} {\bibfnamefont {J.~D.}\ \bibnamefont
  {Witmer}}, \bibinfo {author} {\bibfnamefont {T.~P.}\ \bibnamefont {McKenna}},
  \bibinfo {author} {\bibfnamefont {P.}~\bibnamefont {Arrangoiz-Arriola}},
  \bibinfo {author} {\bibfnamefont {R.~V.}\ \bibnamefont {Laer}}, \bibinfo
  {author} {\bibfnamefont {E.~A.}\ \bibnamefont {Wollack}}, \bibinfo {author}
  {\bibfnamefont {F.}~\bibnamefont {Lin}}, \bibinfo {author} {\bibfnamefont
  {A.~K.-Y.}\ \bibnamefont {Jen}}, \bibinfo {author} {\bibfnamefont
  {J.}~\bibnamefont {Luo}},\ and\ \bibinfo {author} {\bibfnamefont {A.~H.}\
  \bibnamefont {Safavi-Naeini}},\ }\bibfield  {title} {\bibinfo {title} {A
  silicon-organic hybrid platform for quantum microwave-to-optical
  transduction},\ }\href {https://doi.org/10.1088/2058-9565/ab7eed} {\bibfield
  {journal} {\bibinfo  {journal} {Quantum Sci. Technol.}\ }\textbf {\bibinfo
  {volume} {5}},\ \bibinfo {pages} {034004} (\bibinfo {year}
  {2020})}\BibitemShut {NoStop}%
\bibitem [{\citenamefont {Forsch}\ \emph {et~al.}(2020)\citenamefont {Forsch},
  \citenamefont {Stockill}, \citenamefont {Wallucks}, \citenamefont
  {Marinkovi\'{c}}, \citenamefont {G\"{a}rtner}, \citenamefont {Norte},
  \citenamefont {van Otten}, \citenamefont {Fiore}, \citenamefont
  {Srinivasan},\ and\ \citenamefont
  {Gr\"{o}blacher}}]{forsch_microwave--optics_2020}%
  \BibitemOpen
  \bibfield  {author} {\bibinfo {author} {\bibfnamefont {M.}~\bibnamefont
  {Forsch}}, \bibinfo {author} {\bibfnamefont {R.}~\bibnamefont {Stockill}},
  \bibinfo {author} {\bibfnamefont {A.}~\bibnamefont {Wallucks}}, \bibinfo
  {author} {\bibfnamefont {I.}~\bibnamefont {Marinkovi\'{c}}}, \bibinfo
  {author} {\bibfnamefont {C.}~\bibnamefont {G\"{a}rtner}}, \bibinfo {author}
  {\bibfnamefont {R.~A.}\ \bibnamefont {Norte}}, \bibinfo {author}
  {\bibfnamefont {F.}~\bibnamefont {van Otten}}, \bibinfo {author}
  {\bibfnamefont {A.}~\bibnamefont {Fiore}}, \bibinfo {author} {\bibfnamefont
  {K.}~\bibnamefont {Srinivasan}},\ and\ \bibinfo {author} {\bibfnamefont
  {S.}~\bibnamefont {Gr\"{o}blacher}},\ }\bibfield  {title} {\bibinfo {title}
  {Microwave-to-optics conversion using a mechanical oscillator in its quantum
  ground state},\ }\href {https://doi.org/10.1038/s41567-019-0673-7} {\bibfield
   {journal} {\bibinfo  {journal} {Nat. Phys.}\ }\textbf {\bibinfo {volume}
  {16}},\ \bibinfo {pages} {69} (\bibinfo {year} {2020})}\BibitemShut {NoStop}%
\bibitem [{\citenamefont {Fu}\ \emph {et~al.}(2021)\citenamefont {Fu},
  \citenamefont {Xu}, \citenamefont {Liu}, \citenamefont {Zou}, \citenamefont
  {Zhong}, \citenamefont {Han}, \citenamefont {Shen}, \citenamefont {Xu},
  \citenamefont {Cheng}, \citenamefont {Wang}, \citenamefont {Jiang},\ and\
  \citenamefont {Tang}}]{fu_ground-state_2020}%
  \BibitemOpen
  \bibfield  {author} {\bibinfo {author} {\bibfnamefont {W.}~\bibnamefont
  {Fu}}, \bibinfo {author} {\bibfnamefont {M.}~\bibnamefont {Xu}}, \bibinfo
  {author} {\bibfnamefont {X.}~\bibnamefont {Liu}}, \bibinfo {author}
  {\bibfnamefont {C.-L.}\ \bibnamefont {Zou}}, \bibinfo {author} {\bibfnamefont
  {C.}~\bibnamefont {Zhong}}, \bibinfo {author} {\bibfnamefont
  {X.}~\bibnamefont {Han}}, \bibinfo {author} {\bibfnamefont {M.}~\bibnamefont
  {Shen}}, \bibinfo {author} {\bibfnamefont {Y.}~\bibnamefont {Xu}}, \bibinfo
  {author} {\bibfnamefont {R.}~\bibnamefont {Cheng}}, \bibinfo {author}
  {\bibfnamefont {S.}~\bibnamefont {Wang}}, \bibinfo {author} {\bibfnamefont
  {L.}~\bibnamefont {Jiang}},\ and\ \bibinfo {author} {\bibfnamefont {H.~X.}\
  \bibnamefont {Tang}},\ }\bibfield  {title} {\bibinfo {title} {Cavity
  electro-optic circuit for microwave-to-optical conversion in the quantum
  ground state},\ }\href {https://doi.org/10.1103/PhysRevA.103.053504}
  {\bibfield  {journal} {\bibinfo  {journal} {Phys. Rev. A}\ }\textbf {\bibinfo
  {volume} {103}},\ \bibinfo {pages} {053504} (\bibinfo {year}
  {2021})}\BibitemShut {NoStop}%
\bibitem [{\citenamefont {Stockill}\ \emph {et~al.}()\citenamefont {Stockill},
  \citenamefont {Forsch}, \citenamefont {Hijazi}, \citenamefont {Beaudoin},
  \citenamefont {Pantzas}, \citenamefont {Sagnes}, \citenamefont {Braive},\
  and\ \citenamefont {Gr\"{o}blacher}}]{stockill_ultra-low-noise_2021}%
  \BibitemOpen
  \bibfield  {author} {\bibinfo {author} {\bibfnamefont {R.}~\bibnamefont
  {Stockill}}, \bibinfo {author} {\bibfnamefont {M.}~\bibnamefont {Forsch}},
  \bibinfo {author} {\bibfnamefont {F.}~\bibnamefont {Hijazi}}, \bibinfo
  {author} {\bibfnamefont {G.}~\bibnamefont {Beaudoin}}, \bibinfo {author}
  {\bibfnamefont {K.}~\bibnamefont {Pantzas}}, \bibinfo {author} {\bibfnamefont
  {I.}~\bibnamefont {Sagnes}}, \bibinfo {author} {\bibfnamefont
  {R.}~\bibnamefont {Braive}},\ and\ \bibinfo {author} {\bibfnamefont
  {S.}~\bibnamefont {Gr\"{o}blacher}},\ }\bibfield  {title} {\bibinfo {title}
  {Ultra-low-noise {Microwave} to {Optics} {Conversion} in {Gallium}
  {Phosphide}},\ }\Eprint {https://arxiv.org/abs/2107.04433} {arXiv:2107.04433}
  \BibitemShut {NoStop}%
\bibitem [{\citenamefont {Sahu}\ \emph {et~al.}()\citenamefont {Sahu},
  \citenamefont {Hease}, \citenamefont {Rueda}, \citenamefont {Arnold},
  \citenamefont {Qiu},\ and\ \citenamefont {Fink}}]{sahu_quantum-enabled_2021}%
  \BibitemOpen
  \bibfield  {author} {\bibinfo {author} {\bibfnamefont {R.}~\bibnamefont
  {Sahu}}, \bibinfo {author} {\bibfnamefont {W.}~\bibnamefont {Hease}},
  \bibinfo {author} {\bibfnamefont {A.}~\bibnamefont {Rueda}}, \bibinfo
  {author} {\bibfnamefont {G.}~\bibnamefont {Arnold}}, \bibinfo {author}
  {\bibfnamefont {L.}~\bibnamefont {Qiu}},\ and\ \bibinfo {author}
  {\bibfnamefont {J.}~\bibnamefont {Fink}},\ }\bibfield  {title} {\bibinfo
  {title} {Quantum-enabled interface between microwave and telecom light},\
  }\Eprint {https://arxiv.org/abs/2107.08303} {arXiv:2107.08303} \BibitemShut
  {NoStop}%
\bibitem [{\citenamefont {Barzanjeh}\ \emph {et~al.}(2011)\citenamefont
  {Barzanjeh}, \citenamefont {Vitali}, \citenamefont {Tombesi},\ and\
  \citenamefont {Milburn}}]{barzanjeh_entangling_2011}%
  \BibitemOpen
  \bibfield  {author} {\bibinfo {author} {\bibfnamefont {S.}~\bibnamefont
  {Barzanjeh}}, \bibinfo {author} {\bibfnamefont {D.}~\bibnamefont {Vitali}},
  \bibinfo {author} {\bibfnamefont {P.}~\bibnamefont {Tombesi}},\ and\ \bibinfo
  {author} {\bibfnamefont {G.~J.}\ \bibnamefont {Milburn}},\ }\bibfield
  {title} {\bibinfo {title} {Entangling optical and microwave cavity modes by
  means of a nanomechanical resonator},\ }\href
  {https://doi.org/10.1103/PhysRevA.84.042342} {\bibfield  {journal} {\bibinfo
  {journal} {Phys. Rev. A}\ }\textbf {\bibinfo {volume} {84}},\ \bibinfo
  {pages} {042342} (\bibinfo {year} {2011})}\BibitemShut {NoStop}%
\bibitem [{\citenamefont {Zhong}\ \emph {et~al.}(2020)\citenamefont {Zhong},
  \citenamefont {Wang}, \citenamefont {Zou}, \citenamefont {Zhang},
  \citenamefont {Han}, \citenamefont {Fu}, \citenamefont {Xu}, \citenamefont
  {Shankar}, \citenamefont {Devoret}, \citenamefont {Tang},\ and\ \citenamefont
  {Jiang}}]{zhong_proposal_2020}%
  \BibitemOpen
  \bibfield  {author} {\bibinfo {author} {\bibfnamefont {C.}~\bibnamefont
  {Zhong}}, \bibinfo {author} {\bibfnamefont {Z.}~\bibnamefont {Wang}},
  \bibinfo {author} {\bibfnamefont {C.}~\bibnamefont {Zou}}, \bibinfo {author}
  {\bibfnamefont {M.}~\bibnamefont {Zhang}}, \bibinfo {author} {\bibfnamefont
  {X.}~\bibnamefont {Han}}, \bibinfo {author} {\bibfnamefont {W.}~\bibnamefont
  {Fu}}, \bibinfo {author} {\bibfnamefont {M.}~\bibnamefont {Xu}}, \bibinfo
  {author} {\bibfnamefont {S.}~\bibnamefont {Shankar}}, \bibinfo {author}
  {\bibfnamefont {M.~H.}\ \bibnamefont {Devoret}}, \bibinfo {author}
  {\bibfnamefont {H.~X.}\ \bibnamefont {Tang}},\ and\ \bibinfo {author}
  {\bibfnamefont {L.}~\bibnamefont {Jiang}},\ }\bibfield  {title} {\bibinfo
  {title} {Proposal for {Heralded} {Generation} and {Detection} of {Entangled}
  {Microwave}--{Optical}-{Photon} {Pairs}},\ }\href
  {https://doi.org/10.1103/PhysRevLett.124.010511} {\bibfield  {journal}
  {\bibinfo  {journal} {Phys. Rev. Lett.}\ }\textbf {\bibinfo {volume} {124}},\
  \bibinfo {pages} {010511} (\bibinfo {year} {2020})}\BibitemShut {NoStop}%
\bibitem [{\citenamefont {Rau}\ \emph {et~al.}()\citenamefont {Rau},
  \citenamefont {Kyle}, \citenamefont {Kwiatkowski}, \citenamefont {Shojaee},
  \citenamefont {Teufel}, \citenamefont {Lehnert},\ and\ \citenamefont
  {Dennis}}]{rau_entanglement_2021}%
  \BibitemOpen
  \bibfield  {author} {\bibinfo {author} {\bibfnamefont {C.~L.}\ \bibnamefont
  {Rau}}, \bibinfo {author} {\bibfnamefont {A.}~\bibnamefont {Kyle}}, \bibinfo
  {author} {\bibfnamefont {A.}~\bibnamefont {Kwiatkowski}}, \bibinfo {author}
  {\bibfnamefont {E.}~\bibnamefont {Shojaee}}, \bibinfo {author} {\bibfnamefont
  {J.~D.}\ \bibnamefont {Teufel}}, \bibinfo {author} {\bibfnamefont {K.~W.}\
  \bibnamefont {Lehnert}},\ and\ \bibinfo {author} {\bibfnamefont
  {T.}~\bibnamefont {Dennis}},\ }\bibfield  {title} {\bibinfo {title}
  {Entanglement {Thresholds} of {Doubly}-{Parametric} {Quantum}
  {Transducers}},\ }\Eprint {https://arxiv.org/abs/2110.10235}
  {arXiv:2110.10235} \BibitemShut {NoStop}%
\bibitem [{\citenamefont {Higginbotham}\ \emph {et~al.}(2018)\citenamefont
  {Higginbotham}, \citenamefont {Burns}, \citenamefont {Urmey}, \citenamefont
  {Peterson}, \citenamefont {Kampel}, \citenamefont {Brubaker}, \citenamefont
  {Smith}, \citenamefont {Lehnert},\ and\ \citenamefont
  {Regal}}]{higginbotham_harnessing_2018}%
  \BibitemOpen
  \bibfield  {author} {\bibinfo {author} {\bibfnamefont {A.~P.}\ \bibnamefont
  {Higginbotham}}, \bibinfo {author} {\bibfnamefont {P.~S.}\ \bibnamefont
  {Burns}}, \bibinfo {author} {\bibfnamefont {M.~D.}\ \bibnamefont {Urmey}},
  \bibinfo {author} {\bibfnamefont {R.~W.}\ \bibnamefont {Peterson}}, \bibinfo
  {author} {\bibfnamefont {N.~S.}\ \bibnamefont {Kampel}}, \bibinfo {author}
  {\bibfnamefont {B.~M.}\ \bibnamefont {Brubaker}}, \bibinfo {author}
  {\bibfnamefont {G.}~\bibnamefont {Smith}}, \bibinfo {author} {\bibfnamefont
  {K.~W.}\ \bibnamefont {Lehnert}},\ and\ \bibinfo {author} {\bibfnamefont
  {C.~A.}\ \bibnamefont {Regal}},\ }\bibfield  {title} {\bibinfo {title}
  {Harnessing electro-optic correlations in an efficient mechanical
  converter},\ }\href {https://doi.org/10.1038/s41567-018-0210-0} {\bibfield
  {journal} {\bibinfo  {journal} {Nat. Phys.}\ }\textbf {\bibinfo {volume}
  {14}},\ \bibinfo {pages} {1038} (\bibinfo {year} {2018})}\BibitemShut
  {NoStop}%
\bibitem [{\citenamefont {Andrews}\ \emph {et~al.}(2014)\citenamefont
  {Andrews}, \citenamefont {Peterson}, \citenamefont {Purdy}, \citenamefont
  {Cicak}, \citenamefont {Simmonds}, \citenamefont {Regal},\ and\ \citenamefont
  {Lehnert}}]{andrews_bidirectional_2014}%
  \BibitemOpen
  \bibfield  {author} {\bibinfo {author} {\bibfnamefont {R.~W.}\ \bibnamefont
  {Andrews}}, \bibinfo {author} {\bibfnamefont {R.~W.}\ \bibnamefont
  {Peterson}}, \bibinfo {author} {\bibfnamefont {T.~P.}\ \bibnamefont {Purdy}},
  \bibinfo {author} {\bibfnamefont {K.}~\bibnamefont {Cicak}}, \bibinfo
  {author} {\bibfnamefont {R.~W.}\ \bibnamefont {Simmonds}}, \bibinfo {author}
  {\bibfnamefont {C.~A.}\ \bibnamefont {Regal}},\ and\ \bibinfo {author}
  {\bibfnamefont {K.~W.}\ \bibnamefont {Lehnert}},\ }\bibfield  {title}
  {\bibinfo {title} {Bidirectional and efficient conversion between microwave
  and optical light},\ }\href {https://doi.org/10.1038/nphys2911} {\bibfield
  {journal} {\bibinfo  {journal} {Nat. Phys.}\ }\textbf {\bibinfo {volume}
  {10}},\ \bibinfo {pages} {321} (\bibinfo {year} {2014})}\BibitemShut
  {NoStop}%
\bibitem [{\citenamefont {Yu}\ \emph {et~al.}(2014)\citenamefont {Yu},
  \citenamefont {Cicak}, \citenamefont {Kampel}, \citenamefont {Tsaturyan},
  \citenamefont {Purdy}, \citenamefont {Simmonds},\ and\ \citenamefont
  {Regal}}]{yu_phononic_2014}%
  \BibitemOpen
  \bibfield  {author} {\bibinfo {author} {\bibfnamefont {P.-L.}\ \bibnamefont
  {Yu}}, \bibinfo {author} {\bibfnamefont {K.}~\bibnamefont {Cicak}}, \bibinfo
  {author} {\bibfnamefont {N.}~\bibnamefont {Kampel}}, \bibinfo {author}
  {\bibfnamefont {Y.}~\bibnamefont {Tsaturyan}}, \bibinfo {author}
  {\bibfnamefont {T.}~\bibnamefont {Purdy}}, \bibinfo {author} {\bibfnamefont
  {R.}~\bibnamefont {Simmonds}},\ and\ \bibinfo {author} {\bibfnamefont
  {C.}~\bibnamefont {Regal}},\ }\bibfield  {title} {\bibinfo {title} {A
  phononic bandgap shield for high-{Q} membrane microresonators},\ }\href
  {https://doi.org/http://dx.doi.org/10.1063/1.4862031} {\bibfield  {journal}
  {\bibinfo  {journal} {Appl. Phys. Lett.}\ }\textbf {\bibinfo {volume}
  {104}},\ \bibinfo {pages} {023510} (\bibinfo {year} {2014})}\BibitemShut
  {NoStop}%
\bibitem [{\citenamefont {Tsaturyan}\ \emph {et~al.}(2014)\citenamefont
  {Tsaturyan}, \citenamefont {Barg}, \citenamefont {Simonsen}, \citenamefont
  {Villanueva}, \citenamefont {Schmid}, \citenamefont {Schliesser},\ and\
  \citenamefont {Polzik}}]{tsaturyan_demonstration_2014}%
  \BibitemOpen
  \bibfield  {author} {\bibinfo {author} {\bibfnamefont {Y.}~\bibnamefont
  {Tsaturyan}}, \bibinfo {author} {\bibfnamefont {A.}~\bibnamefont {Barg}},
  \bibinfo {author} {\bibfnamefont {A.}~\bibnamefont {Simonsen}}, \bibinfo
  {author} {\bibfnamefont {L.~G.}\ \bibnamefont {Villanueva}}, \bibinfo
  {author} {\bibfnamefont {S.}~\bibnamefont {Schmid}}, \bibinfo {author}
  {\bibfnamefont {A.}~\bibnamefont {Schliesser}},\ and\ \bibinfo {author}
  {\bibfnamefont {E.~S.}\ \bibnamefont {Polzik}},\ }\bibfield  {title}
  {\bibinfo {title} {Demonstration of suppressed phonon tunneling losses in
  phononic bandgap shielded membrane resonators for high-{Q} optomechanics},\
  }\href {https://doi.org/10.1364/OE.22.006810} {\bibfield  {journal} {\bibinfo
   {journal} {Opt. Express}\ }\textbf {\bibinfo {volume} {22}},\ \bibinfo
  {pages} {6810} (\bibinfo {year} {2014})}\BibitemShut {NoStop}%
\bibitem [{\citenamefont {Aspelmeyer}\ \emph {et~al.}(2014)\citenamefont
  {Aspelmeyer}, \citenamefont {Kippenberg},\ and\ \citenamefont
  {Marquardt}}]{aspelmeyer_cavity_2014}%
  \BibitemOpen
  \bibfield  {author} {\bibinfo {author} {\bibfnamefont {M.}~\bibnamefont
  {Aspelmeyer}}, \bibinfo {author} {\bibfnamefont {T.~J.}\ \bibnamefont
  {Kippenberg}},\ and\ \bibinfo {author} {\bibfnamefont {F.}~\bibnamefont
  {Marquardt}},\ }\bibfield  {title} {\bibinfo {title} {Cavity optomechanics},\
  }\href {https://doi.org/10.1103/RevModPhys.86.1391} {\bibfield  {journal}
  {\bibinfo  {journal} {Rev. Mod. Phys.}\ }\textbf {\bibinfo {volume} {86}},\
  \bibinfo {pages} {1391} (\bibinfo {year} {2014})}\BibitemShut {NoStop}%
\bibitem [{\citenamefont {Peterson}\ \emph {et~al.}(2016)\citenamefont
  {Peterson}, \citenamefont {Purdy}, \citenamefont {Kampel}, \citenamefont
  {Andrews}, \citenamefont {Yu}, \citenamefont {Lehnert},\ and\ \citenamefont
  {Regal}}]{peterson_laser_2016}%
  \BibitemOpen
  \bibfield  {author} {\bibinfo {author} {\bibfnamefont {R.~W.}\ \bibnamefont
  {Peterson}}, \bibinfo {author} {\bibfnamefont {T.~P.}\ \bibnamefont {Purdy}},
  \bibinfo {author} {\bibfnamefont {N.~S.}\ \bibnamefont {Kampel}}, \bibinfo
  {author} {\bibfnamefont {R.~W.}\ \bibnamefont {Andrews}}, \bibinfo {author}
  {\bibfnamefont {P.-L.}\ \bibnamefont {Yu}}, \bibinfo {author} {\bibfnamefont
  {K.~W.}\ \bibnamefont {Lehnert}},\ and\ \bibinfo {author} {\bibfnamefont
  {C.~A.}\ \bibnamefont {Regal}},\ }\bibfield  {title} {\bibinfo {title} {Laser
  cooling of a micromechanical membrane to the quantum backaction limit},\
  }\href {https://doi.org/10.1103/PhysRevLett.116.063601} {\bibfield  {journal}
  {\bibinfo  {journal} {Phys. Rev. Lett.}\ }\textbf {\bibinfo {volume} {116}},\
  \bibinfo {pages} {063601} (\bibinfo {year} {2016})}\BibitemShut {NoStop}%
\bibitem [{\citenamefont {Safavi-Naeini}\ \emph {et~al.}(2012)\citenamefont
  {Safavi-Naeini}, \citenamefont {Chan}, \citenamefont {Hill}, \citenamefont
  {Alegre}, \citenamefont {Krause},\ and\ \citenamefont
  {Painter}}]{safavi-naeini_observation_2012}%
  \BibitemOpen
  \bibfield  {author} {\bibinfo {author} {\bibfnamefont {A.~H.}\ \bibnamefont
  {Safavi-Naeini}}, \bibinfo {author} {\bibfnamefont {J.}~\bibnamefont {Chan}},
  \bibinfo {author} {\bibfnamefont {J.~T.}\ \bibnamefont {Hill}}, \bibinfo
  {author} {\bibfnamefont {T.~P.~M.}\ \bibnamefont {Alegre}}, \bibinfo {author}
  {\bibfnamefont {A.}~\bibnamefont {Krause}},\ and\ \bibinfo {author}
  {\bibfnamefont {O.}~\bibnamefont {Painter}},\ }\bibfield  {title} {\bibinfo
  {title} {Observation of {Quantum} {Motion} of a {Nanomechanical}
  {Resonator}},\ }\href {https://doi.org/10.1103/PhysRevLett.108.033602}
  {\bibfield  {journal} {\bibinfo  {journal} {Phys. Rev. Lett.}\ }\textbf
  {\bibinfo {volume} {108}},\ \bibinfo {pages} {033602} (\bibinfo {year}
  {2012})}\BibitemShut {NoStop}%
\bibitem [{\citenamefont {Vlastakis}\ \emph {et~al.}(2013)\citenamefont
  {Vlastakis}, \citenamefont {Kirchmair}, \citenamefont {Leghtas},
  \citenamefont {Nigg}, \citenamefont {Frunzio}, \citenamefont {Girvin},
  \citenamefont {Mirrahimi}, \citenamefont {Devoret},\ and\ \citenamefont
  {Schoelkopf}}]{vlastakis_deterministically_2013}%
  \BibitemOpen
  \bibfield  {author} {\bibinfo {author} {\bibfnamefont {B.}~\bibnamefont
  {Vlastakis}}, \bibinfo {author} {\bibfnamefont {G.}~\bibnamefont
  {Kirchmair}}, \bibinfo {author} {\bibfnamefont {Z.}~\bibnamefont {Leghtas}},
  \bibinfo {author} {\bibfnamefont {S.~E.}\ \bibnamefont {Nigg}}, \bibinfo
  {author} {\bibfnamefont {L.}~\bibnamefont {Frunzio}}, \bibinfo {author}
  {\bibfnamefont {S.~M.}\ \bibnamefont {Girvin}}, \bibinfo {author}
  {\bibfnamefont {M.}~\bibnamefont {Mirrahimi}}, \bibinfo {author}
  {\bibfnamefont {M.~H.}\ \bibnamefont {Devoret}},\ and\ \bibinfo {author}
  {\bibfnamefont {R.~J.}\ \bibnamefont {Schoelkopf}},\ }\bibfield  {title}
  {\bibinfo {title} {Deterministically encoding quantum information using
  100-photon {Schrödinger} cat states},\ }\href
  {https://doi.org/10.1126/science.1243289} {\bibfield  {journal} {\bibinfo
  {journal} {Science}\ }\textbf {\bibinfo {volume} {342}},\ \bibinfo {pages}
  {607} (\bibinfo {year} {2013})}\BibitemShut {NoStop}%
\bibitem [{\citenamefont {Krastanov}\ \emph {et~al.}(2021)\citenamefont
  {Krastanov}, \citenamefont {Raniwala}, \citenamefont {Holzgrafe},
  \citenamefont {Jacobs}, \citenamefont {Lon\v{c}ar}, \citenamefont {Reagor},\
  and\ \citenamefont {Englund}}]{krastanov_optically_2021}%
  \BibitemOpen
  \bibfield  {author} {\bibinfo {author} {\bibfnamefont {S.}~\bibnamefont
  {Krastanov}}, \bibinfo {author} {\bibfnamefont {H.}~\bibnamefont {Raniwala}},
  \bibinfo {author} {\bibfnamefont {J.}~\bibnamefont {Holzgrafe}}, \bibinfo
  {author} {\bibfnamefont {K.}~\bibnamefont {Jacobs}}, \bibinfo {author}
  {\bibfnamefont {M.}~\bibnamefont {Lon\v{c}ar}}, \bibinfo {author}
  {\bibfnamefont {M.~J.}\ \bibnamefont {Reagor}},\ and\ \bibinfo {author}
  {\bibfnamefont {D.~R.}\ \bibnamefont {Englund}},\ }\bibfield  {title}
  {\bibinfo {title} {Optically {Heralded} {Entanglement} of {Superconducting}
  {Systems} in {Quantum} {Networks}},\ }\href
  {https://doi.org/10.1103/PhysRevLett.127.040503} {\bibfield  {journal}
  {\bibinfo  {journal} {Phys. Rev. Lett.}\ }\textbf {\bibinfo {volume} {127}},\
  \bibinfo {pages} {040503} (\bibinfo {year} {2021})}\BibitemShut {NoStop}%
\bibitem [{\citenamefont {Underwood}\ \emph {et~al.}(2015)\citenamefont
  {Underwood}, \citenamefont {Mason}, \citenamefont {Lee}, \citenamefont {Xu},
  \citenamefont {Jiang}, \citenamefont {Shkarin}, \citenamefont {B{\o}rkje},
  \citenamefont {Girvin},\ and\ \citenamefont
  {Harris}}]{underwood_measurement_2015}%
  \BibitemOpen
  \bibfield  {author} {\bibinfo {author} {\bibfnamefont {M.}~\bibnamefont
  {Underwood}}, \bibinfo {author} {\bibfnamefont {D.}~\bibnamefont {Mason}},
  \bibinfo {author} {\bibfnamefont {D.}~\bibnamefont {Lee}}, \bibinfo {author}
  {\bibfnamefont {H.}~\bibnamefont {Xu}}, \bibinfo {author} {\bibfnamefont
  {L.}~\bibnamefont {Jiang}}, \bibinfo {author} {\bibfnamefont {A.~B.}\
  \bibnamefont {Shkarin}}, \bibinfo {author} {\bibfnamefont {K.}~\bibnamefont
  {B{\o}rkje}}, \bibinfo {author} {\bibfnamefont {S.~M.}\ \bibnamefont
  {Girvin}},\ and\ \bibinfo {author} {\bibfnamefont {J.~G.~E.}\ \bibnamefont
  {Harris}},\ }\bibfield  {title} {\bibinfo {title} {Measurement of the
  motional sidebands of a nanogram-scale oscillator in the quantum regime},\
  }\href {https://doi.org/10.1103/PhysRevA.92.061801} {\bibfield  {journal}
  {\bibinfo  {journal} {Phys. Rev. A}\ }\textbf {\bibinfo {volume} {92}},\
  \bibinfo {pages} {061801(R)} (\bibinfo {year} {2015})}\BibitemShut {NoStop}%
\bibitem [{\citenamefont {Jayich}\ \emph {et~al.}(2012)\citenamefont {Jayich},
  \citenamefont {Sankey}, \citenamefont {B{\o}rkje}, \citenamefont {Lee},
  \citenamefont {Yang}, \citenamefont {Underwood}, \citenamefont {Childress},
  \citenamefont {Petrenko}, \citenamefont {Girvin},\ and\ \citenamefont
  {Harris}}]{jayich_cryogenic_2012}%
  \BibitemOpen
  \bibfield  {author} {\bibinfo {author} {\bibfnamefont {A.~M.}\ \bibnamefont
  {Jayich}}, \bibinfo {author} {\bibfnamefont {J.~C.}\ \bibnamefont {Sankey}},
  \bibinfo {author} {\bibfnamefont {K.}~\bibnamefont {B{\o}rkje}}, \bibinfo
  {author} {\bibfnamefont {D.}~\bibnamefont {Lee}}, \bibinfo {author}
  {\bibfnamefont {C.}~\bibnamefont {Yang}}, \bibinfo {author} {\bibfnamefont
  {M.}~\bibnamefont {Underwood}}, \bibinfo {author} {\bibfnamefont
  {L.}~\bibnamefont {Childress}}, \bibinfo {author} {\bibfnamefont
  {A.}~\bibnamefont {Petrenko}}, \bibinfo {author} {\bibfnamefont {S.~M.}\
  \bibnamefont {Girvin}},\ and\ \bibinfo {author} {\bibfnamefont {J.~G.~E.}\
  \bibnamefont {Harris}},\ }\bibfield  {title} {\bibinfo {title} {Cryogenic
  optomechanics with a {Si}$_3${N}$_4$ membrane and classical laser noise},\
  }\href {https://doi.org/10.1088/1367-2630/14/11/115018} {\bibfield  {journal}
  {\bibinfo  {journal} {New J. Phys.}\ }\textbf {\bibinfo {volume} {14}},\
  \bibinfo {pages} {115018} (\bibinfo {year} {2012})}\BibitemShut {NoStop}%
\bibitem [{\citenamefont {Safavi-Naeini}\ \emph {et~al.}(2013)\citenamefont
  {Safavi-Naeini}, \citenamefont {Chan}, \citenamefont {Hill}, \citenamefont
  {Gr{\"o}blacher}, \citenamefont {Miao}, \citenamefont {Chen}, \citenamefont
  {Aspelmeyer},\ and\ \citenamefont {Painter}}]{safavi-naeini_laser_2013}%
  \BibitemOpen
  \bibfield  {author} {\bibinfo {author} {\bibfnamefont {A.~H.}\ \bibnamefont
  {Safavi-Naeini}}, \bibinfo {author} {\bibfnamefont {J.}~\bibnamefont {Chan}},
  \bibinfo {author} {\bibfnamefont {J.~T.}\ \bibnamefont {Hill}}, \bibinfo
  {author} {\bibfnamefont {S.}~\bibnamefont {Gr{\"o}blacher}}, \bibinfo
  {author} {\bibfnamefont {H.}~\bibnamefont {Miao}}, \bibinfo {author}
  {\bibfnamefont {Y.}~\bibnamefont {Chen}}, \bibinfo {author} {\bibfnamefont
  {M.}~\bibnamefont {Aspelmeyer}},\ and\ \bibinfo {author} {\bibfnamefont
  {O.}~\bibnamefont {Painter}},\ }\bibfield  {title} {\bibinfo {title} {Laser
  noise in cavity-optomechanical cooling and thermometry},\ }\href
  {https://doi.org/10.1088/1367-2630/15/3/035007} {\bibfield  {journal}
  {\bibinfo  {journal} {New J. Phys.}\ }\textbf {\bibinfo {volume} {15}},\
  \bibinfo {pages} {035007} (\bibinfo {year} {2013})}\BibitemShut {NoStop}%
\bibitem [{\citenamefont {Delaney}\ \emph {et~al.}()\citenamefont {Delaney},
  \citenamefont {Urmey}, \citenamefont {Mittal}, \citenamefont {Brubaker},
  \citenamefont {Kindem}, \citenamefont {Burns}, \citenamefont {Regal},\ and\
  \citenamefont {Lehnert}}]{delaney_non-destructive_2021}%
  \BibitemOpen
  \bibfield  {author} {\bibinfo {author} {\bibfnamefont {R.~D.}\ \bibnamefont
  {Delaney}}, \bibinfo {author} {\bibfnamefont {M.~D.}\ \bibnamefont {Urmey}},
  \bibinfo {author} {\bibfnamefont {S.}~\bibnamefont {Mittal}}, \bibinfo
  {author} {\bibfnamefont {B.~M.}\ \bibnamefont {Brubaker}}, \bibinfo {author}
  {\bibfnamefont {J.~M.}\ \bibnamefont {Kindem}}, \bibinfo {author}
  {\bibfnamefont {P.~S.}\ \bibnamefont {Burns}}, \bibinfo {author}
  {\bibfnamefont {C.~A.}\ \bibnamefont {Regal}},\ and\ \bibinfo {author}
  {\bibfnamefont {K.~W.}\ \bibnamefont {Lehnert}},\ }\bibfield  {title}
  {\bibinfo {title} {Non-destructive optical readout of a superconducting
  qubit},\ }\Eprint {https://arxiv.org/abs/2110.09539} {arXiv:2110.09539}
  \BibitemShut {NoStop}%
\bibitem [{\citenamefont {Burns}(2019)}]{burns_reducing_2019}%
  \BibitemOpen
  \bibfield  {author} {\bibinfo {author} {\bibfnamefont {P.~S.}\ \bibnamefont
  {Burns}},\ }\emph {\bibinfo {title} {Reducing Added Noise in a
  Microwave-Mechanical-Optical Converter}},\ \href@noop {} {Ph.D. thesis},\
  \bibinfo  {school} {University of Colorado, Boulder} (\bibinfo {year}
  {2019})\BibitemShut {NoStop}%
\bibitem [{\citenamefont {Reagor}\ \emph {et~al.}(2016)\citenamefont {Reagor},
  \citenamefont {Pfaff}, \citenamefont {Axline}, \citenamefont {Heeres},
  \citenamefont {Ofek}, \citenamefont {Sliwa}, \citenamefont {Holland},
  \citenamefont {Wang}, \citenamefont {Blumoff}, \citenamefont {Chou},
  \citenamefont {Hatridge}, \citenamefont {Frunzio}, \citenamefont {Devoret},
  \citenamefont {Jiang},\ and\ \citenamefont
  {Schoelkopf}}]{reagor_quantum_2016}%
  \BibitemOpen
  \bibfield  {author} {\bibinfo {author} {\bibfnamefont {M.}~\bibnamefont
  {Reagor}}, \bibinfo {author} {\bibfnamefont {W.}~\bibnamefont {Pfaff}},
  \bibinfo {author} {\bibfnamefont {C.}~\bibnamefont {Axline}}, \bibinfo
  {author} {\bibfnamefont {R.~W.}\ \bibnamefont {Heeres}}, \bibinfo {author}
  {\bibfnamefont {N.}~\bibnamefont {Ofek}}, \bibinfo {author} {\bibfnamefont
  {K.}~\bibnamefont {Sliwa}}, \bibinfo {author} {\bibfnamefont
  {E.}~\bibnamefont {Holland}}, \bibinfo {author} {\bibfnamefont
  {C.}~\bibnamefont {Wang}}, \bibinfo {author} {\bibfnamefont {J.}~\bibnamefont
  {Blumoff}}, \bibinfo {author} {\bibfnamefont {K.}~\bibnamefont {Chou}},
  \bibinfo {author} {\bibfnamefont {M.~J.}\ \bibnamefont {Hatridge}}, \bibinfo
  {author} {\bibfnamefont {L.}~\bibnamefont {Frunzio}}, \bibinfo {author}
  {\bibfnamefont {M.~H.}\ \bibnamefont {Devoret}}, \bibinfo {author}
  {\bibfnamefont {L.}~\bibnamefont {Jiang}},\ and\ \bibinfo {author}
  {\bibfnamefont {R.~J.}\ \bibnamefont {Schoelkopf}},\ }\bibfield  {title}
  {\bibinfo {title} {Quantum memory with millisecond coherence in circuit
  {QED}},\ }\href {https://doi.org/10.1103/PhysRevB.94.014506} {\bibfield
  {journal} {\bibinfo  {journal} {Phys. Rev. B}\ }\textbf {\bibinfo {volume}
  {94}},\ \bibinfo {pages} {014506} (\bibinfo {year} {2016})}\BibitemShut
  {NoStop}%
\bibitem [{\citenamefont {Galinskiy}\ \emph {et~al.}(2020)\citenamefont
  {Galinskiy}, \citenamefont {Tsaturyan}, \citenamefont {Tsaturyan},
  \citenamefont {Parniak},\ and\ \citenamefont
  {Polzik}}]{galinskiy_phonon_2020}%
  \BibitemOpen
  \bibfield  {author} {\bibinfo {author} {\bibfnamefont {I.}~\bibnamefont
  {Galinskiy}}, \bibinfo {author} {\bibfnamefont {Y.}~\bibnamefont
  {Tsaturyan}}, \bibinfo {author} {\bibfnamefont {Y.}~\bibnamefont
  {Tsaturyan}}, \bibinfo {author} {\bibfnamefont {M.}~\bibnamefont {Parniak}},\
  and\ \bibinfo {author} {\bibfnamefont {E.~S.}\ \bibnamefont {Polzik}},\
  }\bibfield  {title} {\bibinfo {title} {Phonon counting thermometry of an
  ultracoherent membrane resonator near its motional ground state},\ }\href
  {https://doi.org/10.1364/OPTICA.390939} {\bibfield  {journal} {\bibinfo
  {journal} {Optica}\ }\textbf {\bibinfo {volume} {7}},\ \bibinfo {pages} {718}
  (\bibinfo {year} {2020})}\BibitemShut {NoStop}%
\bibitem [{\citenamefont {Kuhn}\ \emph {et~al.}(2014)\citenamefont {Kuhn},
  \citenamefont {Teissier}, \citenamefont {Neuhaus}, \citenamefont {Zerkani},
  \citenamefont {van Brackel}, \citenamefont {Del{\'e}glise}, \citenamefont
  {Briant}, \citenamefont {Cohadon}, \citenamefont {Heidmann}, \citenamefont
  {Michel}, \citenamefont {Pinard}, \citenamefont {Dolique}, \citenamefont
  {Flaminio}, \citenamefont {Ta{\"\i}bi}, \citenamefont {Chartier},\ and\
  \citenamefont {Le~Traon}}]{kuhn_free-space_2014}%
  \BibitemOpen
  \bibfield  {author} {\bibinfo {author} {\bibfnamefont {A.~G.}\ \bibnamefont
  {Kuhn}}, \bibinfo {author} {\bibfnamefont {J.}~\bibnamefont {Teissier}},
  \bibinfo {author} {\bibfnamefont {L.}~\bibnamefont {Neuhaus}}, \bibinfo
  {author} {\bibfnamefont {S.}~\bibnamefont {Zerkani}}, \bibinfo {author}
  {\bibfnamefont {E.}~\bibnamefont {van Brackel}}, \bibinfo {author}
  {\bibfnamefont {S.}~\bibnamefont {Del{\'e}glise}}, \bibinfo {author}
  {\bibfnamefont {T.}~\bibnamefont {Briant}}, \bibinfo {author} {\bibfnamefont
  {P.-F.}\ \bibnamefont {Cohadon}}, \bibinfo {author} {\bibfnamefont
  {A.}~\bibnamefont {Heidmann}}, \bibinfo {author} {\bibfnamefont
  {C.}~\bibnamefont {Michel}}, \bibinfo {author} {\bibfnamefont
  {L.}~\bibnamefont {Pinard}}, \bibinfo {author} {\bibfnamefont
  {V.}~\bibnamefont {Dolique}}, \bibinfo {author} {\bibfnamefont
  {R.}~\bibnamefont {Flaminio}}, \bibinfo {author} {\bibfnamefont
  {R.}~\bibnamefont {Ta{\"\i}bi}}, \bibinfo {author} {\bibfnamefont
  {C.}~\bibnamefont {Chartier}},\ and\ \bibinfo {author} {\bibfnamefont
  {O.}~\bibnamefont {Le~Traon}},\ }\bibfield  {title} {\bibinfo {title}
  {Free-space cavity optomechanics in a cryogenic environment},\ }\href
  {https://doi.org/http://dx.doi.org/10.1063/1.4863666} {\bibfield  {journal}
  {\bibinfo  {journal} {Appl. Phys. Lett.}\ }\textbf {\bibinfo {volume}
  {104}},\ \bibinfo {pages} {044102} (\bibinfo {year} {2014})}\BibitemShut
  {NoStop}%
\bibitem [{\citenamefont {Purdy}\ \emph {et~al.}(2012)\citenamefont {Purdy},
  \citenamefont {Peterson}, \citenamefont {Yu},\ and\ \citenamefont
  {Regal}}]{purdy_cavity_2012}%
  \BibitemOpen
  \bibfield  {author} {\bibinfo {author} {\bibfnamefont {T.~P.}\ \bibnamefont
  {Purdy}}, \bibinfo {author} {\bibfnamefont {R.~W.}\ \bibnamefont {Peterson}},
  \bibinfo {author} {\bibfnamefont {P.-L.}\ \bibnamefont {Yu}},\ and\ \bibinfo
  {author} {\bibfnamefont {C.~A.}\ \bibnamefont {Regal}},\ }\bibfield  {title}
  {\bibinfo {title} {Cavity optomechanics with {Si}$_3${N}$_4$ membranes at
  cryogenic temperatures},\ }\href
  {https://doi.org/10.1088/1367-2630/14/11/115021} {\bibfield  {journal}
  {\bibinfo  {journal} {New J. Phys.}\ }\textbf {\bibinfo {volume} {14}},\
  \bibinfo {pages} {115021} (\bibinfo {year} {2012})}\BibitemShut {NoStop}%
\bibitem [{\citenamefont {Andrews}(2015)}]{andrews_2015}%
  \BibitemOpen
  \bibfield  {author} {\bibinfo {author} {\bibfnamefont {R.~W.}\ \bibnamefont
  {Andrews}},\ }\emph {\bibinfo {title} {Quantum Signal Processing with
  Mechanical Oscillators}},\ \href@noop {} {Ph.D. thesis},\ \bibinfo  {school}
  {University of Colorado, Boulder} (\bibinfo {year} {2015})\BibitemShut
  {NoStop}%
\bibitem [{\citenamefont {Kampel}\ \emph {et~al.}(2017)\citenamefont {Kampel},
  \citenamefont {Peterson}, \citenamefont {Fischer}, \citenamefont {Yu},
  \citenamefont {Cicak}, \citenamefont {Simmonds}, \citenamefont {Lehnert},\
  and\ \citenamefont {Regal}}]{kampel_improving_2017}%
  \BibitemOpen
  \bibfield  {author} {\bibinfo {author} {\bibfnamefont {N.~S.}\ \bibnamefont
  {Kampel}}, \bibinfo {author} {\bibfnamefont {R.~W.}\ \bibnamefont
  {Peterson}}, \bibinfo {author} {\bibfnamefont {R.}~\bibnamefont {Fischer}},
  \bibinfo {author} {\bibfnamefont {P.-L.}\ \bibnamefont {Yu}}, \bibinfo
  {author} {\bibfnamefont {K.}~\bibnamefont {Cicak}}, \bibinfo {author}
  {\bibfnamefont {R.~W.}\ \bibnamefont {Simmonds}}, \bibinfo {author}
  {\bibfnamefont {K.~W.}\ \bibnamefont {Lehnert}},\ and\ \bibinfo {author}
  {\bibfnamefont {C.~A.}\ \bibnamefont {Regal}},\ }\bibfield  {title} {\bibinfo
  {title} {Improving {Broadband} {Displacement} {Detection} with {Quantum}
  {Correlations}},\ }\href {https://doi.org/10.1103/PhysRevX.7.021008}
  {\bibfield  {journal} {\bibinfo  {journal} {Phys. Rev. X}\ }\textbf {\bibinfo
  {volume} {7}},\ \bibinfo {pages} {021008} (\bibinfo {year}
  {2017})}\BibitemShut {NoStop}%
\bibitem [{\citenamefont {Pl\"{o}{\ss}l}\ and\ \citenamefont
  {Kr\"{a}uter}(1999)}]{waferBonding}%
  \BibitemOpen
  \bibfield  {author} {\bibinfo {author} {\bibfnamefont {A.}~\bibnamefont
  {Pl\"{o}{\ss}l}}\ and\ \bibinfo {author} {\bibfnamefont {G.}~\bibnamefont
  {Kr\"{a}uter}},\ }\bibfield  {title} {\bibinfo {title} {Wafer direct bonding:
  tailoring adhesion between brittle materials},\ }\href
  {https://doi.org/https://doi.org/10.1016/S0927-796X(98)00017-5} {\bibfield
  {journal} {\bibinfo  {journal} {Mater. Sci. Eng. R Rep.}\ }\textbf {\bibinfo
  {volume} {25}},\ \bibinfo {pages} {1} (\bibinfo {year} {1999})}\BibitemShut
  {NoStop}%
\bibitem [{\citenamefont {Caves}(1982)}]{caves_quantum_1982}%
  \BibitemOpen
  \bibfield  {author} {\bibinfo {author} {\bibfnamefont {C.~M.}\ \bibnamefont
  {Caves}},\ }\bibfield  {title} {\bibinfo {title} {Quantum limits on noise in
  linear amplifiers},\ }\href {https://doi.org/10.1103/PhysRevD.26.1817}
  {\bibfield  {journal} {\bibinfo  {journal} {Phys. Rev. D}\ }\textbf {\bibinfo
  {volume} {26}},\ \bibinfo {pages} {1817} (\bibinfo {year}
  {1982})}\BibitemShut {NoStop}%
\bibitem [{\citenamefont {Gao}\ \emph {et~al.}(2007)\citenamefont {Gao},
  \citenamefont {Zmuidzinas}, \citenamefont {Mazin}, \citenamefont {LeDuc},\
  and\ \citenamefont {Day}}]{gao_noise_2007}%
  \BibitemOpen
  \bibfield  {author} {\bibinfo {author} {\bibfnamefont {J.}~\bibnamefont
  {Gao}}, \bibinfo {author} {\bibfnamefont {J.}~\bibnamefont {Zmuidzinas}},
  \bibinfo {author} {\bibfnamefont {B.~A.}\ \bibnamefont {Mazin}}, \bibinfo
  {author} {\bibfnamefont {H.~G.}\ \bibnamefont {LeDuc}},\ and\ \bibinfo
  {author} {\bibfnamefont {P.~K.}\ \bibnamefont {Day}},\ }\bibfield  {title}
  {\bibinfo {title} {Noise properties of superconducting coplanar waveguide
  microwave resonators},\ }\href {https://doi.org/10.1063/1.2711770} {\bibfield
   {journal} {\bibinfo  {journal} {Appl. Phys. Lett.}\ }\textbf {\bibinfo
  {volume} {90}},\ \bibinfo {pages} {102507} (\bibinfo {year}
  {2007})}\BibitemShut {NoStop}%
\bibitem [{\citenamefont {Yu}(2015)}]{yu_quantum_2015}%
  \BibitemOpen
  \bibfield  {author} {\bibinfo {author} {\bibfnamefont {P.-L.}\ \bibnamefont
  {Yu}},\ }\emph {\bibinfo {title} {Quantum {Optomechanics} with {Engineered}
  {Membrane} {Resonators}}},\ \href@noop {} {Ph.D. thesis},\ \bibinfo  {school}
  {University of Colorado, Boulder} (\bibinfo {year} {2015})\BibitemShut
  {NoStop}%
\end{thebibliography}

%

\end{document}